\documentclass[11pt]{article}
\usepackage[margin=1in]{geometry}
\usepackage{times}

\usepackage{xcolor}
\usepackage[colorlinks=true,citecolor=blue,linkcolor=blue]{hyperref}

\hypersetup{
    colorlinks,
    linkcolor={red!50!black},
    citecolor={blue!50!black},
    urlcolor={blue!80!black}
}

\usepackage[]{amsmath,amssymb,amsfonts,latexsym,amsthm,enumerate,xcolor,cite,bbm}
\usepackage{fullpage}
\usepackage[nameinlink, noabbrev, capitalize]{cleveref}
\usepackage{comment}
\usepackage{nicefrac}
\usepackage{centernot}
\crefname{prop}{Proposition}{Propositions}
\crefname{ineq}{inequality}{inequalities}
\creflabelformat{ineq}{#2(#1)#3}

\newtheorem{theorem}{Theorem}
\newtheorem{lemma}{Lemma}
\newtheorem{proposition}{Proposition}
\newtheorem{claim}{Claim}
\newtheorem{fact}{Fact}

\newtheorem{corollary}{Corollary}
\newtheorem{definition}{Definition}

\newtheorem{remark}{Remark}

\crefname{THM}{Theorem}{Theorems}

\usepackage{subcaption}
\usepackage{graphicx}
\usepackage{tikz}
\usepackage{color, colortbl}
\definecolor{LightCyan}{rgb}{0.88,1,1}
\definecolor{Gray}{gray}{0.9}
\usetikzlibrary{positioning}
\usetikzlibrary{decorations.pathmorphing}
\usetikzlibrary{automata,positioning}
\usetikzlibrary{arrows}
\usetikzlibrary{arrows.meta}
\usetikzlibrary{calc}

\newcommand{\E}{\mathbb{E}}
\newcommand{\N}{\mathbb{N}}
\newcommand{\Z}{\mathbf{Z}}
\newcommand{\Zponelong}{\mathbf{Z}^1_{[p_1]}}
\newcommand{\Zptwolong}{\mathbf{Z}^2_{[p_2]}}
\newcommand{\Zone}{\underline{\mathbf{Z}}^1}

\newcommand{\Ztwo}{\underline{\mathbf{Z}}^2}

\newcommand{\C}{\mathbf{C}}
\newcommand{\R}{\mathbf{R}}

\newcommand{\poly}{\operatorname{poly}}

\newcommand{\supp}{\text{support}}
\newcommand{\pr}{{\prime}}
\newcommand{\prpr}{{\prime\prime}}
\newcommand{\U}{\mathbf{U}}
\newcommand{\X}{\mathbf{X}}
\newcommand{\Y}{\mathbf{Y}}

\newcommand{\A}{\mathbf{A}}
\newcommand{\B}{\mathbf{B}}
\newcommand{\W}{\mathbf{W}}
\newcommand{\V}{\mathbf{V}}

\newcommand{\T}{\mathbf{T}}

\newcommand{\guv}{\mathsf{GUV}}
\newcommand{\SL}{\mathsf{SL}}

\newcommand{\zo}{\{0,1\}}

\newcommand{\sExt}{\mathsf{sExt}}

\newcommand{\cond}{\mathsf{Cond}}
\newcommand{\sCond}{\mathsf{sCond}}
\newcommand{\nmCond}{\mathsf{nmCond}}

\newcommand{\Q}{\mathbf{Q}}

\newcommand{\eps}{\varepsilon}

\renewcommand{\S}{\mathbf{S}}
\usepackage{subfiles}
\newcommand{\dobib}{
    \bibliographystyle{alpha}
    \bibliography{references} 
}

\begin{document}
\renewcommand{\dobib}{}

\title{Improved Condensers for Chor-Goldreich Sources}
\author{Jesse Goodman\thanks{Supported by a Simons Investigator Award (\#409864, David Zuckerman).}\\ The University of Texas at Austin\\ \texttt{jpmgoodman@utexas.edu} \and Xin Li\thanks{Supported by NSF CAREER Award CCF-1845349 and NSF Award CCF-2127575.}\\ Johns Hopkins University\\ \texttt{lixints@cs.jhu.edu} \and David Zuckerman\thanks{Supported by NSF Grant CCF-2312573 and a Simons Investigator Award (\#409864).}\\ The University of Texas at Austin\\ \texttt{diz@cs.utexas.edu}}

\begin{titlepage}
\maketitle
\begin{abstract}

One of the earliest models of weak randomness is the Chor-Goldreich (CG) source.
A $(t,n,k)$-CG source is a sequence of random variables
$\mathbf{X}=(\mathbf{X}_1,\dots,\mathbf{X}_t) \sim (\{0,1\}^n)^t$, where each $\mathbf{X}_i$ has min-entropy $k$ conditioned on any fixing of $\mathbf{X}_1,\dots,\mathbf{X}_{i-1}$.
Chor and Goldreich proved that there is no deterministic way to extract randomness from such a source.
Nevertheless, Doron, Moshkovitz, Oh, and Zuckerman showed that there is a deterministic way to condense a CG source into a string with small entropy gap.
They gave applications of such a condenser to simulating randomized algorithms with small error and to certain cryptographic tasks.
They studied the case where the block length $n$ and entropy rate $k/n$ are both constant.

We study the much more general setting where the block length can be arbitrarily large, and the entropy rate can be arbitrarily small. We construct the first explicit condenser for CG sources in this setting, and it can be instantiated in a number of different ways. When the entropy rate of the CG source is constant, our condenser requires just a constant number of blocks $t$ to produce an output with entropy rate $0.9$, say. In the low entropy regime, using $t=\operatorname{poly}(n)$ blocks, our condenser can achieve output entropy rate $0.9$ even if each block has just $1$ bit of min-entropy. Moreover, these condensers have exponentially small error.

Finally, we provide strong existential and impossibility results. For our existential result, we show that a random function is a seedless condenser (with surprisingly strong parameters) for any small family of sources. As a corollary, we get new existential results for seeded condensers and condensers for CG sources. For our impossibility result, we show the latter result is nearly tight, by giving a simple proof that the output of any condenser for CG sources must inherit the entropy gap of (one block of) its input.

\dobib

\end{abstract}
\thispagestyle{empty}
\end{titlepage}
\pagenumbering{arabic}
\newpage

\maketitle

\tableofcontents

\section{Introduction}

Randomness is extremely useful in computing, yet it is difficult or expensive to obtain high-quality randomness.  It is therefore important to understand what can be done with low-quality, or weak, random sources.  Researchers have studied models of weak random sources for decades.  One of the earliest models is the Chor-Goldreich (CG) source \cite{chor1988unbiased}, which generalized the related Santha-Vazirani source \cite{santha1986generating}.

\begin{definition}
The \emph{min-entropy} of a random variable $\X$ is given by $H_\infty(\X) = \min_{x \in \supp(\X)} \log_2 (\frac{1}{\Pr[\X=x]})$. We say $\X$ is an $(n,k)$ source if $\X$ is over $\zo^n$ and has min-entropy $H_\infty(\X) \geq k$.
\end{definition}

\begin{definition}
    A random variable $\X=(\X_1,\dots,\X_t)\sim(\zo^n)^t$ is called a $(t,n,k)$-CG source if for all $i\in[t]$ and all $(x_1,\dots,x_{i-1}) \in (\zo^n)^{i-1}$, it holds that $H_\infty(\X_i | \X_1=x_1,\dots,\X_{i-1}=x_{i-1}) \geq k$. Each $\X_i$ is called a \emph{block}.
\end{definition}

We would like to make use of a CG source knowing only the parameters $t$, $n$, and $k$.
That is, our algorithms should work for all $(t,n,k)$-CG sources; an adversary can pick a $(t,n,k)$-CG source after seeing our algorithm.

The most natural way to use a weak source is to convert it into high quality randomness.
However, generalizing the argument by Santha and Vazirani, Chor and Goldreich showed that it is impossible to deterministically extract even one nearly-uniform bit from a CG source (if $k \le n-1$).
It was therefore accepted by the community that one needed to add more randomness, either in the form of a random seed or a second CG source, to do anything useful.

That changed recently when Doron, Moshkovitz, Oh, and Zuckerman \cite{doron2023almost} showed how to deterministically condense a CG source.
Specifically, they showed how to efficiently output a string $\Z\sim\zo^m$ with small \emph{entropy gap}, defined as $g:=m-H_\infty(\Z)$.
(Strictly speaking, their condenser only outputs a string that is close in variation distance to a distribution with small entropy gap.)

Distributions with small entropy gap are useful in certain applications.
They can be used to simulate algorithms with small error probability.
They are also useful for unpredictability applications in cryptography.
For example, they can be used as the input for a one-way function, and as the key to generate message authentication codes.
Note that seeded extractors are not so useful in these applications, since cycling over seeds is not realistic in a cryptographic setting.
For more on the utility of small entropy gap, see the work of Doron et al.\ \cite{doron2023almost}.

Thus, CG sources are intermediate in the following sense.
A very general source, such as an $(n,k)$-source (which is a CG source with $t=1$), does not admit any deterministic condensing.
Other, less general sources such as affine sources admit deterministic extraction.
CG sources are one of the few models where we can do something extremely useful deterministically, even though we can't extract a single random bit.

Doron et al.\ construct their deterministic condenser by using the CG source to take a random walk on a lossless expander. They show that for any constant block length $n$, constant entropy rate $k/n$, and constant error $\eps$, they can output a string that contains a constant fraction of the original entropy, and has a constant entropy gap.

In this paper, we study whether their results can be generalized to the case of a small number $t$ of long blocks, as well as to subconstant entropy rate.
This is natural and important for a few reasons.
First, small $t$ allows for much more general sources; indeed, 
$t=1$ gives the most general model of an $(n,k)$-source.
It is interesting to find the most general model of a weak source where we can condense deterministically, and CG sources with few blocks seem like a natural candidate.
Second, such CG sources often appear as intermediate objects in extractor constructions,
where they are often called block sources.
Third, long blocks seem even more likely to model natural defective random sources.
It allows for more short-range correlations, and if there aren't too many long-range correlations then it should be a CG source with long blocks.

It appears hard to generalize the techniques of \cite{doron2023almost} to work for long blocks.
This is because known constructions of lossless expanders are not good enough.
First, to obtain results for any entropy rate, Doron et al.\ had to use a two-level construction, where one level relied on a brute force construction of a small lossless expander.  For long block lengths this is infeasible.

Second, for longer blocks, one could try higher degree lossless expanders,
such as those by Guruswami, Umans, and Vadhan \cite{guruswami2009unbalanced}.
However, the price of their extremely good lossless expansion is that the entropy gap becomes too large.

We study deterministic condensers for CG sources with few large blocks, and obtain improved results. Before describing our constructions, we briefly mention that we show the entropy gap $g'$ in the output of any condenser for CG sources must always be at least the entropy gap $g=n-k$ of the last block $\X_t$ of the CG source. Thus, our goal is to ideally achieve $g'=O(g)$, while preserving almost all of the entropy.

\subsection{Our results}\label{subsec:our-results}

\subsubsection*{Explicit constructions}

For our main theorem, we construct the first explicit condenser for Chor-Goldreich sources that can be instantiated with any block length $n$, any min-entropy $k$, and any error $\eps$. We present the general version of our condenser below, and then proceed to highlight two interesting instantiations.

\begin{theorem}[Explicit condensers for CG sources]\label{thm:main:intro:explicit-condensers-CG-sources}
For any \(\alpha>0\), there is a constant \(C\geq1\) such that the following holds. For all \(t,n\in\N\) and \(\delta,\eps>0\), there is an explicit condenser \(\cond:(\zo^n)^t\to\zo^m\) for \((t,n,k=\delta n = n-g)\)-CG sources with output length \(m=k^\pr+g^\pr\), output entropy \(k^\pr\geq(1-\alpha)kt\), output gap \(g^\pr\leq C\cdot(1/\delta)^C\cdot(g+\log(1/\eps))\), and error \(\eps\).    
\end{theorem}

Thus, our explicit condenser is able to preserve \(99\%\) of the min-entropy, while achieving a gap that is only \(\poly(1/\delta)\) times larger than the gap \(g\) of a single block. Moreover, there is no restriction on how the input parameters can be set, and we highlight two interesting settings below.

We first consider the case where the entropy rate \(\delta\) is constant, as in \cite{doron2023almost}. Here, we obtain qualitatively similar results, but ours works for arbitrarily large blocks (instead of constant-sized blocks) and has exponentially small error. Moreover, we only need the number of blocks \(t\) to be a large enough constant to output entropy rate \(0.9\). This constant is a polynomial in \(1/\delta\).

\begin{corollary}\label{cor:main:intro:explicit-condensers-CG-sources:high-entropy}
    For any constant \(\delta>0\), there exists a constant \(C>0\) such that the following holds. For any \(t,n\in\N\), there exists an explicit condenser \(\cond:(\zo^n)^t\to\zo^{k^\pr+g^\pr}\) for \((t,n,k:=\delta n)\)-CG sources, which has output entropy \(k^\pr\geq0.99kt\), output gap \(g^\pr\leq Cn\), and error \(\eps=2^{-n}\).
\end{corollary}

Next, we dramatically improve the entropy requirement from \(k=0.01n\) to just \(k=1\), while the entropy gap grows by just a polynomial factor. As a result, we only need a polynomial number of blocks \(t\) to output entropy rate \(0.9\).

\begin{corollary}\label{cor:main:intro:explicit-condensers-CG-sources:low-entropy}
There exists a universal constant \(C>0\) such that the following holds. For any \(t,n\in\N\), there exists an explicit condenser \(\cond:(\zo^n)^t\to\zo^{k^\pr+g^\pr}\) for \((t,n,k:=1)\)-CG sources, which has output entropy \(k^\pr\geq0.99kt\), output gap \(g^\pr\leq n^C\), and error \(\eps=2^{-n}\).
\end{corollary}

In fact, looking at \cref{thm:main:intro:explicit-condensers-CG-sources}, our condenser can even handle CG sources that have min-entropy $k\ll1$, while achieving error \(\eps\ll2^{-n}\). However, it is worth pointing out that this result is only useful when the stated output gap \(g^\pr\) is less than \(tg\), since this is the original entropy gap in the input CG source.

Overall, as we mentioned, our condensers work for smaller entropy rates and larger blocks than those in \cite{doron2023almost}. Moreover, our condensers achieve exponentially small error, while the constructions in \cite{doron2023almost} have constant error. Nevertheless, their condenser does have some advantages over ours.
First, their condenser works in an online manner, and ours doesn't. Second, they analyzed their condenser for almost-CG sources, and we haven't. That said, our condensers do extend to at least one notion of ``almost,'' as we describe next.

\begin{remark}[Explicit condensers for almost CG sources]\label{rem:main:intro:explicit-condensers-CG-sources:extensions}
Our explicit condensers can also be extended to certain notions of almost CG sources, such as \emph{suffix-friendly} almost CG sources, as defined in \cite{doron2023almost}. This is because such sources can be reduced to standard block sources, simply by grouping together blocks. While such a reduction will produce uneven block lengths (unlike standard CG sources), our constructions can easily be adapted to handle this more general setting.
\end{remark}

\subsubsection*{Existential results}

We complement our explicit constructions with strong existential results. For our main existential result, we show that a random function is a seedless condenser (with surprisingly strong parameters) for any small family of sources. Throughout, we use capital letters to denote exponential versions of lower-case letters.\footnote{For example, \(L:=2^\ell, K:=2^k\), and so on.}

\begin{theorem}[Existential results for any small family]\label{thm:main:intro:existential-results:seedless-condenser:small-family}

There exist universal constants \(C,c>0\) such that the following holds. Let \(\mathcal{X}\) be a family of \((n,k)\)-sources. For any \(\ell\in[0,k]\) and \(g>0\) such that \(m:=k-\ell+g\) is an integer, and any \(\eps\in(0,1]\), the following holds. If \(|\mathcal{X}|\leq2^{c\eps K\psi}\),
where
\[
\psi=\max\left\{g-\frac{1}{\lfloor L\rfloor}\log(1/\eps)-C,\quad g-\frac{1}{\lfloor L\rfloor}\log(C2^gg/\eps)\frac{C2^g}{g}\right\},
\]
then there exists a condenser \(\cond:\zo^n\to\zo^m\) for \(\mathcal{X}\) with loss \(\ell\), gap \(g\), and error \(\eps\).
\end{theorem}

The above can be viewed as a condenser version of the classic result that there exist good seedless extractors for any small family of sources. In fact, it strictly generalizes it.\footnote{This is because the extractor case corresponds to the case where the error is \(\eps/2\) and the gap is \(g=\eps/2\), as this implies an error of \(\eps\) and a gap of \(0\).} Overall, this result shows that condensers can handle much larger families of sources than extractors, while outputting much more of the original min-entropy. In particular, the classical existential result for extractors only works for families of size \(2^{\Omega(\eps^2K)}\), and requires the extractor to lose \(\ell=2\log(1/\eps)\) bits of min-entropy. The above result shows that condensers can handle families of size up to \(2^{\Omega(g\eps K)}\), provided the gap is of the form \(g=O(\frac{1}{L}\log(1/\eps))\). This means that allowing just \(g=1\) bit of gap can significantly increase the size of the family that can be handled, while decreasing the loss to \(\ell=\log\log(1/\eps)+O(1)\). Furthermore, the loss can be decreased all the way to \(\ell=0\), at the price of a slightly larger gap \(g = O(\log(1/\eps))\).\footnote{In fact, note that this gap can be reduced to \(g=1\log(1/\eps)+O(1)\) if we only wish to handle families of size \(2^{\Omega(\eps K)}\).}

As an immediate corollary, we get improved existential results for seeded condensers.

\begin{corollary}[Existential results for seeded condensers]\label{thm:main:intro:existential-results:seeded-condenser}

There is a universal constant \(C\geq1\) such that the following holds. There exists a seeded condenser \(\sCond:\zo^n\times\zo^d\to\zo^m\) for \((n,k)\)-sources with output length \(m=k+d-\ell+g\), error \(\eps\), loss \(\ell\), and gap
\[
g\leq\frac{1}{\lfloor L\rfloor}\log(1/\eps) + C,
\]
provided that \(d\geq\log(\frac{n-k}{\eps})+C\).
\end{corollary}

We note that we can improve the seed length requirement to \(d\geq\log(\frac{n-k}{\eps g})\), if one is willing to increase the gap to \(g=\frac{2}{\lfloor L\rfloor}\log(1/\eps)+C\).\footnote{Moreover, \cref{thm:main:intro:existential-results:seedless-condenser:small-family} can be used to give a more general version of the above result, which recovers known existential results for seeded extractors, but we only present the above version for simplicity.} Previously, a work of Aviv and Ta-Shma \cite{aviv2019entropy} established similar existential bounds for seeded condensers, but in the lossless regime \(\ell=0\) their result required entropy gap \(g=O(\frac{\log(1/\eps)}{\eps})\), while we only require \(g=O(\log(1/\eps))\).\footnote{It is worth noting that they focused on \emph{strong} seeded condensers, while we focus on standard seeded condensers, since our result is just a corollary of our existential seedless condensers (\cref{thm:main:intro:existential-results:seedless-condenser:small-family}), for which there is no notion of ``strong.'' However, it should be relatively straightforward to extend our result to obtain strong seeded condensers, using standard tricks.}

For our last existential result, we show the existence of good condensers for Chor-Goldreich sources. Since the number of such sources is very large, we cannot apply \cref{thm:main:intro:existential-results:seedless-condenser:small-family} to obtain this result. Instead, we show that one can iteratively condense CG sources using seeded condensers (in the spirit of \cite{nisan1996randomness}, but with a correlated seed). Then, we plug in the seeded condensers from \cref{thm:main:intro:existential-results:seeded-condenser} to obtain the following, which we take some time to digest immediately after.

\begin{corollary}[Existential results for CG sources]\label{thm:main:intro:existential-results:CG-sources}

There is a constant \(C\geq1\) such that the following hold.

\begin{itemize}
    \item \textbf{Two blocks:} There exists a condenser \(\cond:(\zo^n)^2\to\zo^m\) for \((2,n,k=n-g)\)-CG sources with output length \(m=2k-\ell+g\), error \(\eps\), loss \(\ell\), and gap
    \[
    g^\pr\leq g + \frac{1}{\lfloor L\rfloor}\left(g+\log(1/\eps)\right) + C,
    \]
    provided that \(k\geq\log(g/\eps)+C\).

    \item \textbf{More than two blocks:} There exists a condenser \(\cond:(\zo^n)^t\to\zo^m\) for \((t,n,k=n-g)\)-CG sources with output length \(m=kt+g^\pr\), error \(\eps\), loss \(\ell=0\), and gap
    \[
    g^\pr\leq g+2^{C(\log^*t)^2}\cdot(g+\log(1/\eps)+C\log^*t),
    \]
    provided that \(k\geq\log(g/\eps)+C\).
    
    On the other hand, if \(m=kt-\ell+g^\pr\) and the loss is \(\ell=2(\log^*t)^2\), then one can obtain gap \[
    g^\pr\leq g + C\cdot2^{-\log^*t}\cdot(g+\log(1/\eps))+C\log^*t,
    \]
    provided that \(k\geq\log(g/\eps)+2\log^*t+C\).
\end{itemize}    
\end{corollary}

Thus, it is possible to condense Chor-Goldreich sources, even when there are just \(t=2\) blocks with logarithmic min-entropy. In the multi-block setting \(t>2\), we obtain a full tradeoff between the loss \(\ell\) and gap \(g^\pr\) (\cref{subsec:existential-block-source-condensers}), but only highlight the extreme regimes above, for simplicity. In particular, the above shows that in the \emph{lossless} regime \(\ell=0\), one can condense from multi-block CG sources with a modest multiplicative blow-up of \(2^{O((\log^\ast t)^2)}\) in the gap (where \(\log^\ast\) denotes the iterated logarithm). On the other hand, if one is willing to lose a little more min-entropy, this blow-up can be improved to an additive \(O(\log^\ast t)\). Moreover, we note that at the expense of a significantly greater loss in min-entropy, it is possible to blow-up the gap by an additive constant (with no dependence on \(t\)), and refer the reader to \cref{subsec:existential:multi-block:iterative-condensing} for more.

\subsubsection*{Impossibility results}

Finally, we show a lower bound, which says that the gap in the CG source must propagate to the output.

\begin{theorem}[Impossibility results for CG sources]\label{thm:main:intro:impossibility-results:CG-sources}
Fix any \(0\leq g\leq m\leq n\in\N\) and \(\eps\in[0,1)\).
    For every function \(\cond:(\zo^n)^t\to\zo^m\), there exists a \((t,n,n-g)\)-CG source \(\X\) such that \(\cond(\X)\) is \(\eps\)-far from every \((m,m-g+c_\eps)\)-source, where \(c_\eps:=\log(\frac{1}{1-\eps})\).\footnote{We remark that \(c_\eps\) is an unavoidable term, since sources with \(0\) min-entropy are still \(\eps\)-close to min-entropy \(c_\eps\).}
\end{theorem}

This impossibility result is a strengthening of the fact that it is impossible to condense general \((n,k)\)-sources, and was independently obtained by Chattopadhyay, Gurumukhani, and Ringach \cite{CGR24}.\footnote{Beyond this impossibility result, there is little overlap between our two works, which will both appear at FOCS 2024. This is because we focus on explicit condensers for CG sources, whereas \cite{CGR24} focuses on existential and impossibility results for almost CG sources, and other more general models.}

\paragraph{Organization} The rest of this paper is organized as follows. We start with an overview of our techniques in \cref{sec:overview}. Then, after some preliminaries in \cref{sec:prelims}, we provide a collection of (mostly new) tools and tricks around block sources in \cref{sec:block-source-basics}, which we'll use throughout the paper. In \cref{sec:explicit-constructions}, we provide our main explicit condenser for Chor-Goldreich sources, and prove \cref{thm:main:intro:explicit-condensers-CG-sources}. Following this, we provide our existential results in \cref{sec:existential} and our impossibility results in \cref{sec:impossibility}. Finally, we conclude with some open problems in \cref{sec:conclusions}.

\dobib

\section{Overview of our techniques}\label{sec:overview}

To begin, we give an informal overview of the techniques used in our constructions and proofs.

\subsection{Explicit constructions}

As discussed in the introduction, it seems difficult to extend the techniques of Doron et al. \cite{doron2023almost} to obtain a condenser that can handle CG sources with long blocks. This is because their construction involves the use of excellent lossless expanders, which we don't know how to explicitly construct. They get around this problem by considering constant block length \(n\), which allows them to obtain the lossless expanders via a brute-force search. But since we want to work with a larger block length, this is no longer possible. Thus, we need a new idea.

\subsubsection*{High-level plan}
Our idea is to return to a classical paradigm in the construction of seedless extractors for independent sources, and show that it can be adapted to get seedless condensers for Chor-Goldreich sources. Intuitively, this makes sense given that CG sources are a natural generalization of independent sources, and condensers are a natural generalization of extractors.

The well-known paradigm that we use involves taking a single independent source, expanding it into a table where one row is uniform (or has high entropy), and gradually collapsing that table (with the help of the other independent sources) until all that remains is that one good row \cite{ta1996extracting,rao2009extractors,barak20122,li2013extractors,cohen2016local}. Our goal is to extend this paradigm so that it still works even if the sources are not truly independent sources, but blocks coming from a CG source.

In order to make this happen, the core tool that we use is a simple observation, which says that every seeded condenser (and thus seeded extractor) still works even if its seed is ``CG-correlated'' with the source. In more detail, suppose that a seeded condenser was expecting to receive an \((n,k)\)-source \(\X\) and independent seed \(\Y\sim\zo^d\) as input, but instead received an \((n,k)\)-source \(\X\) and a correlated seed \(\Y\sim\zo^d\), which is only guaranteed have min-entropy \(d-g\) on each fixing of \(\X\). The core tool we use says that the error of the seeded condenser blows up from \(\eps\to\eps2^g\), while its output gap \(g^\pr\) blows up from \(g^\pr\to g^\pr+g\). This key observation has appeared a few times in prior work, with slightly weaker parameters \cite{ben2019two} or in a slightly different context \cite{ball2022randomness}. We record it as \cref{lem:seeded-condensers-work-on-block-sources}.

By combining the paradigm for extracting from independent sources with the above tool, we now have a very high-level plan for condensing CG sources with long blocks. However, several challenges arise along the way, since the known tools for collapsing the table (such as non-malleable extractors and mergers)\footnote{Here, a non-malleable extractor can roughly be thought of as a seeded extractor \(\sExt:\zo^n\times\zo^d\to\zo^m\) that offers an additional ``robustness'' guarantee. This robustness guarantee says that the output of \(\sExt\) not only looks uniform, but also looks \emph{independent} of (an output produced by) an additional call to \(\sExt\) on the same source and a correlated seed. Such an object can be used to break the correlation between pairs of rows in the table, while keeping the good row looking uniform (or high-entropy).} cannot be ported over in a black-box manner. Instead, we must construct our own non-malleable condensers (for CG sources) from scratch, and we do so via a simple composition of seeded extractors. With this high-level plan in mind, we proceed with a more detailed description of our condenser.

\subsubsection*{A detailed description of our condenser}

Given a CG source (or block source)\footnote{Recall that a block source is a generalization of a CG source in that the blocks need not have the same length.} \(\X=(\X_1,\dots,\X_t)\), we work backwards starting with $\X_t$, and try to extract (almost) all of the entropy out of $\X$ while preserving a small entropy gap. To get the condenser, we first convert the block $\X_t$ into a \emph{somewhere high-entropy source}, which is a table with some number of rows such that at least one row has high entropy rate. If the entropy rate of $\X_t$ is relatively large (e.g., any constant), we can use well-known somewhere condensers \cite{barak2010simulating,raz2005extractors}, which produce a small (constant) number of rows with exponentially small error.

Our next step is to use the other blocks to reduce the number of rows in this table, while preserving almost all of the entropy. If the blocks were independent, prior work shows that we could eventually reduce the table to a single row that is close to uniform (which gives an extractor). However, when we only have a block source, for technical reasons we'll soon explain, this is no longer possible and eventually we get one row with large entropy and small entropy gap. This gives a condenser.

\paragraph{The non-malleable condenser} The key ingredient in achieving this is a new merger (or \emph{non-malleable condenser}, similar to those defined in \cite{DBLP:conf/stoc/Li12, DBLP:conf/tcc/Li15}), that we design to merge two rows in the table while using a few additional blocks. Our final condenser is obtained by repeatedly applying this merger, until the number of rows in the table reduces to one. To illustrate our ideas, let's consider the simplified case of merging two rows in the table (say $\Y_1, \Y_2$), where at least one row is uniform (but we don't know which one). Our basic merger works as follows. First take two other blocks (say $\X_1, \X_2$), and take two slices: $\Z_1$ from $\Y_1$ and $\Z_2$ from $\Y_2$. We use a standard seeded extractor $\sExt$ which works even when the seed only has entropy rate $0.9$ \cite{guruswami2009unbalanced}, and compute $\W_1=\sExt(\X_2, \Z_1), \W_2=\sExt(\X_2, \Z_2)$. In this step we make sure that the sizes of these random variables satisfy
\[|\W_1| \gg |\W_2| \gg |\Z_2| \gg |\Z_1|.\]
Next we apply the seeded extractor again and compute $\S_1=\sExt(\X_1, \W_1)$ and $\S_2=\sExt(\X_1, \W_2)$, where $|\S_1|=|\S_2|$. Finally, we output  $\V= \S_1 \oplus \S_2$, where $\oplus$ denotes bit-wise XOR.

\paragraph{The analysis} For the analysis, let us first consider the case where all the blocks are \emph{independent}. We have two cases. If $\Y_1$ is the uniform row, then $\Z_1$ is uniform and therefore $\W_1$ is uniform (ignoring the error of the extractor). Since $|\W_1| \gg |\W_2| \gg |\Z_2|$, we can fix $\Z_2$ and $\W_2$, and conditioned on this fixing, (with high probability) $\W_1$ still has entropy rate say $>0.9$. Notice at this point, $\X_1$ is still independent of $\W_1$. Now we can further fix $\S_2=\sExt(\X_1, \W_2)$, which is a deterministic function of $\X_1$ since $\W_2$ is fixed. Conditioned on this fixing, $\X_1$ still has good entropy as long as the size of $\S_2$ is not too large. Hence, $\S_1=\sExt(\X_1, \W_1)$ is close to uniform and so is $\V= \S_1 \oplus \S_2$.

In the other case, $\Y_2$ is the uniform row, and thus $\Z_2$ is uniform.\footnote{Note that we are only analyzing the case where at least one of \(\Y_1,\Y_2\) are uniform, since this is true for some pair of consecutive rows in the table, and we don't care what happens when we merge other pairs of consecutive rows.} Now we first fix $\Z_1$ and $\W_1$. Notice that since $|\Z_2| \gg |\Z_1|
$, conditioned on this fixing $\Z_2$ still has entropy rate $>0.9$. Furthermore when $\Z_1$ is fixed, $\W_1$ is a deterministic function of $\X_2$. Thus conditioned on the further fixing of $\W_1$, $\X_2$ and $\Z_2$ are still independent while $\X_2$ still has good entropy as long as the size of $\W_1$ is not too large. Therefore $\W_2=\sExt(\X_2, \Z_2)$ is close to uniform even conditioned on $\W_1$. Now, we can further fix $\S_1=\sExt(\X_1, \W_1)$, which is a deterministic function of $\X_1$ since $\W_1$ is fixed. Conditioned on this fixing, $\X_1$ is still independent of $\W_2$ and still has good entropy as long as the size of $\S_1$ is not too large. Hence, $\S_2=\sExt(\X_1, \W_2)$ is close to uniform and so is $\V= \S_1 \oplus \S_2$. 

\paragraph{Extending the analysis to correlated blocks} Now let us see what happens if the blocks are not independent, but rather form a block source. We will again use the core observation that for any seeded extractor, if the seed and the source form a block source, then the output of the extractor becomes a source that suffers roughly the same entropy gap as the seed.

With this property in hand, our previous analysis can go through with a few modifications. Most importantly, some of the random variables in $\{\W_1, \W_2, \S_1, \S_2, \V\}$ will no longer be uniform, since the entropy gap of the seed will be inherited in the output when we apply a seeded extractor. In addition, we need to set the errors in the extractors appropriately so that the blow up factor $2^g$ in the error can be absorbed. Finally, in the analysis, when we fix certain random variables (e.g., $\Z_1, \Z_2, \W_1, \W_2$), this may affect other blocks besides the block from which the random variable is produced, because now the blocks are no longer independent. However, as long as we keep the sizes of the random variables relatively small compared to the entropy in each block, after conditioning the blocks still have enough entropy left and thus they are still (close to) a block source.

\paragraph{Tracking the gap} Notice that if initially the ``good" row in $\Y_1, \Y_2$ has some entropy gap $g^\pr$, then the final entropy gap $g^\prpr$ of our basic merger will be a constant factor larger than $g^\pr$, due to our conditioning argument and the requirement that the seed used in the extractor has entropy rate $>0.9$. Therefore when we repeatedly apply the basic merger, the entropy gap will increase by a constant factor at each step. As a result, the final entropy gap will become larger than $g$, the entropy gap of each block in the CG source. To see this, consider the case where $k=\delta n$ for some constant $\delta>0$ and we start with a somewhere high-entropy condenser as in \cite{barak2010simulating,raz2005extractors}. If we boost the entropy rate to say $0.99$, then the initial entropy gap of the good row will be $\poly(\delta)g$, since the length of each row in the table is $\poly(\delta)n$. However, the table itself also has $\poly(1/\delta)$ rows, so by the above analysis, we eventually get an entropy gap of $C^{\log(\poly(1/\delta))} \poly(\delta) g = \poly(1/\delta) g$, since each time we use the non-malleable condenser, we halve the number of rows in the table, but blow up the gap by a constant factor \(C\). Thus if \(\delta\) was originally a constant, the final entropy gap is \(\poly(1/\delta)g = O(g)\). Note that this is as expected since our impossibility result shows that an entropy gap of $g$ is necessary.

This also results in another modification we must make in our constructions. Specifically, when the entropy gap becomes large in the process of repeated merging, in order to obtain a seed for the extractor with entropy rate $>0.9$, it is no longer enough to just use the entropy from one block. Rather, at this point we need to use the concatenation of several blocks as the source in the seeded extractor to get sufficient entropy. In doing so, we need a slightly larger seed length over time, but this does not drastically change any of the parameters. In fact, since the seed length must grow anyway, we can (for free) force the error of each merging step to be half the error of the prior step, resulting in a geometric series of errors. As a result, the overall error of the condenser is simply the error of the first (somewhere-condensing) step, which is just \(2^{-\poly(\delta)n}\) if we start with a \((t,n,\delta n)\)-CG source.

\paragraph{Pre- and post-processing}

Finally, we have just two loose ends that we need to tie up. First, the discussion above assumed that we started with a \((t,n,\delta n)\)-CG source for some constant \(\delta>0\), so that we could apply the somewhere-condensers of \cite{barak2010simulating,raz2005extractors} to create the initial table. However, what if we want to condense from CG sources with sub-constant \(\delta\)? As it turns out (and is well-known), these somewhere-condensers can actually handle an input source of min-entropy \(\delta n = n^{0.99}\), and thus the construction can still be applied even if we start off with a CG-source with \(\delta n = n^{0.99}\). More importantly, if we start off with a \((t,n,\delta n)\)-CG source with \(\delta n \ll n^{0.99}\), we can always turn it into a \((t/b,nb,\delta nb)\)-CG source with \(\delta nb\geq(nb)^{0.99}\), via a \emph{pre-processing step}, where we simply group the blocks into ``super-blocks'' containing sufficiently many blocks \(b\) each. This will slightly impact the parameters of our condenser, but not enough to be noticeable (when compared to the impact of the other steps).

Second, the discussion above gave a detailed overview of how we can condense the CG source into a string \(\Z\) with high entropy rate, but what if this was done using a relatively small number of blocks in the CG source, and most of the entropy in the CG source still remains (i.e., in the unused blocks \(\X^\star\))? To deal with this, we append a simple \emph{post-processing} step to our condenser. As it turns out, since we already have obtained a (perhaps short) string with high entropy rate, it is relatively easy to condense the rest of the min-entropy out of the CG source. Indeed, since the entropy rate is so high, we can use our core tool that a seeded extractor can handle CG-correlated seeds, and suffer very little loss. Thus, a first attempt to get the rest of the min-entropy out may involve calling a seeded extractor with \(\X^\star\) as the source and \(\Z\) as the seed. However, it may be the case that \(\X^\star\) is extremely long compared to \(\Z\), which would make this approach fail. Instead, the right approach is to use classic block-source extraction framework of Nisan and Zuckerman \cite{nisan1996randomness}, or rather a slight generalization that works for condensers and a CG-correlated seed. With this approach, we can successfully condense the rest of the min-entropy out of \(\X^\star\), even if \(\Z\) is very tiny in comparison.

\subsection{Existential results}

Next, we briefly discuss the ideas that go into our existential results. As a reminder, our main result shows that there exist great seedless condensers for any small enough family \(\mathcal{X}\) of sources. As a corollary (i.e., by picking the appropriate family \(\mathcal{X}\)), we immediately get our existential results for seeded condensers. Then, by plugging these seeded condensers into the (slight generalization of the) block-source extraction framework described in the paragraph above, we immediately get our existential results for CG sources. Thus, all that remains is to show that there exist great seedless condensers for any small family of sources.

In order to show the above, we show that a random function \(f:\zo^n\to\zo^m\) is, with high probability, a great seedless condenser for a single source \(\X\sim\zo^n\) (and apply a union bound over all \(\X\in\mathcal{X}\)). As it turns out, if one wishes to get good parameters, this is quite nontrivial to show.

The overall approach is as follows. First, we recall that a random variable \(f(\X)\sim\zo^m\) is \(\eps\)-close (in statistical distance) to min-entropy \(k^\pr\) iff for every \(S\subseteq\zo^m\), it holds that
\[
\Pr[f(\X)\in S]\leq |S|\cdot2^{-k^\pr}+\eps.
\]
Thus, it is tempting to fix a set \(S\), show that the above is true with high probability over \(f\), and then union bound over all \(S\subseteq\zo^m\). However, there are simply too many sets \(S\) for this to yield good parameters.

As a second approach, one may recall a classical lemma (in, e.g., \cite[Lemma 6.2]{guruswami2009unbalanced}), which says that if you want to ensure that \(f(\X)\) is \(\eps\)-close to min-entropy \(k^\pr\), it is enough to show that there exists no \emph{small} set \(S\subseteq\zo^m\) of size \(\leq \eps 2^{k^\pr}\) such that
\[
\Pr[f(\X)\in S]\geq\eps.
\]
This is much better, since we have greatly reduced the number of sets \(S\subseteq\zo^m\) that we ultimately need to union bound over. However, we can still do even better.

The key realization (which is inspired by existence proofs for lossless condensers) is that we can specify \(S\subseteq\zo^m\) by instead specifying its \emph{preimage} \(f^{-1}(S)\). Thus, instead of counting sets from \(\zo^m\) (for the union bound), we can count sets from \(\supp(\X)\). This is much better when \(\supp(\X)\ll 2^m\), which happens when we are targetting a regime where the gap of the condenser will need to exceed the loss (e.g., the lossless regime) and \(\X\) is flat.\footnote{As a reminder, a flat source is uniform over its support.} But what if \(\X\) is not flat? When talking about \emph{seeded condensers}, one can often assume that \(\X\) is flat for free. But this is not true for \emph{seedless condensers} (for an explanation why, see \cref{subsec:seedless-condenser-any-small-family}).

In order to deal with an \((n,k)\)-source \(\X\) that may not be flat, we break its support into two parts \(X_1,X_2\). We pick some threshold \(T\) and let \(X_1\) contain the heaviest \(T\) elements in \(\supp(\X)\), while \(X_2\) contains the rest. Then, instead of analyzing the performance of \(f\) on the entirety of \(\X\), we analyze it on the subdistributions of \(\X\) over \(X_1\) and \(X_2\) (and make sure that the images of \(X_1\) and \(X_2\) do not interact too much). If we pick the threshold \(T\) correctly, then the subdistribution on \(X_1\) will look roughly flat, while the subdistribution on \(X_2\) has much higher entropy than \(\X\). This is exactly what we want, because the former allows us to safely count tests via their preimages in \(X_1\), while the latter allows us to safely count tests by picking them from \(\zo^m\) (since \(f\) will be nowhere close to the lossless regime for the subdistribution on \(X_2\), as it has much higher min-entropy than \(\X\)). All that remains is to ensure that the images of \(X_1\) and \(X_2\) do not interact too much, which follows without too much additional trouble.

\subsection{Impossibility results}

Finally, our impossibility result for condensing CG sources is a simple extension and generalization of the well-known impossibility result for extracting from CG sources \cite{chor1988unbiased}, which uses backwards induction on the blocks. Indeed, the latter can be viewed as a special case of the former.

\dobib

\section{Preliminaries}\label{sec:prelims}

Before we dive into our main proofs, we collect some preliminaries that will be used throughout the paper.

\paragraph{Notation} We adopt the convention that capital letters denote the exponential version of lower-case letters. For example, \(N:=2^n,D:=2^d\), and so on. Given a string \(x\in\zo^n\), we let \(x_i\) denote the value it holds at its \(i^\text{th}\) index, and for a set \(S\subseteq[n]\) we let \(x_S\) denote \((x_i)_{i\in S}\) (concatenated in increasing order of \(i\)). We also use \(x_{<i}\) as shorthand for \(x_{[1,i-1]}\), and we define \(x_{\leq i},x_{>i}\), and \(x_{\geq i}\) similarly. All logs are base \(2\), unless otherwise noted. In particular, we write \(\log():=\log_2()\) and \(\ln():=\log_e()\).

\subsection{Probability}

We use bold letters, such as \(\X\), to refer to random variables (which we often call \emph{sources}). We let \(\U_n\) denote the uniform random variable over \(\zo^n\), and more generally say that a random variable \(\X\) is \emph{flat} if it is uniform over its support. Furthermore, if \(\supp(\X)\subseteq V\), we say that \(\X\) is supported on \(V\) and denote this by \(\X\sim V\). Finally, for any two random variables \(\X,\Y\) defined over the same space, and \(y\in\supp(\Y)\), we let \((\X\mid\Y=y)\) denote a random variable that hits \(x\) with probability \(\Pr[\X=x\mid\Y=y]\).

\subsubsection*{Statistical distance}
Next, we introduce a standard way to measure the distance between two random variables.

\begin{definition}[Statistical distance]
The \emph{statistical distance} between random variables \(\X,\Y\sim V\) is defined
\[
|\X-\Y|:=\max_{S\subseteq V}\Pr[\X\in S]-\Pr[\Y\in S]=\frac{1}{2}\sum_{v\in V}|\Pr[\X=v]-\Pr[\Y=v]|.
\]
We say that \(\X,\Y\) are \(\eps\)-close and write \(\X\approx_{\eps}\Y\) iff \(|\X-\Y|\leq\eps\). If \(\X,\Y\) are \(0\)-close then we write \(\X\equiv\Y\). If \(\X,\Y\) are not \(\eps\)-close, we say they are \(\eps\)-far and write \(\X\not\approx_\eps\Y\).
\end{definition}

Statistical distance is a metric, which means that it satisfies the triangle inequality.

\begin{fact}[Triangle inequality]\label{fact:triangle-inequality}
For any random variables \(\X,\Y,\Z\sim V\),
\[
|\X-\Z|\leq|\X-\Y|+|\Y-\Z|.
\]
\end{fact}

Throughout this paper, we will often want to bound the statistical distance between random variables. A classic tool for this is the following.

\begin{fact}[Data-processing inequality]\label{fact:data-processing}
For any random variables \(\X,\Y\sim V\) and function \(f:V\to W\),
\[
|\X-\Y|\geq|f(\X)-f(\Y)|.
\]
\end{fact}

Another tool that is useful for bounding statistical distance is the coupling lemma:

\begin{lemma}[Coupling lemma]\label{lem:coupling}
For any two random variables \(\X,\Y\sim V\), the following holds. For every pair of jointly distributed random variables \((\X^\pr,\Y^\pr)\) with \(\X^\pr\equiv\X\) and \(\Y^\pr\equiv\Y\), it holds that
\[
|\X-\Y|\leq\Pr[\X^\pr\neq\Y^\pr].
\]
Moreover, there exists a pair of jointly distributed random variables \((\X^\star,\Y^\star)\) with \(\X^\star\equiv\X\) and \(\Y^\star\equiv\Y\) such that
\[
|\X-\Y|=\Pr[\X^\star\neq\Y^\star].
\]
\end{lemma}

\subsubsection*{Convex combinations}

We will also frequently use the notion of convex combinations. We say \(\X\) is a convex combination of distributions from \(\mathcal{Y}\) if there exist probabilities \(\{p_i\}\) summing to \(1\) and distributions \(\Y_i\in\mathcal{Y}\) such that \(\X=\sum_ip_i\Y_i\), meaning that \(\X\) samples from \(\Y\) with probability \(p_i\). The following fact will be quite useful.

\begin{fact}\label{fact:convex-combo-error-to-closeness}
Let \(\X\sim V\) and \(\A\sim W\) be (arbitrarily correlated) random variables, and let \(\mathcal{X}\) be a family of random variables over \(V\). Suppose that \(\Pr_{a\sim\A}[\X\notin\mathcal{X}\mid\A=a]\leq\eps\). Then \(\X\) is \(\eps\)-close to a convex combination of random variables from \(\mathcal{X}\).
\end{fact}
\begin{proof}
    For every fixed \(a\) such that \((\X\mid\A=a)\in\mathcal{X}\), define \(\Y^a:=(\X\mid\A=a)\). For all other \(a\), define \(\Y^a\) to be an arbitrary member of \(\mathcal{X}\). Consider the convex combination \(\Y^\star:=\sum_a\Pr[\A=a]\cdot\Y^a\). Clearly, it is a convex combination of distributions from \(\mathcal{X}\). It is also straightforward to verify \(\Y^\star\approx_\eps\X\).
\end{proof}

\subsubsection*{Concentration bounds}

Finally, we will use the following version of the multiplicative Chernoff bound, which works even if we only know an upper bound on the expectation of the random variable of interest.

\begin{theorem}[Chernoff bound]\label{thm:chernoff-bound}
Let \(\X_1,\dots,\X_n\) be a sequence of independent random variables, where each \(\X_i\sim\{0,p_i\}\) for some \(p_i\in[0,1]\), and let \(\X:=\sum_i\X_i\) denote their sum. Then for any \(\delta>0\) and \(\mu\geq\E[\X]\),
\[
\Pr[\X\geq(1+\delta)\mu]\leq\left(\frac{e^\delta}{(1+\delta)^{(1+\delta)}}\right)^\mu.
\]
\end{theorem}

\subsection{Entropy}\label{subsec:entropy}

In extractor theory, the standard way to measure the randomness content of a source is via its \emph{min-entropy}.

\begin{definition}[Min-entropy]
    The \emph{min-entropy} of a random variable \(\X\sim\zo^n\) is defined
    \[
    H_\infty(\X):=\min_{x\in\supp(\X)}\log\left(\frac{1}{\Pr[\X=x]}\right),
    \]
    while its min-entropy \emph{gap} is defined as \(n-H_\infty(\X)\).\footnote{For convenience, from here on out, whenever we say ``entropy'' we really mean ``min-entropy.''}
\end{definition}

It is often the case that a random variable \(\X\sim\zo^n\) does not exactly have high min-entropy, but is (statistically) close to a random variable that does. In many applications, this is just as good as \(\X\) having high min-entropy itself, and as a result, this notion has earned its own name: \emph{smooth min-entropy}. In order to formally introduce this definition, we first let \(\mathcal{B}_\eps(\X)\) denote the set of random variables \(\Y\sim\zo^n\) that are \(\eps\)-close to \(\X\) in statistical distance. Then, we define smooth min-entropy as follows.

\begin{definition}[Smooth min-entropy]\label{def:smooth-min-entropy}
    The \emph{\(\eps\)-smooth min-entropy} of a random variable \(\X\sim\zo^n\) is defined as
    \[
H_\infty^\eps(\X):=\sup_{\Y\in\mathcal{B}_\eps(\X)}H_\infty(\Y)=\max_{\Y\in\mathcal{B}_\eps(\X)}H_\infty(\Y).
    \]
\end{definition}

Looking at this definition, a few remarks are in order. First, we note that we were able to replace the supremum with a maximum due to standard tools from analysis.\footnote{In particular, one can argue that \(\mathcal{B}_\eps(\X)\) is closed and bounded, and by the Heine-Borel theorem for finite-dimensional normed vector spaces, it is also compact. Then, since the min-entropy function \(H_\infty()\) is continuous, \(H_\infty(\mathcal{B}_\eps(\X))\) is also compact (and therefore closed and bounded). It follows that \(\sup H_\infty(\mathcal{B}_\eps(\X))\in H_\infty(\mathcal{B}_\eps(\X))\), allowing us to replace \(\sup\) with \(\max\).} (In doing so, we know there always exists some distribution \(\Y\approx_\eps\X\) such that \(H_\infty(\Y)=H^\eps_\infty(\X)\).) Next, in order to highlight that smooth min-entropy is a weaker notion than standard min-entropy (and thus easier to obtain), we point out that there are other well-studied notions of entropy that imply much better guarantees on the former than the latter.\footnote{Consider the R\'{e}nyi entropy of a random variable, defined \(H_2(\X):=\log\left(\frac{1}{\sum_x\Pr[\X=x]^2}\right)\). Comparing this to min-entropy, we have \(H_\infty(\X)\geq\frac{1}{2}H_2(\X)\), but if we compare this to smooth min-entropy, it is known that \(H_\infty^\eps(\X)\geq H_2(\X)-\log(1/\eps)\) \cite[Lemma 4.2]{renner2004smooth}. A similar connection was used in \cite{doron2023almost} in order to use the \(\ell_q\) norm as a proxy for smooth min-entropy.} Finally, we mention two strange artifacts of the above definition, which distinguish it from other notions of entropy: First, note that a constant random variable has \(\eps\)-smooth min-entropy \(c_\eps:=\log(\frac{1}{1-\eps})\), which is \(>0\) for \(\eps>0\). Second, notice that when \(\eps\) is large, the smooth min-entropy of \(\X\) can actually depend on the \emph{ambient space} on which \(\X\) was defined! While this may seem concerning at first, one may find comfort in thinking of smooth min-entropy simply as convenient shorthand for the expression in \cref{def:smooth-min-entropy}, instead of as a true ``entropy.''

Next, we record a very useful characterization of smooth min-entropy, which will be used throughout. This has appeared a few times in prior work, albeit in slightly different forms (see, e.g., \cite[Lemma 2.2]{zuckerman2007linear} or \cite[Lemma 6.2]{guruswami2009unbalanced}).

\begin{lemma}[A characterization of smooth min-entropy]\label{cor:characterization-corollary}
    For any \(\X\sim\zo^n\) and \(k\leq n\),
    \[
    H_\infty^\eps(\X)\geq k\iff\forall S:\Pr[\X\in S]\leq|S|\cdot2^{-k}+\eps.
    \]
\end{lemma}
    \begin{proof}
    (\(\implies\)) Let \(\X^\pr\sim \zo^n\) be a source of min-entropy at least \(k\) such that \(\X^\pr\approx_\eps\X\). Then for any \(S\),
    \[
    \Pr[\X\in S] \leq \Pr[\X^\pr\in S]+\eps\leq|S|\cdot2^{-k}+\eps.
    \]
    (\(\impliedby\)) Let \(\mathsf{Heavy}\) be the set of elements that \(\X\) assigns probability \(>2^{-k}\), and let \(\mathsf{Light}:=\zo^n\setminus\mathsf{Heavy}\). Notice that since \(n\geq k\), we have \(\Pr[\X\in\mathsf{Heavy}]-2^{-k}\cdot|\mathsf{Heavy}|\leq 2^{-k}\cdot|\mathsf{Light}|-\Pr[\X\in\mathsf{Light}]\). In other words, there is a way to shift the excess weight that \(\X\) assigns to \(\mathsf{Heavy}\) onto \(\mathsf{Light}\) without going over probability \(2^{-k}\) on any of these elements. Let \(\X^\pr\sim\zo^n\) denote this new source, and note \(H_\infty(\X^\pr)=k\). By our construction of \(\X^\pr\) and the hypothesis, we have
    \begin{align*}
    |\X-\X^\pr|&=\max_S|\Pr[\X\in S]-\Pr[\X^\pr\in S]|\\&=\Pr[\X\in\mathsf{Heavy}]-\Pr[\X^\pr\in\mathsf{Heavy}]\\
    &\leq|\mathsf{Heavy}|\cdot2^{-k}+\eps - |\mathsf{Heavy}|\cdot2^{-k}\\
    &\leq\eps,
    \end{align*}
    as desired.
\end{proof}

In fact, notice that the proof of the lemma above actually proved the following stronger result.

\begin{lemma}[A characterization of smooth min-entropy]\label{lemma:brand-new-characterization}
    For any \(\X\sim\zo^n\), \(0\leq k\leq n\), and \(\eps\in[0,1]\), the following holds. If we define
    \(\mathsf{Heavy}:=\left\{x\in\zo^n : \Pr[\X=x] > 2^{-k}\right\}\), then we have the equivalence 
    \[
    H_\infty^\eps(\X)\geq k \iff \Pr[\X\in\mathsf{Heavy}] \leq |\mathsf{Heavy}|\cdot2^{-k} + \eps.
    \]
\end{lemma}

Finally, we record one technical lemma that will be useful later on.

\begin{claim}\label{claim:hard-closeness-smoothness}
    Consider any random variables \(\A,\A^\pr\sim A\) and \(\B,\B^\pr\sim B\) such that \((\A,\B)\approx_\eps(\A^\pr,\B^\pr)\). Then
    \[
    \Pr_{a\sim\A}\left[H^{\gamma}_\infty(\B\mid\A=a)<k\right]\leq\Pr_{a\sim\A^\pr}[H^{\gamma/2}_\infty(\B^\pr\mid\A^\pr=a)<k]+4\eps/\gamma+\eps.
    \]
\end{claim}
\begin{proof}
    Let \(\mathsf{BAD}:=\{a : H_\infty^{\gamma}(\B\mid\A=a)<k\}\), let \(\mathsf{BAD}^\pr:=\{a : H_\infty^{\gamma/2}(\B^\pr\mid\A^\pr=a)<k\}\), and define \(S:=\mathsf{BAD}\setminus\mathsf{BAD}^\pr\). Note that
    \begin{align*}
    \Pr_{a\sim\A}[H_\infty^{\gamma}(\B\mid\A=a)<k]&=\Pr_{a\sim\A}[a\in\mathsf{BAD}]\\
    &\leq\Pr_{a\sim\A}[a\in S]+\Pr_{a\sim\A}[a\in\mathsf{BAD}^\pr]\\
    &\leq\Pr_{a\sim\A}[a\in S]+\Pr_{a\sim\A^\pr}[a\in\mathsf{BAD}^\pr]+\eps\\
    &=\Pr_{a\sim\A}[a\in S]+\Pr_{a\sim\A^\pr}[H_\infty^{\gamma/2}(\B^\pr\mid\A^\pr=a)<k]+\eps,
    \end{align*}
    and thus all that remains is to bound \(\Pr_{a\sim\A}[a\in S]\). Towards this end, notice that for every \(a\in S\), it holds that \(H_\infty^{\gamma}(\B\mid\A=a)<k\) and \(H_\infty^{\gamma/2}(\B^\pr\mid\A^\pr=a)\geq k\). In other words, \((\B^\pr\mid\A^\pr=a)\) is \((\gamma/2)\)-close to the family \(\mathcal{X}\) of sources with min-entropy at least \(k\), yet \((\B\mid\A=a)\) has distance \(>\gamma\) from this same family. By the triangle inequality, this means \((\B\mid\A=a)\) has distance \(>\gamma/2\) from \((\B^\pr\mid\A^\pr=a)\).

Now, define \(p:=\Pr[\A\in S]\), and partition \(S\) into subsets \(X_1,X_2\) such that \(\Pr[\A=a]\geq\Pr[\A^\pr=a]\) for all \(a\in X_1\), and \(\Pr[\A=a]<\Pr[\A^\pr=a]\) for all \(a\in X_2\). Since \(X_2,X_2\) is a partition, it must hold that either \(\Pr[\A\in X_1]\geq p/2\) or \(\Pr[\A\in X_2]\geq p/2\). Suppose the former is true, and recall that for all \(a\in X_1\subseteq S\), it holds that \((\B\mid\A=a)\not\approx_{\gamma/2}(\B^\pr\mid\A^\pr=a)\). By definition of statistical distance, this means that for every \(a\in X_1\) there is a set \(Q_a\) such that \(\Pr[(\B\mid\A=a)\in Q_a]-\Pr[(\B^\pr\mid\A^\pr=a)\in Q_a]\geq\gamma/2\). Thus
\begin{align*}
|(\A,\B)-(\A^\pr,\B^\pr)|&\geq\sum_{a\in X_1,b\in Q_a}\Pr[(\A,\B)=(a,b)]-\Pr[(\A^\pr,\B^\pr)=(a,b)]\\
&=\sum_{a\in X_1,b\in Q_a}\Pr[\A=a]\cdot\Pr[\B=b\mid\A=a]-\Pr[\A^\pr=a]\cdot\Pr[\B^\pr=b\mid\A^\pr=a]\\
&\geq\sum_{a\in X_1,b\in Q_a}\Pr[\A=a]\cdot(\Pr[\B=b\mid\A=a]-\Pr[\B^\pr=b\mid\A^\pr=a])\\
&=\sum_{a\in X_1}\Pr[\A=a]\sum_{b\in Q_a}(\Pr[\B=b\mid\A=a]-\Pr[\B^\pr=b\mid\A^\pr=a])\\
&\geq\frac{\gamma}{2}\sum_{a\in X_1}\Pr[\A=a]\geq p\gamma/4.
\end{align*}
Consider now the case that \(\Pr[\A\in X_2]\geq p_2\), and recall that for all \(a\in X_2\subseteq S\) it holds that \((\B\mid\A=a)\not\approx_{\gamma/2}(\B^\pr\mid\A^\pr=a)\). By definition of statistical distance, for every \(a\in X_2\) there is a set \(Q_a\) such that \(\Pr[(\B^\pr\mid\A^\pr=a)\in Q_a]-\Pr[(\B\mid\A=a)\in Q_a]\geq\gamma/2\). By definition of \(X_2\), we know that \(\Pr[\A^\pr=a]>\Pr[\A=a]\) for all \(a\in X_2\), which implies that \(\Pr[\A^\pr\in X_2]\geq p/2\). Thus
\begin{align*}
|(\A,\B)-(\A^\pr,\B^\pr)|&\geq\sum_{a\in X_2,b\in Q_a} \Pr[(\A^\pr,\B^\pr)=(a,b)]-\Pr[(\A,\B)=(a,b)]\\
&=\sum_{a\in X_2,b\in Q_a}\Pr[\A^\pr=a]\cdot\Pr[\B^\pr=b\mid\A^\pr=a]-\Pr[\A=a]\cdot\Pr[\B=b\mid\A=a]\\
&\geq\sum_{a\in X_2}\Pr[\A^\pr=a]\sum_{b\in Q_a}(\Pr[\B^\pr=b\mid\A^\pr=a]-\Pr[\B=b\mid\A=a])\\
&\geq\frac{\gamma}{2}\sum_{a\in X_2}\Pr[\A^\pr=a]\geq p\gamma/4.
\end{align*}
Thus we see that no matter what, \(|(\A,\B)-(\A^\pr,\B^\pr)|\geq p\gamma/4\). And since this statistical distance is at most \(\eps\) by the hypothesis, we get that \(p\leq 4\eps/\gamma\). Since \(p\) was defined to be \(\Pr_{a\sim\A}[a\in S]\), the result follows.
\end{proof}

\subsection{Condensers}

At last, we are ready to present a formal definition for the main objects of study in this paper.

\begin{definition}[Condenser]\label{def:condenser}
    Let \(\mathcal{X}\) be a family of \((n,k)\)-sources. A function \(\cond:\zo^n\to\zo^m\) is called a \emph{condenser} for \(\mathcal{X}\) with error \(\eps\), loss \(\ell\in[0,k]\), and gap \(g\), if \(m=k-\ell+g\) and for every \(\X\in\mathcal{X}\),
    \[
    H_\infty^\eps(\cond(\X))\geq k-\ell.
    \]
    We call \(k^\pr:=k-\ell\) the \emph{output entropy} of the condenser.
\end{definition}

Note that after specifying the family \(\mathcal{X}\) and the error \(\eps\) of the condenser, there are many equivalent ways to describe the remaining parameters. In particular, one may choose to specify its loss and gap, or its output entropy and gap, or its loss and output length, and so on (and the other parameters can be inferred).\footnote{When there is a notion of ``gap'' in the input, we often refer to the gap of the condenser as the ``output gap'' to avoid confusion.} Our choice will often depend on whichever feels the most appropriate in context. One important note, however, is that the output entropy simply describes \emph{a lower bound on} the actual (smooth) min-entropy of the output, while the output gap describes an upper bound on the actual gap. Indeed, the parameters of the condenser should not change as you plug in different sources from \(\mathcal{X}\)!

Now, while the above definition seems to describe ``deterministic'' or ``seedless'' condensers, it is easy to see that it also captures seeded condensers, simply by setting \(\mathcal{X}\) to consist of all sources of the form \((\X,\Y)\), where \(\X\) is an \((n,k)\)-source and \(\Y\) is an independent \((d,d)\)-source. Still, it is helpful to introduce a separate (perhaps redundant) definition, which makes it easier to refer to their parameters.

\begin{definition}[Seeded condenser]
    A function \(\sCond:\zo^n\times\zo^d\to\zo^m\) is an \((n,k)\times(d,d)\to_\eps(m,k^\pr)\) \emph{seeded condenser} if for any \((n,k)\)-source \(\X\), it holds that \(H_\infty^\eps(\sCond(\X,\U_d))\geq k^\pr\).
\end{definition}

Next, note that condensers (as put forth in \cref{def:condenser}) strictly generalize extractors, which correspond to the case where \(g=0\). As a result, the same is true of seeded condensers and seeded extractors. Still, it will be handy to record a separate definition of these objects, for ease of reference.

\begin{definition}[Seeded extractor]
    A function \(\sExt:\zo^n\times\zo^d\to\zo^m\) is a \((k,\eps)\)-\emph{seeded extractor} if for any \((n,k)\)-source \(\X\), it holds that \(\sExt(\X,\U_d)\approx_\eps\U_m\).
\end{definition}

Note that such a seeded extractor is automatically a \((n,k)\times(d,d)\to_\eps(m,m)\)-seeded condenser.

Finally, we record a useful ``trick,'' which can be thought of as a trivial condenser. In the world of extractors, it is well-known that you can shorten the output length ``for free,'' simply by taking a prefix (i.e., this operation won't harm the other parameters of the extractor). In the world of condensers, this may harm the overall output entropy \emph{rate} \(k^\pr/m\), but it cannot harm the absolute \emph{gap}.

\begin{fact}\label{fact:prefix-entropy}
If \(\X\sim\zo^n\) has min-entropy gap \(\leq g\), its prefix \(\X_{[p]}\) of length \(p\) has min-entropy gap \(\leq g\).
\end{fact}
\begin{proof}
Let \(x\) be the most likely element hit by \(\X_{[p]}\), and suppose it is hit with probability \(2^{-\ell}\). Conditioned on \(\X_{[p]}=x\), there is some element \(y\in\zo^{n-p}\) hit by \(\X_{[p+1,n]}\) with probability at least \(2^{-(n-p)}\). This means \(\X\) hits \((x,y)\) with probability at least \(2^{-\ell-(n-p)}\). Since \(\X\) has min-entropy gap \(\leq g\), this means that \(p-\ell\leq g\), and the result follows.
\end{proof}

\dobib

\section{Basics of block sources}\label{sec:block-source-basics}

In this section, we'll introduce some definitions, facts, and tools related to CG sources and block sources. Many of the tools we develop here are new, and they find good use throughout the rest of the paper.

\subsection{Definitions}

First, recall that an \((n,k)\)-source is simply a random variable \(\X\sim\zo^n\) with min-entropy at least \(k\). Chor-Goldreich sources generalize \((n,k)\)-sources in the following way:

\begin{definition}[CG sources]\label{def:CG-sources}
A source \(\X\sim(\zo^n)^t\) is called a \((t,n,k)\)-Chor-Goldreich source if
\[
H_\infty(\X_i\mid\X_{<i}=x)\geq k
\]
for all \(i\in[t]\) and \(x\in(\zo^n)^{i-1}\).
\end{definition}

Note that a \((1,n,k)\)-Chor-Goldreich source is exactly an \((n,k)\)-source. A \((t,1,k)\)-Chor-Goldreich source, on the other hand, is known as a \emph{Santha-Vazirani source} \cite{santha1986generating}. Next, the following allows us to assume that every Chor-Goldreich source has some nice structure.

\begin{fact}
If \(\X\sim(\zo^n)^t\) is a \((t,n,k)\)-Chor-Goldreich source, then it is a convex combination of \((t,n,k)\)-Chor-Goldreich sources \(\X^\pr\sim(\zo^n)^t\) such that for any \(i\in[t]\) and \(x\in(\zo^n)^{i-1}\),
\[
(\X_i\mid\X_{<i}=x)
\]
is a flat \((n,k)\)-source.
\end{fact}
\begin{proof}
    It is well-known that any \((n,k)\)-source (with \(k\) an integer) is a convex combination of flat \((n,k)\)-sources \cite[Lemma 6.10]{vadhan2012pseudorandomness}. Iteratively apply this to blocks \(\X_1,\dots,\X_t\), using the fact that under any conditioning on \(\X_{<i}\), the block \(\X_i\) is still an \((n,k)\)-source (by definition of Chor-Goldreich source).
\end{proof}

In our constructions, we'll often need to work with a generalization of CG sources, where the block lengths are uneven. These are called \emph{block sources}.

\begin{definition}[Block sources]\label{def:block-sources}
    A source \(\X=(\X_1,\dots,\X_t)\) is called an \(((n_1,k_1),(n_2,k_2),\dots,(n_t,k_t))\)-block source if each \(\X_i\) is over \(n_i\) bits, and
    \[
    H_\infty(\X_i\mid\X_{<i}=x)\geq k_i
    \]
    for all \(i\in[t]\) and \(x\in\zo^{n_1}\times\zo^{n_2}\times\dots\times\zo^{n_{i-1}}\).
\end{definition}

Note that a \((t,n,k)\)-CG source is just a block source with \(n_1=\dots=n_t=n\) and \(k_1=\dots=k_t=k\).

To streamline our proofs, it will be convenient to take this generalization two steps further. We use the following definition, which generalizes block sources by only requiring each block \(\X_i\) to be \emph{close} to having a min-entropy guarantee, and only requiring this closeness to hold for \emph{most} fixings of the prefix \(\X_{<i}\).

\begin{definition}[Almost block sources]
    A source \(\X=(\X_1,\dots,\X_t)\) is called an \(((\eta_1, \gamma_1),\dots,(\eta_t,\gamma_t))\)-almost \(((n_1,k_1),\dots,(n_t,k_t))\)-block source if each \(\X_i\) is over \(n_i\) bits, and for every \(i\in[t]\) it holds that
    \[
    \Pr_{x\sim\X_{<i}}\left[(\X_i\mid\X_{<i}=x)\text{ is \(\gamma_i\)-close to an \((n_i,k_i)\)-source}\right]\geq1-\eta_i.
    \]
\end{definition}

This notion is a generalization of the first type of almost block sources studied in \cite[Definition 1.3]{doron2023almost} (which correspond to the special case where \(\gamma_1=\dots=\gamma_t=\gamma\) and \(\eta_1=\dots=\eta_t=0\)), and a specialization of the third type of almost block sources studied in \cite[Definition 8.3]{doron2023almost} (with \(\lambda=0\)).

\subsection{Almost block sources are close to block sources}

As it turns out, it is not too difficult to show that an almost block source is close to a true block source.

\begin{lemma}\label{lem:almost-means-close}
If \(\X=(\X_1,\dots,\X_t)\) is an \(((\eta_1, \gamma_1),\dots,(\eta_t,\gamma_t))\)-almost \(((n_1,k_1),\dots,(n_t,k_t))\)-block source, then \(\X\) is \(\eps\)-close to an \(((n_1,k_1),\dots,(n_t,k_t))\)-block source \(\X^\star\), where \(\eps=\sum_{i\in[t]}(\eta_i + \gamma_i)\).
\end{lemma}

The key tool is the following, which can be viewed as a tightness result for a key lemma (on amplifying statistical distance) of Chattopadhyay, Goodman, and Zuckerman \cite[Lemma 1, ECCC version]{ourSpaceComplexity}. More formally, it can be viewed as a ``local-to-global'' closeness result for sequences of correlated random variables. It also generalizes the classic fact that a sequence of \emph{independent} random variables, each close to uniform, is itself (relatively) close to uniform (e.g., \cite[Proposition 2.11]{rao2009extractors}).

\begin{lemma}\label{lem:no-distance-amplification}
    Let \(\X\sim V_1\times\dots\times V_t\) and \(\Y\sim V_1\times\dots\times V_t\) each be a sequence of (not necessarily independent) random variables. Suppose that for every \(i\in[t]\) and \(v\in\supp(\X_{<i})\cap\supp(\Y_{<i})\),
    \[
    |(\X_i\mid\X_{<i}=v)-(\Y_i\mid\Y_{<i}=v)|\leq\eps_i.
    \]
    Then \[|\X-\Y|\leq\sum_{i\in[t]}\eps_i.\]
\end{lemma}

We first prove the key tool above.

\begin{proof}
We proceed via a coupling argument. Namely, we will define jointly distributed random variables \(\X^\pr,\Y^\pr\sim V_1\times\dots\times V_t\) such that \(\X^\pr\equiv\X,\Y^\pr\equiv\Y\), and so that it is easy to get a good upper bound on \(\Pr[\X^\pr\neq\Y^\pr]\). The result will then follow by the first part of the coupling lemma (\cref{lem:coupling}).

In order to actually construct \(\X^\pr,\Y^\pr\), we will use the second part of the coupling lemma (\cref{lem:coupling}). In more detail, we define these random variables iteratively (from \(i=1,2,\dots,t\)), as follows. For every \(i\in\{1,2,\dots,t\}\), we will define a new pair of jointly distributed random variables \((\X_i^\pr,\Y_i^\pr)\) such that for every \((u,v)\in\supp(\X_{<i}^\pr,\Y_{<i}^\pr)\), all of the following bullet points hold:
\begin{itemize}
    \item \(\left(\X_i^\pr\mid(\X_{<i}^\pr,\Y_{<i}^\pr)=(u,v)\right)\equiv\left(\X_i\mid\X_{<i}=u\right)\).
    \item \(\left(\Y_i^\pr\mid(\X_{<i}^\pr,\Y_{<i}^\pr)=(u,v)\right)\equiv\left(\Y_i\mid\Y_{<i}=v\right)\).
    \item \(\Pr[\X_i^\pr\neq\Y_i^\pr\mid(\X_{<i}^\pr,\Y_{<i}^\pr)=(u,v)]=|(\X_i\mid\X_{<i}=u)-(\Y_i\mid\Y_{<i}=v)|\).
\end{itemize}
We now show (via induction on \(i\)) that such \((\X_i^\pr,\Y_i^\pr)\) exist, and that \(\X_{\leq i}^\pr\equiv\X_{\leq i}\) and \(\Y_{\leq i}^\pr\equiv\Y_{\leq i}\).

When \(i=1\), we know that such random variables \((\X_1^\pr,\Y_1^\pr)\) exist via the second part of the coupling lemma (\cref{lem:coupling}).\footnote{Formally, note that when \(i=1\), the phrase ``for every \((u,v)\in\supp(\X_{<i}^\pr,\Y_{<i}^\pr)\)'' is removed, and there is no conditioning.} Furthermore, the bullet points tell us that \(\X_{\leq 1}^\pr\equiv\X_{\leq 1}\) and \(\Y_{\leq 1}^\pr\equiv\Y_{\leq 1}\).

When \(i>1\), we assume (via the induction hypothesis) that \(\X_{<i}^\pr\equiv\X_{<i}\) and \(\Y_{<i}^\pr\equiv\Y_{<i}\). As a result, \((u,v)\in\supp(\X_{<i}^\pr,\Y_{<i}^\pr)\) implies that both \(u\in\supp(\X_{<i})\) and \(v\in\supp(\Y_{<i})\), and so \((\X_i\mid\X_{<i}=u)\) and \((\Y_i\mid\Y_{<i}=v)\) are well-defined. Thus, the second part of the coupling lemma (\cref{lem:coupling}) once again tells us that there exist random variables \((\X_i^\pr,\Y_i^\pr)\) satisfying all three bullets. Furthermore, we assert that \(\X_{\leq i}^\pr\equiv\X_{\leq i}\) and \(\Y_{\leq i}^\pr\equiv\Y_{\leq i}\). To see why the former holds, note that for all \(x\in\supp(\X_{\leq i}^\pr)\),
\begin{align*}
    \Pr[\X_{\leq i}^\pr=x]&=\Pr[\X_{<i}^\pr=x_{<i}]\cdot\Pr[\X_i^\pr=x_i\mid\X_{<i}^\pr=x_{<i}]\\
    &=\Pr[\X_{<i}=x_{<i}]\cdot\Pr[\X_i^\pr=x_i\mid\X_{<i}^\pr=x_{<i}],
\end{align*}
since the induction hypothesis tells us that \(\X_{<i}^\pr\equiv\X_{<i}\). Then, by the law of total probability,
\begin{align*}
    \Pr[\X_i^\pr=x_i\mid\X_{<i}^\pr=x_{<i}]&=\sum_{y\in\supp(\Y_{<i}^\pr\mid\X_{<i}^\pr=x_{<i})}\Pr[\X_i^\pr=x_i\land\Y_{<i}^\pr=y\mid\X_{<i}^\pr=x_{<i}]\\
    &=\sum_{y\in\supp(\Y_{<i}^\pr\mid\X_{<i}^\pr=x_{<i})}\Pr[\Y_{<i}^\pr=y\mid\X_{<i}^\pr=x_{<i}]\cdot\Pr[\X_i^\pr=x_i\mid\X_{<i^\pr}=x_{<i},\Y^\pr_{<i}=y]\\
    &=\Pr[\X_i=x_i\mid\X_{<i}=x_{<i}]\cdot\sum_{y\in\supp(\Y_{<i}^\pr\mid\X_{<i}^\pr=x_{<i})}\Pr[\Y_{<i}^\pr=y\mid\X_{<i}^\pr=x_{<i}]\\
    &=\Pr[\X_i=x_i\mid\X_{<i}=x_{<i}],
\end{align*}
where the penultimate equality follows from the first bullet above. Thus
\begin{align*}
    \Pr[\X_{\leq i}^\pr=x]&=\Pr[\X_{<i}=x_{<i}]\cdot\Pr[\X_i^\pr=x_i\mid\X_{<i}^\pr=x_{<i}]\\
    &=\Pr[\X_{<i}=x_{<i}]\cdot\Pr[\X_i=x_i\mid\X_{<i}=x_{<i}]\\
    &=\Pr[\X_{\leq i}=x].
\end{align*}
As a result, we have that \(\X_{\leq i}^\pr\equiv\X_{\leq i}\), and an identical argument shows that \(\Y_{\leq i}^\pr\equiv\Y_{\leq i}\).

Finally, we now have joint random variables \(\X^\pr,\Y^\pr\) such that \(\X^\pr\equiv\X\) and \(\Y^\pr\equiv\Y\). Thus, by the first part of the coupling lemma (\cref{lem:coupling}), we know that
\[
|\X-\Y|\leq\Pr[\X^\pr\neq\Y^\pr],
\]
and so all that remains it to upper bound this probability. To do so, note that
\begin{align*}
\Pr[\X^\pr\neq\Y^\pr]&=\sum_{i\in[t]}\Pr[\X_i^\pr\neq\Y_i^\pr\land\X_{<i}^\pr=\Y_{<i}^\pr]\\
&=\sum_{i\in[t]}\sum_{(v,v)\in\supp(\X_{<i}^\pr,\Y_{<i}^\pr)}\Pr[\X_i^\pr\neq\Y_i^\pr\land(\X_{<i}^\pr,\Y_{<i}^\pr)=(v,v)]\\
&=\sum_{i\in[t]}\sum_{(v,v)\in\supp(\X_{<i}^\pr,\Y_{<i}^\pr)}\Pr[(\X_{<i}^\pr,\Y_{<i}^\pr)=(v,v)]\cdot\Pr[\X_i^\pr\neq\Y_i^\pr\mid(\X_{<i}^\pr,\Y_{<i}^\pr)=(v,v)]\\
&=\sum_{i\in[t]}\sum_{(v,v)\in\supp(\X_{<i}^\pr,\Y_{<i}^\pr)}\Pr[(\X_{<i}^\pr,\Y_{<i}^\pr)=(v,v)]\cdot\left|\left(\X_i\mid\X_{<i}=v\right)-\left(\Y_i\mid\Y_{<i}=v\right)\right|\\
&\leq\sum_{i\in[t]}\eps_i\cdot\sum_{(v,v)\in\supp(\X_{<i}^\pr,\Y_{<i}^\pr)}\Pr[(\X_{<i}^\pr,\Y_{<i}^\pr)=(v,v)]\\
&\leq\sum_{i\in[t]}\eps_i,
\end{align*}
where the last equality follows from the third bullet point above, and the penultimate inequality follows from the lemma hypothesis, since
\[
(v,v)\in\supp(\X_{<i}^\pr,\Y_{<i}^\pr)\implies v\in\supp(\X_{<i}^\pr)\cap\supp(\Y_{<i}^\pr)\implies v\in\supp(\X_{<i})\cap\supp(\Y_{<i}).
\]
Thus
\[
|\X-\Y|\leq\Pr[\X^\pr\neq\Y^\pr]\leq\sum_{i\in[t]}\eps_i,
\]
as desired.
\end{proof}

With this tool in hand, it is now easy to show that almost block sources are close to true block sources.

\begin{proof}[Proof of \cref{lem:almost-means-close}]

Let \(\X=(\X_1,\dots,\X_t)\) be an \(((\eta_1, \gamma_1),\dots,(\eta_t,\gamma_t))\)-almost \(((n_1,k_1),\dots,(n_t,k_t))\)-block source. We first show how to ``zero out'' the \(\eta_i\) terms, and then the \(\gamma_i\) terms.

\paragraph{Zeroing out the \(\eta_i\) terms in \(\X\).} We start by defining, for every \(i\in[t]\), the set
\[
\mathsf{Good}_i:=\{x\in\supp(\X_{<i}) : \left(\X_i\mid\X_{<i}=x\right)\text{ is \(\gamma_i\)-close to an \((n_i,k_i)\)-source}\}.
\]
Then, we define a source \(\X^\pr=(\X_1^\pr,\dots,\X_t^\pr)\) such that for every \(i\in\{1,2,\dots,t\}\) and \(x\in\supp(\X_{<i}^\pr)\),
\[
\left(\X_i^\pr\mid\X_{<i}^\pr=x\right)\equiv
\begin{cases}
    \left(\X_i\mid\X_{<i}=x\right)&\text{if }x\in\mathsf{Good}_i,\\
    \U_{n_i}&\text{otherwise}.
\end{cases}
\]
It is immediate that \(\X^\pr\) is an \(\left((0,\gamma_1),\dots,(0,\gamma_t)\right)\)-almost \(\left((n_1,k_1),\dots,(n_t,k_t)\right)\)-block source.

Furthermore, observe that for any \(x\in\zo^{n_1}\times\dots\times\zo^{n_t}\) with \(x_{<i}\in\mathsf{Good}_i\) for all \(i\in[t]\), \[\Pr[\X^\pr=x]=\Pr[\X=x].\] Thus, for any set \(S\subseteq\zo^{n_1}\times\dots\times\zo^{n_t}\), we have that
\begin{align*}
\Pr[\X\in S]&\leq\Pr\left[\X\in S\land\X_{<i}\in\mathsf{Good}_i, \forall i\in[t]\right] + \Pr\left[\exists i\in[t] : \X_{<i}\notin\mathsf{Good}_i\right]\\
&\leq\Pr\left[\X^\pr\in S\land\X_{<i}^\pr\in\mathsf{Good}_i,\forall i\in[t]\right] + \sum_{i\in[t]}\Pr[\X_{<i}\notin\mathsf{Good}_i]\\
&\leq\Pr[\X^\pr\in S] + \sum_{i\in[t]}\eta_i.
\end{align*}
In other words, \(\X^\pr\) is \(\left(\sum_{i\in[t]}\eta_i\right)\)-close to \(\X\).

\paragraph{Zeroing out the \(\gamma_i\) terms in \(\X^\pr\).} Since \(\X^\pr\) is an \(\left((0,\gamma_1),\dots,(0,\gamma_t)\right)\)-almost \(\left((n_1,k_1),\dots,(n_t,k_t)\right)\)-block source, we know that for every \(i\in[t]\) and \(x\in\supp(\X^\pr_{<i})\), it holds that \(\left(\X_i^\pr\mid\X_{<i}^\pr=x\right)\) is \(\gamma_i\)-close to an \((n_i,k_i)\)-source, which we will call \(\Z_i^{(x)}\). Using this, we define a new source \(\X^\star=(\X_1^\star,\dots,\X_t^\star)\) such that for every \(i\in\{1,2,\dots,t\}\) and \(x\in\supp(\X^\star_{<i})\),
\[
\left(\X_i^\star\mid\X_{<i}^\star=x\right)\equiv
\begin{cases}
    \Z_i^{(x)}&\text{if }x\in\supp(\X_{<i}^\pr),\\
    \U_{n_i}&\text{otherwise}.
\end{cases}
\]
It is immediate that \(\X^\star\) is an \(\left((0,0),\dots,(0,0)\right)\)-almost \(\left((n_1,k_1),\dots,(n_t,k_t)\right)\)-block source; or in other words, an \(\left((n_1,k_1),\dots,(n_t,k_t)\right)\)-block source.

Furthermore, observe that for any \(i\in[t]\) and \(x\in\supp(\X^\pr_{<i})\cap\supp(\X^\star_{<i})\),
\[
\left|(\X_i^\pr\mid\X_{<i}^\pr=x)-(\X_i^\star\mid\X_{<i}^\star=x)\right|\leq\gamma_i.
\]
As a result, \cref{lem:no-distance-amplification} immediately tells us that \(\X^\star\) is \(\left(\sum_{i\in[t]}\gamma_i\right)\)-close to \(\X^\pr\).

\paragraph{Wrapping up} By a standard application of the triangle inequality, we get that \(\X^\star\) is \(\eps\)-close to \(\X\), where \(\eps=\sum_{i\in[t]}(\eta_i+\gamma_i)\). Since \(\X^\star\) is an \(\left((n_1,k_1),\dots,(n_t,k_t)\right)\)-block source, this completes the proof.
\end{proof}

\subsection{Keeping a block source fresh while fixing correlated randomness}

In extractor theory, the situation often arises that you have a collection of independent random variables \(\X_1,\dots,\X_t\), and additional random variables \(\X_1^\pr,\dots,\X_t^\pr\) where each \(\X_i^\pr\) is a deterministic function \(\X_i\). The latter variables often get in the way of the analysis, and the goal is usually to condition (``fix'') them to constant values, while keeping the entropy and independence in \(\X_1,\dots,\X_t\). The classic tool used for this is the chain rule for min-entropy.

\begin{lemma}[Min-entropy chain rule \cite{maurer1997privacy}]\label{lem:chain-rule}
For any random variables \(\X\sim X\) and \(\Y\sim Y\),
\[
\Pr_{y\sim\Y}\left[H_\infty(\X\mid\Y=y)\geq H_\infty(\X)-\log(|Y|)-\log(1/\eps)\right]\geq1-\eps.
\]
\end{lemma}

Indeed, as long as the entropy in each \(\X_i\) is larger than the length (support size) of each \(\X_i^\pr\), the above lemma can be used to fix \(\X_1^\pr,\dots,\X_t^\pr\) without losing the independence of \(\X_1,\dots,\X_t\) or too much entropy. But what if \(\X_1,\dots,\X_t,\X_1^\pr,\dots,\X_t^\pr\) have correlations among them? As we will see, this situation will frequently arise in our analysis of CG sources. In this section, we establish a formal way to deal with this. We prove the following, which shows how to keep a block source ``fresh'' (looking like a block source) while fixing a series of correlated random variables.

\begin{lemma}\label{lem:fixing-randomness-against-block}
Let \(\X=(\X_1,\X_2,\dots,\X_t)\) be an \(((n_1,k_1),(n_2,k_2),\dots,(n_t,k_t))\)-block source, and let \(\X^\pr=(\X^\pr_1,\X_2^\pr,\dots,\X_t^\pr)\) be another sequence of (possibly correlated) random variables satisfying the following.
\begin{itemize}
    \item \(\X_i^\pr\) is supported on a set of size at most \(2^{n_i^\pr}\), for every \(i\in[\tau]\).
    \item The random variables \(\left(\X_i\mid\X_{<i}=x,\X^\pr_{\geq i}=x^\pr\right)\) and \((\X^\pr_{<i}\mid\X_{<i}=x,\X^\pr_{\geq i}=x^\pr)\) are independent, for every \(i\in[t]\), \(x\in\zo^{n_1}\times\dots\times\zo^{n_{i-1}}\), \(x^\pr\in\zo^{n_i^\pr}\times\dots\times\zo^{n_t^\pr}\).
\end{itemize}
Then
\[
\Pr_{x^\pr\sim\X^\pr}\left[(\X\mid\X^\pr=x^\pr)\text{ is not \(t\sqrt{\eps}\)-close to an \(((n_1,\ell_1),(n_2,\ell_2),\dots,(n_t,\ell_t))\)-block source} \right]\leq t\sqrt{\eps},
\]
where each \(\ell_i:=k_i-\sum_{j\geq i}n_j^\pr-\log(1/\eps)\).
\end{lemma}

When we construct our condenser, it is crucial that the entropy loss on \(\X_i\) only comes from \(\X_j^\pr,j\geq i\).

\begin{proof}
Pick any index \(i\in[t]\), and define \(\ell_i:=k_i-\sum_{j\geq i}n_j^\pr-\log(1/\eps)\). Note that for every fixed \(x\), \((\X_i\mid\X_{<i}=x)\) has min-entropy at least \(k_i\) (since it is a block source), and thus the min-entropy chain rule (\cref{lem:chain-rule}) tells us that
\[
\Pr_{x^\pr\sim\X^\pr_{\geq i}}[H_\infty(\X_i\mid\X_{<i}=x,\X^\pr_{\geq i}=x^\pr)<\ell_i]\leq\eps.
\]
By the independence guaranteed in the second bullet of the lemma, we know that for any fixed \(x,x^\star\), the distributions \(\left(\X_i\mid\X_{<i}=x,\X^\pr=x^\star\right)\) and \(\left(\X_i\mid\X_{<i}=x,\X^\pr_{\geq i}=x^\star_{\geq i}\right)\) are identical. Thus we know that
\[
\Pr_{x^\star\sim\X^\pr}[H_\infty(\X_i\mid\X_{<i}=x,\X^\pr=x^\star)<\ell_i]\leq\eps
\]
for every fixed \(x\). As a result, we have that
\[
\Pr_{\substack{x^\star\sim\X^\pr\\x\sim\X_{<i}}}[H_\infty(\X_i\mid\X_{<i}=x,\X^\pr=x^\star)<\ell_i]\leq\eps.
\]
Using an averaging argument, this gives
\[
\Pr_{x^\star\sim\X^\pr}\left[\Pr_{x\sim\X_{<i}}[H_\infty(\X_i\mid\X_{<i}=x,\X^\pr=x^\star)<\ell_i]\geq\sqrt{\eps}\right]\leq\sqrt{\eps},
\]
and a union bound tells us
\[
\Pr_{x^\star\sim\X^\pr}\left[\exists i \in [t]: \Pr_{x\sim\X_{<i}}[H_\infty(\X_i\mid\X_{<i}=x,\X^\pr=x^\star)<\ell_i]\geq\sqrt{\eps}\right]\leq t\sqrt{\eps},
\]
In other words, we get that \(\X\) becomes an \(\sqrt{\eps}\)-almost \(((n_1,\ell_1),(n_2,\ell_2),\dots,(n_t,\ell_t))\)-block source except with probability at most \(t\sqrt{\eps}\) over fixing \(\X^\pr=x^\star\). Applying \cref{lem:almost-means-close} completes the proof.
\end{proof}

\subsection{Seeded condensers automatically work for two-block sources}

A core tool we use is the fact that seeded condensers can be used on block sources, while suffering just a small loss in parameters. This observation has been made in prior work, with slightly weaker parameters \cite[Lemma 28]{ben2019two}, or using a slightly different language \cite[Proof of Theorem 4.4]{ball2022randomness}.

\begin{lemma}\label{lem:seeded-condensers-work-on-block-sources}
    Let \(\sCond:\zo^n\times\zo^d\to\zo^m\) be a seeded \((n,k)\to_\eps(m,k^\pr)\) condenser. Then for any \(((n,k),(d,d-g))\)-block source \((\X,\Y)\), it holds that \(H_\infty^{2^g\eps}(\sCond(\X,\Y))\geq k^\pr-g\).
\end{lemma}

In other words, the output loses \(g\) bits of entropy, and the error blows up by a factor of \(2^g\).

\begin{proof}
Let \((\X,\Y)\sim\zo^n\times\zo^d\) be an \(((n,k),(d,d-g))\) block source, and let \(\Y^\ast\sim\zo^d\) be an independent uniform random variable. Notice that for any fixed \(x,S\) we have
\[
\Pr[\sCond(x,(\Y\mid\X=x))\in S]\leq2^g\cdot\Pr[\sCond(x,\Y^\ast)\in S],
\]
since if we define \(S_x:=\{y : \sCond(x,y)\in S\}\) then \(\Pr[\sCond(x,\Y^\ast)\in S]=2^{-d}|S_x|\) and \(\Pr[\sCond(x,(\Y\mid\X=x))\in S]\leq 2^{-(d-g)}|S_x|\). Thus we have
\begin{align*}
\Pr[\sCond(\X,\Y)\in S]&=\sum_x\Pr[\X=x]\cdot\Pr[\sCond(x,(\Y\mid\X=x))\in S]\\
&\leq2^g\sum_x\Pr[\X=x]\cdot\Pr[\sCond(x,\Y^\ast)\in S]\\
&=2^g\Pr[\sCond(\X,\Y^\ast)\in S].
\end{align*}
Since \(\sCond\) is a seeded \((n,k)\to_\eps(m,k^\pr)\) condenser, the above expression is at most
\begin{align*}
&\leq2^g\cdot(|S|\cdot2^{-k^\pr}+\eps)\\
&=|S|\cdot2^{-k^\pr+g}+2^g\eps.
\end{align*}
The result now follows by the standard characterization of smooth min-entropy (\cref{cor:characterization-corollary}).
\end{proof}

\subsection{Iterative condensing of multi-block sources}

Finally, the following generalizes well-known block-source extraction and condensing results, such as in \cite{nisan1996randomness,doron2023almost}. For example, if instantiated with an \(((n_1,k_1),\dots,(n_{t-1},k_{t-1}),(n_t,n_t))\)-block-source and seeded condensers with gap \(0\) (i.e., seeded extractors), then you get well-known results about extracting from block sources with a small seed (which is constant for constant error). We will use this framework in both our explicit and existential constructions, in order to handle sources with a very large number of blocks.

\begin{lemma}\label{lem:iterated-condensing-framework}
Consider a sequence of functions \(\sCond_1,\sCond_2,\dots,\sCond_{t-1}\), where each \(\sCond_i:\zo^{n_i}\times\zo^{m_{i+1}}\to\zo^{m_i}\) is a seeded \((n_i,k_i)\to_{\eps_i}(m_i,m_i-g_i)\) condenser. Furthermore, consider any pair of nonnegative real numbers \((n_t,k_t)\) such that \(m_t=n_t\), and define \(g_t:=n_t-k_t\) and \(\eps_t:=0\).

Now, define a function \(\cond^\pr:\zo^{n_1}\times\zo^{n_2}\times\dots\times\zo^{n_t}\to\zo^{m_1}\times\zo^{m_2}\times\dots\times\zo^{m_t}\) as \(\cond^\pr(x_1,x_2,\dots,x_t):=(y_1,y_2,\dots,y_t)\), where \(y_t:=x_t\), and for all other \(i\in[t-1]\),
\begin{align*}
y_i := \sCond_i(x_i,y_{i+1}).
\end{align*}
Then the function \(\cond:\zo^{n_1}\times\zo^{n_2}\times\dots\times\zo^{n_t}\to\zo^{m_1}\) defined as \(\cond(x_1,x_2,\dots,x_t):=y_1\) is a condenser for \(((n_1,k_1),(n_2,k_2),\dots,(n_t,k_t))\)-block sources with output gap \(g:=\sum_{i\in[t]}g_i\) and error \(\eps:=\sum_{i\in[t]}\eps_i\cdot 2^{\sum_{j\in(i,t]}g_j}\).
\end{lemma}

\begin{proof}
Let \(\X=(\X_1,\X_2,\dots,\X_t)\) be an arbitrary \(((n_1,k_1),(n_2,k_2),\dots,(n_t,k_t))\)-block source, and define \(\Y=(\Y_1,\Y_2,\dots,\Y_t):=\cond^\pr(\X_1,\X_2,\dots,\X_t)\) as in the lemma statement. We will prove a stronger claim than in the lemma, and show that for every \(a\in[t]\) and \(x\in\supp(\X_{<a})\),
\begin{center}
\((\Y_a\mid\X_{<a}=x)\) is \(\eps^\pr_a\)-close to an \((m_a,m_a-g^\pr_a)\)-source,
\end{center}
where \(\eps^\pr_a:=\sum_{i\in[a,t]}\eps_i\cdot2^{\sum_{j\in(i,t]}g_j}\), and \(g^\pr_a:=\sum_{i\in[a,t]}g_i\).\footnote{We also note that when \(a=1\), the expression \((\Y_a\mid\X_{<a}=x)\) should be interpreted as just \(\Y_1\).}

The proof will proceed via backwards induction on \(a\). We start by noting the claim is easy when \(a=t\). Indeed, recall that \(\Y_t:=\X_t\), and that \(\X\) is an \(((n_1,k_1),(n_2,k_2),\dots,(n_t,k_t))\)-block source. This means that for every fixing of \(\X_{<t}\), it holds that \(\X_t\) (and thus \(\Y_t\)) is an \((n_t,k_t)\)-source. In other words, every \((\Y_t\mid\X_{<t}=x)\) is \(\eps^\pr_t\)-close to an \((m_t,m_t-g^\pr_t)\)-source, since \(\eps^\pr_t=0\) and \((m_t,m_t-g^\pr_t)=(n_t,k_t)\).

Next, consider any \(1\leq a<t\) and \(x\in\supp(\X_{<a})\), and assume the claim holds for \(a+1\). Recall that \(\Y_a:=\sCond_a(\X_a,\Y_{a+1})\), and thus
\[
(\Y_a\mid\X_{<a}=x)=(\sCond_a(\X_a,\Y_{a+1})\mid\X_{<a}=x).
\]

Now, since \(\X\) is a block source, we know that \((\X_a\mid\X_{<a}=x)\) is an \((n_a,k_a)\)-source. Furthermore, the induction hypothesis tells us that for every \(x^\pr\in\supp(\X_a\mid\X_{<a}=x)\), it holds that \(\left(\Y_{a+1}\mid\X_{<a}=x,\X_a^\pr=x^\pr\right)\) is \(\eps^\pr_{a+1}\)-close to an \((m_{a+1},m_{a+1}-g^\pr_{a+1})\)-source. This means that the source
\[
\left((\X_a,\Y_{a+1})\mid\X_{<a}=x\right)
\]
is an \(\left((0,0),(0,\eps^\pr_{a+1})\right)\)-almost \(\left((n_a,k_a),(m_{a+1},m_{a+1}-g^\pr_{a+1})\right)\)-block source. And by \cref{lem:almost-means-close}, this means it is \(\eps^\pr_{a+1}\)-close to some \((\left((n_a,k_a),(m_{a+1},m_{a+1}-g^\pr_{a+1})\right)\)-block source \((\X_a^\star,\Y_{a+1}^\star)\). Thus, by a standard application of the data-processing inequality (\cref{fact:data-processing}), we have that
\begin{align*}
    \left(\Y_a\mid\X_{<a}=x\right)&=\left(\sCond_a(\X_a,\Y_{a+1})\mid\X_{<a}=x\right)\\
    &\approx_{\eps^\pr_{a+1}}\sCond_a(\X_a^\star,\Y_{a+1}^\star).
\end{align*}
Now, using the fact that seeded condensers automatically work for block sources (\cref{lem:seeded-condensers-work-on-block-sources}), we get that \(\sCond_a(\X_a^\star,\Y_{a+1}^\star)\) is \(\left(2^{g^\pr_{a+1}}\eps_a\right)\)-close to some \((m_a,m_a-g_a-g^\pr_{a+1})\)-source. Thus, the triangle inequality tells us \((\Y_a\mid\X_{<a}=x)\) is \(\left(\eps^\pr_{a+1}+2^{g^\pr_{a+1}}\eps_a\right)\)-close to an \((m_a,m_a-g_a-g^\pr_{a+1})\)-source. And by definition,
\begin{align*}
    \eps^\pr_{a+1} + 2^{g^\pr_{a+1}}\eps_a&=\left(\sum_{i\in[a+1,t]}\eps_i\cdot2^{\sum_{j\in(i,t]}g_j}\right) + \left(2^{\sum_{j\in[a+1,t]}g_j}\eps_a\right)=\sum_{i\in[a,t]}\eps_i\cdot2^{\sum_{j\in(i,t]}g_j}=\eps^\pr_a,
\end{align*}
and
\begin{align*}
g_a+g^\pr_{a+1}&=g_a+\sum_{i\in[a+1,t]}g_i=\sum_{i\in[a,t]}g_i=g^\pr_a.
\end{align*}
Thus, we get that \(\left(\Y_a\mid\X_{<a}=x\right)\) is \(\eps^\pr_a\)-close to an \((m_a,m_a-g_a^\pr)\)-source, as desired.

To conclude, we now know that for all \(a\in[t]\) and \(x\in\supp(\X_{<a})\),
\begin{center}
    \(\left(\Y_a\mid\X_{<a}=x\right)\) is \(\eps^\pr\)-close to an \((m_a,m_a-g^\pr)\)-source,
\end{center}
where \(\eps^\pr=\sum_{i\in[a,t]}\eps_i\cdot2^{\sum_{j\in(i,t]}g_j}\) and \(g_a^\pr=\sum_{i\in[a,t]}g_i\). This completes the proof, since the lemma statement corresponds to the special case where \(a=1\).
\end{proof}

\dobib

\section{Explicit constructions}\label{sec:explicit-constructions}

We are now ready to build our condenser for Chor-Goldreich sources, and ultimately prove \cref{thm:main:intro:explicit-condensers-CG-sources}.

\subsection{Somewhere-condensers from non-malleable condensers}\label{subsec:purification}

Our condenser will be built by expanding the last block of the CG source into a \emph{somewhere-random} source, and iteratively purifying it until we are left with just a single row that has high entropy. To make things formal, we'll need some definitions.

\begin{definition}[Somewhere-\(\ell\)-sources]
A source \(\Y\sim(\zo^m)^D\) is called a \emph{somewhere-\(\ell\)-source} if there exists some \(i\in[D]\) such that \(\Y_i\) has min-entropy at least \(\ell\).
\end{definition}

\begin{definition}[Somewhere-condensers for CG sources]
A function \(\sCond:(\zo^n)^t\to(\zo^w)^D\) is a somewhere-\(\ell\)-condenser for \((t,n,k)\)-CG sources with error \(\eps\) if for any \((t,n,k)\)-CG source \(\X\sim(\zo^n)^t\), \(\sCond(\X)\) is \(\eps\)-close to a convex combination of somewhere-\(\ell\)-sources.
\end{definition}

\begin{definition}[Non-malleable condensers for block sources]\label{def:non-malleable-condensers-for-block-sources}
A function \(\nmCond:\zo^n\times\zo^n\times\zo^w\times[2]\to\zo^m\) is a non-malleable condenser (with advice) for \(((n,k),(n,k),(w,\ell))\)-block sources with error \(\eps\) and output entropy \(r\) if the following holds. For any \(\X,\Y\sim\zo^n\) and \(\Z_1,\Z_2\sim\zo^w\) such that at least one of the sequences \((\X,\Y,\Z_1)\) and \((\X,\Y,\Z_2)\) is an \(((n,k),(n,k),(w,\ell))\)-block source,
\[
\nmCond(\X,\Y,\Z_1,1)\oplus\nmCond(\X,\Y,\Z_2,2)
\]
is \(\eps\)-close to an \((m,r)\)-source.
\end{definition}

In our first key lemma, we show how a non-malleable condenser can be used to improve the quality of a somewhere-condenser. We prove the following, which we will eventually apply iteratively.

\begin{lemma}[Purifying a somewhere-condenser]\label{lem:purify}
Suppose you have the following objects.
\begin{itemize}
    \item \(\sCond:(\zo^n)^t\to(\zo^w)^{2^d}\) a somewhere-\(\ell\)-condenser for \((t,n,k)\)-CG sources with error \(\eps_1\).
    \item \(\nmCond:\zo^{nb}\times\zo^{nb}\times\zo^w\times[2]\to\zo^m\) a non-malleable condenser (with advice) for \(((nb,kb-d-\log(1/\eps_2)),(nb,kb-d-\log(1/\eps_2)),(w,\ell))\)-block sources, which has error \(\eps_2\) and output entropy \(r\).
\end{itemize}
Consider the function \(\sCond^\star:(\zo^n)^{b}\times(\zo^n)^b\times(\zo^n)^t\to(\zo^m)^{2^{d-1}}\) whose \(i^\text{th}\) output is
\[
\sCond^\star_i(X,Y,Z):=\nmCond(X,Y,\sCond(Z)_{2i-1},1) \oplus \nmCond(X,Y,\sCond(Z)_{2i},2).
\]
Then, \(\sCond^\star\) is a somewhere-\(r\)-condenser for \((2b+t,n,k)\)-CG sources with error \(\eps=\eps_1+4\sqrt{\eps_2}+\eps_2\).
\end{lemma}

The core technical claim we use is the following.

\begin{claim}\label{cl:core-technical-claim-somewhere-condensers}
    Let \(\X,\Y\sim\zo^n\) and \(\Z:=(\Z_1,\dots,\Z_D)\sim(\zo^w)^D\) be random variables such that:
    \begin{itemize}
        \item \((\X,\Y)\) is an \(((n,k),(n,k))\)-block source.
        \item \(\forall x,y\in\zo^n\), \((\Z\mid\X=x,\Y=y)\) is \(\eps_1\)-close to a convex combination of somewhere-\(\ell\)-sources.
    \end{itemize}
    Then \((\X,\Y,\Z)\) is \((\eps_1+4\sqrt{\eps_2})\)-close to a convex combination of sources of the form \((\X^\pr,\Y^\pr,\Z^\pr)\), where:
    \begin{itemize}
        \item \(\exists i\in[D]\) s.t. \((\X^\pr,\Y^\pr,\Z_i^\pr)\) is an \(((n,k-d-\log(1/\eps_2)),(n,k-d-\log(1/\eps_2)),(w,\ell))\)-block source.
    \end{itemize}
\end{claim}

\begin{proof}
By the lemma hypothesis, we know that for every fixed \(x,y\), \((\Z\mid\X=x,\Y=y)\) is \(\eps_1\)-close to a convex combination of somewhere-\(\ell\)-sources. This means that for every fixed \(x,y\), there is some convex combination of the form \(\R^{x,y}:=\sum_{i\in[D]}p_i^{x,y}\cdot\R^{x,y,i}\) such that
\begin{align}\label{eq:piecewise-close-somewhere-source}
(x,y,(\Z\mid\X=x,\Y=y))\approx_{\eps_1}(x,y,\R^{x,y}),
\end{align}
where we may assume that each \(\R^{x,y,i}\sim(\zo^w)^D\) is not only a somewhere-\(\ell\)-source, but in fact has its entropy in its \(i^\text{th}\) row. That is, \(\R^{x,y,i}_i\sim\zo^w\) has min-entropy at least \(\ell\). Moreover, since \cref{eq:piecewise-close-somewhere-source} is true for all fixed \(x,y\), it follows that
\[
(\X,\Y,\Z)\approx_{\eps_1}(\X,\Y,\R^{\X,\Y}).
\]
We henceforth focus on \((\X,\Y,\R^{\X,\Y})\). Towards this end, recall that \(\R^{\X,\Y}\) is a convex combination of the form \(\R^{\X,\Y}=\sum_{i\in[D]}p_i^{\X,\Y}\cdot\R^{\X,\Y,i}\), where for every fixed \(x,y\) it holds that \(\R^{x,y,i}\sim\zo^w\) has min-entropy at least \(\ell\). Notice that because of this structure, we can equivalently sample \((\X,\Y,\R^{\X,\Y})\) as follows. First, define a new random variable \(\A\sim[D]\) that depends on \(\X,\Y\) in the following way: for every fixed \(x,y\), define
\[
\Pr[\A=i\mid\X=x,\Y=y]:=p_i^{x,y}.
\]
Then, for every fixed \(x,y\) define a new random variable \(\T^{x,y}\sim(\zo^w)^D\) independent of \(\X,\Y\)\footnote{By this, we mean \(\T^{x,y}\) is independent of \(\X,\Y\). Later, we will use \(\T^{\X,\Y}\), which is of course not independent of \(\X,\Y\). However, the independence assumption tells us that \((\T^{\X,\Y}\mid\X=x,\Y=y)\equiv(\T^{x,y}\mid\X=x,\Y=y)\equiv\T^{x,y}\).} such that
\[
(\T^{x,y}\mid\A=i)\equiv\R^{x,y,i}.
\]
This means that the random variable \((\T^{x,y}\mid\A=i)\sim(\zo^w)^D\) has entropy at least \(\ell\) in its \(i^\text{th}\) row: in other words, \((\T^{x,y}_i\mid\A=i)\sim\zo^w\) has entropy at least \(\ell\) for all fixed \(x,y,i\). Given these definitions, it is straightforward to verify that
\[
(\X,\Y,\R^{\X,\Y})\equiv(\X,\Y,\T^{\X,\Y}).
\]
This is useful, because the latter three random variables are defined in the same space as another random variable \(\A\sim[D]\), which has the property that \((\T^{\X,\Y}_i\mid\A=i)\sim\zo^w\) has min-entropy at least \(\ell\) for all \(i\). Moreover, recall that \((\X,\Y)\) is an \(((n,k),(n,k))\)-block source. Thus we can apply our lemma on fixing randomness against block sources (\cref{lem:fixing-randomness-against-block}) to get
\[
\Pr_{i\sim\A}[(\X,\Y\mid\A=i)\text{ is not \(2\sqrt{\eps_2}\)-close to an \(((n,k^\pr),(n,k^\pr))\)-block source}]\leq2\sqrt{\eps_2},
\]
where \(k^\pr=k-d-\log(1/\eps_2)\). Thus we get that upon fixing \(\A=i\), both of the following hold (except with probability at most \(2\sqrt{\eps_2}\)):

\begin{itemize}
    \item \((\X,\Y\mid\A=i)\) is \(2\sqrt{\eps_2}\)-close to an \(((n,k^\pr),(n,k^\pr))\)-block source, and
    \item \((\T^{\X,\Y}_i\mid\A=i)\sim\zo^w\) has min-entropy at least \(\ell\). In fact, for all fixed \(x,y\), it remains true that \((\T^{\X,\Y}_i\mid\A=i,\X=x,\Y=y)\sim\zo^w\) has min-entropy at least \(\ell\).
\end{itemize}
Applying a standard fact about convex combinations (\cref{fact:convex-combo-error-to-closeness}), we therefore get that \((\X,\Y,\T^{\X,\Y})\sim\zo^n\times\zo^n\times(\zo^w)^D\) is \(2\sqrt{\eps_2}\)-close to a convex combination of distributions of the form \((\X^\star,\Y^\star,\Z^\star)\sim\zo^n\times\zo^n\times(\zo^w)^D\) satisfying:
\begin{itemize}
    \item \((\X^\star,\Y^\star)\sim\zo^n\times\zo^n\) is \(2\sqrt{\eps_2}\)-close to an \(((n,k^\pr),(n,k^\pr))\)-block source, and
    \item \(\Z^\star\sim(\zo^w)^D\) admits some \(i\in[D]\) such that \(H_\infty(\Z^\star_i\mid\X^\star=x,\Y^\star=y)\geq\ell\) for all \(x,y\).
\end{itemize}
Finally, let \((\X^{\star\star},\Y^{\star\star})\) be the \(((n,k^\pr),(n,k^\pr))\)-block source that \((\X^\star,\Y^\star)\) is \(2\sqrt{\eps_2}\)-close to, and define a new random variable \(\Z^{\star\star}\) as follows:
\[
(\Z^{\star\star}\mid\X^{\star\star}=x,\Y^{\star\star}=y) \equiv \begin{cases} 
      (\Z^\star\mid\X^\star=x,\Y^\star=y) & \text{if } (x,y)\in\supp(\X^\star,\Y^\star),\\
      \U & \text{otherwise.}
   \end{cases}
\]
It is straightforward to verify the following about \((\X^{\star\star},\Y^{\star\star},\Z^{\star\star})\sim\zo^n\times\zo^n\times(\zo^w)^D\).
\begin{itemize}
\item \((\X^{\star\star},\Y^{\star\star})\) is an \(((n,k^\pr),(n,k^\pr))\)-block source.
\item There exists some \(i\in[D]\) such that \(H_\infty(\Z^{\star\star}_i\mid\X^{\star\star}=x,\Y^{\star\star}=y)\geq\ell\) for all \(x,y\).
    \item \((\X^{\star\star},\Y^{\star\star},\Z^{\star\star})\approx_{2\sqrt{\eps_2}}(\X^\star,\Y^\star,\Z^{\star\star})\).
\end{itemize}
Note that the first two conditions in fact imply that there exists some \(i\in[D]\) such that \((\X^{\star\star},\Y^{\star\star},\Z^{\star\star}_i)\) is an \(((n,k^\pr),(n,k^\pr),(w,\ell))\)-block source, where recall that \(k^\pr=k-d-\log(1/\eps_2)\). Thus \((\X^{\star\star},\Y^{\star\star},\Z^{\star\star})\) has the exact structure we were originally looking for. To summarize, recall that \((\X,\Y,\Z)\approx_{\eps_1}(\X,\Y,\R^{\X,\Y})\equiv(\X,\Y,\T^{\X,\Y})\), and the latter is \(2\sqrt{\eps_2}\)-close to a convex combination of distributions \((\X^\star,\Y^\star,\Z^\star)\) of the form specified above, and each of these is \(2\sqrt{\eps_2}\)-close to a distribution \((\X^{\star\star},\Y^{\star\star},\Z^{\star\star})\) of the desired structure. Applying the triangle inequality (\cref{fact:triangle-inequality}), we immediately get that \((\X,\Y,\Z)\) is \((\eps_1 + 4\sqrt{\eps_2})\)-close to a convex combination of distributions \((\X^{\star\star},\Y^{\star\star},\Z^{\star\star})\) of the desired form.
\end{proof}

Given the above claim, it is now straightforward to show that a non-malleable condenser can be used to purify a somewhere-condenser.

\begin{proof}[Proof of \cref{lem:purify}]
  Let \(\B\sim(\zo^n)^{2b+t}\) be a \((2b+t,n,k)\)-CG source. Observe that we can parse it as an \(\left((nb,kb),(nb,kb),(nt,kt)\right)\)-block source \((\X,\Y,\Z)\sim(\zo^n)^b\times(\zo^n)^b\times(\zo^n)^t\), with the additional property that for every fixed \(x,y\), \((\Z\mid\X=x,\Y=y)\) is a \((t,n,k)\)-CG source. The goal is to show \(\sCond^\star(\X,\Y,\Z)\) is \(\eps\)-close to a somewhere-\(r\)-source \(\A\sim(\zo^m)^{2^{d-1}}\). Recalling the definition of \(\sCond^\star\), this means we must show that the random variable
  \begin{align*}
      \T:=\Bigg(\quad\quad\nmCond(\X,\Y,\sCond_1(\Z),1)&\oplus\nmCond(\X,\Y,\sCond_2(\Z),2),\\
\nmCond(\X,\Y,\sCond_3(\Z),1)&\oplus\nmCond(\X,\Y,\sCond_4(\Z),2),\\
&\ \ \vdots\\
\nmCond(\X,\Y,\sCond_{D-1}(\Z),1)&\oplus\nmCond(\X,\Y,\sCond_D(\Z),2) \quad\quad\Bigg)
  \end{align*}
  is \(\eps\)-close to a convex combination of somewhere-\(r\)-sources. Towards this end, define for each \(i\in[D]\) a random variable \(\W_i:=\sCond_i(\Z)\sim\zo^w\), and let \(\W:=(\W_1,\W_2,\dots,\W_D)\). We can rewrite \(\T\) as
  \begin{align*}
\T:=\Big(\quad\quad\nmCond(\X,\Y,\W_1,1)&\oplus\nmCond(\X,\Y,\W_2,2),\\
\nmCond(\X,\Y,\W_3,1)&\oplus\nmCond(\X,\Y,\W_4,2),\\
&\ \ \vdots\\
\nmCond(\X,\Y,\W_{D-1},1)&\oplus\nmCond(\X,\Y,\W_D,2) \quad\quad\Big)
  \end{align*}

  Now, recall that \(\Z\) is a \((t,n,k)\)-CG source, even conditioned on any fixing of \(\X=x,\Y=y\). Furthermore, recall that \(\sCond\) is a somewhere-\(\ell\)-condenser for \((t,n,k)\)-CG sources with error \(\eps_1\). We can therefore say the following about the random variables \(\X,\Y,\W\):
  \begin{itemize}
      \item \((\X,\Y)\) is an \(((nb,kb),(nb,kb))\)-block source.
      \item \(\forall x,y\in\zo^{nb}\), \((\W\mid\X=x,\Y=y)\) is \(\eps_1\)-close to a convex combination of somewhere-\(\ell\)-sources.
  \end{itemize}
Applying our core technical claim (\cref{cl:core-technical-claim-somewhere-condensers}), we know that \((\X,\Y,\W)\) is \((\eps_1+4\sqrt{\eps_2})\)-close to a convex combination of sources of the form \((\X^\pr,\Y^\pr,\W^\pr)\) where:
\begin{itemize}
    \item \(\exists i\in[D]\) such that \((\X^\pr,\Y^\pr,\W^\pr_i)\) is an \(\left((nb,kb-d-\log(1/\eps_2)),(nb,kb-d-\log(1/\eps_2)),(w,\ell)\right)\)-block source.
\end{itemize}

By a straightforward application of the data-processing inequality (\cref{fact:data-processing}), this means that \(\T\) is \((\eps_1+4\sqrt{\eps_2})\)-close to a convex combination of distributions of the form
  \begin{align*}
\T^\pr=\Big(\quad\quad\nmCond(\X^\pr,\Y^\pr,\W_1^\pr,1)&\oplus\nmCond(\X^\pr,\Y^\pr,\W_2^\pr,2),\\
\nmCond(\X^\pr,\Y^\pr,\W_3^\pr,1)&\oplus\nmCond(\X^\pr,\Y^\pr,\W_4^\pr,2),\\
&\ \ \vdots\\
\nmCond(\X^\pr,\Y^\pr,\W_{D-1}^\pr,1)&\oplus\nmCond(\X^\pr,\Y^\pr,\W_D^\pr,2) \quad\quad\Big),
  \end{align*}
where \((\X^\pr,\Y^\pr,\W^\pr)\) have the guarantee that there exists some \(i\in[D]\) such that \((\X^\pr,\Y^\pr,\W_i^\pr)\) is an \(\left((nb,kb-d-\log(1/\eps_2)),(nb,kb-d-\log(1/\eps_2)),(w,\ell)\right)\)-block source. Call this the ``good'' index \(i\). Now, for all \(i\in[D/2]\), define
\[
\R_i^\pr:=\nmCond(\X^\pr,\Y^\pr,\W_{2i-1}^\pr)\oplus\nmCond(\X^\pr,\Y^\pr,\W_{2i}^\pr))
\]
so that we may write
\[
\T^\pr=(\R_1^\pr,\R_2^\pr,\dots,\R_{D/2}^\pr).
\]
If \(i^\star\) denotes the ``good'' index, then the definition of non-malleable condensers (\cref{def:non-malleable-condensers-for-block-sources}) tells us that \(\R^\pr_{i^\star}\) is \(\eps_2\)-close to an \((m,r)\)-source \(\R^\prpr_{i^\star}\sim\zo^m\). Furthermore, we can define a random variable \(\R^\prpr_{-i^\star}\sim(\zo^m)^{D/2-1}\) such that for every fixed \(r\in\zo^m\),
\[
(\R_{-i^\star}^\prpr\mid\R_{i^\star}^\prpr=r)\equiv \begin{cases} 
      (\R_{-i^\star}^\pr\mid\R_{i^\star}^\pr=r) & \text{if }r\in\supp(\R_{i^\star}^\pr),\\
      \U & \text{otherwise.}
   \end{cases}
\]
Then, if we define \(\T^\prpr:=(\R_{i^\star}^\prpr,\R_{-i^\star}^\prpr)\sim(\zo^m)^{D/2}\), it is straightforward to verify that \(\T^\prpr\approx_{\eps_2}\T^\pr\), and moreover \(\T^\prpr_{i^\star}\) is an \((m,r)\)-source. In other words, \(\T^\prpr\) is a somewhere-\(r\)-source, and thus \(\T^\pr\) is \(\eps_2\)-close to a somewhere-\(r\)-source. Recall that at the beginning, we showed that \(\T\) is \((\eps_1+4\sqrt{\eps_2})\)-close to a convex combination of such sources \(\T^\pr\), and we now know that each such source \(\T^\pr\) is \(\eps_2\)-close to a somewhere-\(r\)-source \(\T^\prpr\). As a result, it immediately follows that \(\T\) is \((\eps_1+4\sqrt{\eps_2}+\eps_2)\)-close to a convex combination of somewhere-\(r\)-sources, as desired.
\end{proof}

\subsection{Non-malleable condensers from seeded extractors}\label{subsec:non-malleable-from-seeded-extractors}

While it is known how to explicitly construct somewhere-condensers, it is not known how to construct non-malleable condensers for block sources. In this section, we show how to use basic seeded extractors to construct them. Later, we'll instantiate the recipe below in order to obtain our non-malleable condensers.

\begin{lemma}[Non-malleable condensers from seeded extractors]\label{lem:non-malleable-from-seeded}
Suppose you have the following objects.
\begin{itemize}
    \item \(\sExt_1:\zo^n\times\zo^{p_1}\to\zo^{d_1}\) a \((k_0,\eps_1)\)-seeded extractor.
    \item \(\sExt_1^\pr:\zo^n\times\zo^{d_1}\to\zo^m\) a \((k_0,\eps_1^\pr)\)-seeded extractor.
    \item \(\sExt_2:\zo^n\times\zo^{p_2}\to\zo^{d_2}\) a \((k_0,\eps_2)\)-seeded extractor.
    \item \(\sExt^\pr_2:\zo^n\times\zo^{d_2}\to\zo^m\) a \((k_0,\eps_2^\pr)\)-seeded extractor.
\end{itemize}
Consider the function \(\nmCond:\zo^n\times\zo^n\times\zo^w\times[2]\to\zo^m\) defined as
\[
\nmCond(X,Y,Z,b):=\sExt^\pr_b(X,\sExt_b(Y,Z_{[p_b]}))
\]
Then \(\nmCond\) is a non-malleable condenser (with advice) for \(((n,k),(n,k),(w,w-g))\)-block sources with output entropy \(m-(g+2d_1+p_2+\log(1/\eps_1)+\log(1/\eps_2) )\) and error \(2^{g+p_2+3}\eps_1^{1/4}+2^{g+4}\eps_2^{1/4}\), as long as:

\begin{itemize}
    \item \(k\geq k_0+m+2d_1+d_2+p_2+\log(1/\eps_1)+\log(1/\eps_2)\)
    \item \(\eps_1=\eps_1^\pr\) and \(\eps_2^\pr=\eps_2\cdot2^{-2d_1}\)
\end{itemize}
\end{lemma}

\begin{proof}
Consider any \(\X,\Y\sim\zo^n\) and \(\Z^1,\Z^2\sim\zo^w\) such that either \((\X,\Y,\Z^1)\) or \((\X,\Y,\Z^2)\) is an \(((n,k),(n,k),(w,w-g))\)-block source. Unwrapping the definition of \(\nmCond\), the goal is to show that
\begin{align*}
\sExt_1^\pr(\X,\sExt_1(\Y,\Zponelong))\oplus\sExt_2^\pr(\X,\sExt_2(\Y,\Zptwolong))
\end{align*}
is \(\eps\)-close to an \((m,r)\)-source. We must show this to be true in two cases: the case where \((\X,\Y,\Z^1)\) is the block source, and the case where \((\X,\Y,\Z^2)\) is the block source. We proceed with each case separately. But for both cases, it will be useful to define the following auxiliary random variables:
\begin{align*}
    \Zone&:=\Zponelong&&\Ztwo:=\Zptwolong\\
    \W_1&:=\sExt_1(\Y,\Zone)&&\W_2:=\sExt_2(\Y,\Ztwo)\\
    \S_1&:=\sExt^\pr_1(\X,\W_1)&&\S_2:=\sExt_2^\pr(\X,\W_2)
\end{align*}
With this notation, the goal is simply to show that \(\S_1\oplus\S_2\) is \(\eps\)-close to an \((m,r)\)-source. Let's get started.

\paragraph{Case 1.} In this case, we assume that \((\X,\Y,\Z^1)\) is the block source. In order to show that \(\S_1\oplus\S_2\) is \(\eps\)-close an \((m,r)\)-source, the idea is to find a sequence of fixings that will force \(\S_2\) to become a constant, but under which \(\S_1\) can be shown to have high min-entropy. In order to fix \(\S_2\), we will actually fix the entire sequence of random variables \((\S_2,\W_2,\Ztwo)\), and argue that the sequence \((\X,\Y,\Zone)\) maintains its structure.

To make things more formal, let's start by better understanding the structure of \((\X,\Y,\Zone)\). Recall that in this case, we assumed that \((\X,\Y,\Z^1)\) is an \(((n,k),(n,k),(w,w-g))\)-block source. This means that for every fixing of \(\X,\Y\), \(\Z^1\) still has min-entropy at least \(w-g\). And if \(\Z^1\) has min-entropy at least \(w-g\), it is not hard to show that its prefix \(\Zone=\Zponelong\) of length \(p_1\) has entropy at least \(p_1-g\) (\cref{fact:prefix-entropy}). This tells us that \((\X,\Y,\Zone)\) is a \(((n,k),(n,k),(p_1,p_1-g))\)-block source.

Next, let's better understand the structure of \((\S_2,\W_2,\Ztwo)\), and how it relates to \((\X,\Y,\Zone)\). Towards this end, first note that \(\S_2,\W_2\) and \(\Ztwo\) are supported on sets of size \(2^m,2^{d_2}\) and \(2^{p_2}\), respectively. Then, observe the following independence relationships between \((\X,\Y,\Zone)\) and \((\S_2,\W_2,\Ztwo)\):

\begin{itemize}
    \item Upon fixing \(\X,\Y,\Ztwo\), the random variables \(\S_2,\W_2\) become a constant. As a result, we know that \((\Zone\mid\X=x,\Y=y,\Ztwo=z_2)\) and \((\S_2,\W_2\mid\X=x,\Y=y,\Ztwo=z_2)\) are independent, \(\forall x,y,z_2\).
    \item Upon fixing \(\X,\W_2,\Ztwo\), the random variable \(\S_2\) becomes a constant. As a result, we know that \((\Y\mid\X=x,\W_2=w_2,\Ztwo=z_2)\) and \((\S_2\mid\X=x,\W_2=w_2,\Ztwo=z_2)\) are independent, \(\forall x,w_2,z_2\).
\end{itemize}

Because of these independence relationships between \((\X,\Y,\Zone)\) and \((\S_2,\W_2,\Ztwo)\), it turns out that we can safely fix the latter sequence without severely affecting the structure of the former. In particular, by combining the above observations with our lemma on fixing randomness against block sources (\cref{lem:fixing-randomness-against-block}), we immediately get the following, for any \(\gamma>0\).
\begin{align}\label{eq:very-complicated-probability}
\Pr_{(s_2,w_2,z_2)\sim(\S_2,\W_2,\Ztwo)}\Big[(\X,\Y,\Zone\mid\S_2=s_2,\W_2=w_2,&\ \Ztwo=z_2)\text{ is not \(3\sqrt{\gamma}\)-close to an}\\&\text{\(((n,k^\pr),(n,k^\pr),(p_1,\ell^\pr))\)-block source}\Big]\leq3\sqrt{\gamma},\notag
\end{align}
where \(k^\pr=k-(m+d_2+p_2+\log(1/\gamma))\) and \(\ell^\pr=p_1-(g+p_2+\log(1/\gamma))\). The reason why this bound will be useful is because it says that with high probability over fixing \(\S_2,\W_2,\Ztwo\), it follows that \((\X,\Y,\Zone)\) is still a block source, and of course \(\S_2\) is also a constant. Thus, if we can just show that \(\S_1=\sExt_1^\pr(\X,\sExt_1(\Y,\Zone))\) has high entropy whenever \((\X,\Y,\Zone)\) is a block source, we will be done, since we just needed \(\S_1\oplus\S_2\) to have high entropy, and this is true if \(\S_1\) has high entropy and \(\S_2\) is constant.

More formally, consider an arbitrary \(((n,k^\pr),(n,k^\pr),(p_1,\ell^\pr))\)-block source \((\A,\B,\C)\). Let's analyze what \(\sExt_1^\pr(\A,\sExt_1(\B,\C))\) looks like. By definition of block source, we know that for every \(a\), it holds that \((\B,\C\mid\A=a)\) is an \(((n,k^\pr),(p_1,\ell^\pr))\)-block source. Furthermore, recall that \(\sExt_1:\zo^n\times\zo^{p_1}\to\zo^{d_1}\) is a \((k_0,\eps_1)\)-seeded extractor, and thus it is trivially a seeded \((n,k_0)\to_{\eps_1}(d_1,d_1)\) condenser. Since every seeded condenser also works for block sources (\cref{lem:seeded-condensers-work-on-block-sources}), it follows that \(\sExt_1(\B,\C\mid\A=a)\) is \((2^{p_1-\ell^\pr}\eps_1)\)-close to a source \(\Q_a\sim\zo^{d_1}\) with min-entropy at least \(d_1-(p_1-\ell^\pr)\), provided that \(k^\pr\geq k_0\). Since it holds that \((a,\sExt_1(\B,\C\mid\A=a))\approx_{2^{p_1-\ell^\pr}\eps_1}(a,\Q_a)\) for every fixed \(a\), it follows that the random variables \((\A,\sExt_1(\B,\C))\) and \((\A,\Q_\A)\) enjoy the same statistical distance bound. And by a straightforward application of the data-processing inequality (\cref{fact:data-processing}), it also follows that
\[
\sExt_1^\pr(\A,\sExt_1(\B,\C))\approx_{2^{p_1-\ell^\pr}\eps_1}\sExt_1^\pr(\A,\Q_\A).
\]
Moreover, observe that \((\A,\Q_\A)\) is in fact a \(((n,k^\pr),(d_1,d_1-(p_1-\ell^\pr)))\)-block source. Repeating an identical analysis to what was done above, we can thus conclude that
\[
\sExt_1^\pr(\A,\Q_\A)\approx_{2^{p_1-\ell^\pr}\eps_1^\pr}\R^\star,
\]
where \(\R^\star\sim\zo^m\) is some source with min-entropy at least \(m-(p_1-\ell^\pr)\), provided that \(k^\pr\geq k_0\).

To summarize, we get that for any \(((n,k^\pr),(n,k^\pr),(p_1,\ell^\pr))\)-block source \((\A,\B,\C)\),
\[
\sExt_1^\pr(\A,\sExt_1(\B,\C))\approx_{2^{p_1-\ell^\pr}(\eps_1+\eps_1^\pr)}\R^\star,
\]
where \(\R^\star\) is some source with min-entropy at least \(m-(p_1-\ell^\pr)\), provided that \(k^\pr\geq k_0\). Moreover, it is straightforward to see that if \((\A,\B,\C)\) is actually only \(\xi\)-close to a block source \((\A^\star,\B^\star,\C^\star)\) of the above type, then the data-processing inequality (\cref{fact:data-processing}) tells us that \(\sExt_1^\pr(\A,\sExt_1(\B,\C))\) is \(\xi\)-close to \(\sExt_1^\pr(\A^\star,\sExt_1(\B^\star,\C^\star))\), which we showed above to be \(2^{p_1-\ell^\pr}(\eps_1+\eps_1^\pr)\)-close to \(\R^\star\).

Thus, we get that for any \((\A,\B,\C)\) that is \(\xi\)-close to an \(((n,k^\pr),(n,k^\pr),(p_1,\ell^\pr))\)-block source,
\[
\sExt_1^\pr(\A,\sExt_1(\B,\C))\approx_{2^{p_1-\ell^\pr}(\eps_1+\eps_1^\pr)+\xi}\R^\star,
\]
where \(\R^\star\) is an \((m,m-(p_1-\ell^\pr))\)-source, provided that \(k^\pr\geq k_0\).

Put differently, if \(\sExt_1^\pr(\A,\sExt_1(\B,\C))\) is \emph{not} \((2^{p_1-\ell^\pr}(\eps_1+\eps_1^\pr)+\xi)\)-close to any such source \(\R^\star\), then we know that \((\A,\B,\C)\) is not \(\xi\)-close to an \(((n,k^\pr),(n,k^\pr),(p_1,\ell^\pr))\)-block source. Thus, if we define \(g^\pr:=p_1-\ell^\pr\), \(\xi^\pr:=2^{p_1-\ell^\pr}(\eps_1+\eps_1^\pr)+\xi\), and \(\xi=3\sqrt{\gamma}\), we can combine the above with \cref{eq:very-complicated-probability} to obtain
\begin{align*}
&\Pr_{(s_2,w_2,z_2)\sim(\S_2,\W_2,\Ztwo)}\Big[(\S_1\oplus\S_2\mid\S_2=s_2,\W_2=w_2,\Ztwo=z_2)\text{ is not \(\xi^\pr\)-close to an \((m,m-g^\pr)\)-source}\Big]\\
&=\Pr_{(s_2,w_2,z_2)}\Big[(\S_1\oplus s_2\mid\S_2=s_2,\W_2=w_2,\Ztwo=z_2)\text{ is not \(\xi^\pr\)-close to an \((m,m-g^\pr)\)-source}\Big]\\
&=\Pr_{(s_2,w_2,z_2)}\Big[(\S_1\mid\S_2=s_2,\W_2=w_2,\Ztwo=z_2)\text{ is not \(\xi^\pr\)-close to an \((m,m-g^\pr)\)-source}\Big]\\
&=\Pr_{(s_2,w_2,z_2)}\Big[\left(\sExt_1^\pr(\X,\sExt_1(\Y,\Zone)\right)\mid\S_2=s_2,\W_2=w_2,\Ztwo=z_2)\text{ is not}\\&\quad\quad\quad\quad\quad\quad\quad\quad\quad\quad\quad\quad\quad\quad\quad\quad\quad\quad\quad\quad\quad\quad\text{\(\xi^\pr\)-close to an \((m,m-g^\pr)\)-source}\Big]\\
&\leq\Pr_{(s_2,w_2,z_2)}\Big[(\X,\Y,\Zone\mid\S_2=s_2,\W_2=w_2,\Ztwo=z_2)\text{ is not}\\&\quad\quad\quad\quad\quad\quad\quad\quad\quad\quad\quad\quad\quad\quad\quad\quad\quad\ \ \text{\(\xi\)-close to an \(((n,k^\pr),(n,k^\pr),(p_1,\ell^\pr))\)-block source}\Big]\\
&\leq\xi.
\end{align*}
To summarize, we've shown that there exists a random variable \(\mathbf{V}:=(\S_2,\W_2,\Ztwo)\) such that
\[
\Pr_{v\sim\mathbf{V}}\left[(\S_1\oplus\S_2\mid\mathbf{V}=v)\text{ is not \(\xi^\pr\)-close to an \((m,m-g^\pr)\)-source}\right]\leq\xi.
\]
By a standard fact about convex combinations (\cref{fact:convex-combo-error-to-closeness}), it immediately follows that \(\S_1\oplus\S_2\) is \(\xi\)-close to a convex combination of sources that \emph{are} \(\xi^\pr\)-close to an \((m,m-g^\pr)\)-source. As such, it holds that \(\S_1\oplus\S_2\) is \((\xi+\xi^\pr)\)-close to a convex combination of \((m,m-g^\pr)\)-sources. Since a convex combination of \((m,m-g^\pr)\)-sources is, itself, an \((m,m-g^\pr)\)-source, we conclude that \(\S_1\oplus\S_2\) is \((\xi+\xi^\pr)\)-close to an \((m,m-g^\pr)\)-source. Finally, recall that
\[
\xi+\xi^\pr = 6\sqrt{\gamma} +2^{p_1-\ell^\pr}(\eps_1+\eps_1^\pr) = 6\sqrt{\gamma} +2^{g+p_2+\log(1/\gamma)}(\eps_1+\eps_1^\pr),
\]
for any \(\gamma>0\). If we set \(\eps_1=\eps_1^\pr\) and \(\gamma=(2\eps_1)^{1/2}\), this is at most \(2^{g+p_2+3}\eps_1^{1/4}\). Furthermore, recall that
\[
g^\pr=p_1-\ell^\pr=g+p_2+\log(1/\gamma)\leq g+p_2+\log(1/\eps_1)
\]
Recall that to make everything work, we needed \(k^\pr\geq k_0\), and plugging in our definition of \(k^\pr\) from before, this requirement becomes (no worse than)
\[
k\geq k_0 + m + d_2 + p_2 + \log(1/\gamma)=k_0+m+d_2+p_2+\log(1/\eps_1).
\]
Thus, we get that the output of the non-malleable condenser is \(2^{g+p_2+3}\eps_1^{1/4}\)-close to an \((m,m-(g+p_2+\log(1/\eps_1)))\)-source, as long as \(k\geq k_0+m+d_2+p_2+\log(1/\eps_1)\) and \(\eps_1=\eps_1^\pr\).

\paragraph{Case 2.} We now proceed to the second case, where we assume that \((\X,\Y,\Z^2)\) is the block source. In order to show that \(\S_1\oplus\S_2\) is \(\eps\)-close to an \((m,r)\)-source, we now seek a sequence of fixings that will force \(\S_1\) to be constant, but under which \(\S_2\) can be shown to have high min-entropy.

To start, recall that \((\X,\Y,\Z^2)\) is an \(((n,k),(n,k),(w,w-g))\)-block source. Using an identical argument to the one appearing at the beginning of the previous case, we know this implies that \((\X,\Y,\Ztwo)\) is an \(((n,k),(n,k),(p_2,p_2-g))\)-block source. Next, we'd like to argue that \((\X,\W_2)\) is close to a block source \((\X^\pr,\W_2^\pr)\). For technical reasons that we will soon see, we actually need a slightly more involved result. In particular, we need to show there is a sequence \((\X^\pr,\W_2^\pr,\S_1^\pr,\W_1^\pr)\) such that:
\begin{itemize}
    \item \((\X^\pr,\W_2^\pr,\S_1^\pr,\W_1^\pr)\) is close to \((\X,\W_2,\S_1,\W_1)\),
    \item \((\X^\pr,\W_2^\pr)\) is a block source, and
    \item \(\S_1^\pr\) is constant upon any fixing of \(\X^\pr,\W_1^\pr\).
\end{itemize}

We start by constructing \((\X^\pr,\W_2^\pr)\). To do so, first recall that \((\X,\Y,\Ztwo)\) is a \(((n,k),(n,k),(p_2,p_2-g))\)-block source. This means that for every fixed \(x\), \((\Y,\Ztwo\mid\X=x)\) is an \(((n,k),(p_2,p_2-g))\)-block source. Now, recall that \(\sExt:\zo^n\times\zo^{p_2}\to\zo^{d_2}\) is a \((k_0,\eps_2)\)-seeded extractor, and is therefore also a seeded \((n,k_0)\to_{\eps_2}(d_2,d_2)\) condenser. Since every seeded condenser also works for block sources (\cref{lem:seeded-condensers-work-on-block-sources}), it follows that \(\sExt_2(\Y,\Ztwo\mid\X=x)\) is \(2^{g}\eps_2\)-close to a source \(\Q_x\sim\zo^{d_2}\) with min-entropy at least \(d_2-g\), provided that \(k\geq k_0\). Since it holds that \[(x,\sExt_2(\Y,\Ztwo\mid\X=x))\approx_{2^g\eps_2}(x,\Q_x)\] for every fixed \(x\), it follows that \((\X,\W_2)=(\X,\sExt_2(\Y,\Ztwo))\approx_{2^g\eps_2}(\X,\Q_\X)\). Moreover, observe that \((\X,\Q_\X)\) is an \(((n,k),(d_2,d_2-g))\)-block source. We define \((\X^\pr,\W_2^\pr)=(\X,\Q_\X)\).

Next, let us proceed with constructing \(\S_1^\pr,\W_1^\pr\). This is not too difficult. First, we define \(\W_1^\pr\) by asserting that for every fixed \(x,w_2\),
\[
(\W_1^\pr\mid\X^\pr=x,\W_2^\pr=w_2) \equiv \begin{cases} 
      (\W_1\mid\X=x,\W_2=w_2) & \text{if }(x,w_2)\in\supp(\X,\W_2)\\
      \U & \text{otherwise.}
   \end{cases}
\]
Then, we define \(\S_1^\pr:=\sExt_1^\pr(\X^\pr,\W_1^\pr)\). This trivially satisfies the condition that \(\S_1^\pr\) is constant upon any fixing of \(\X^\pr,\W_1^\pr\). Moreover, recall from above that \((\X^\pr,\W_2^\pr)\) is an \(((n,k),(d_2,d_2-g))\)-block source. Thus all that remains is to show that \((\X^\pr,\W_2^\pr,\S_1^\pr,\W_1^\pr)\) is close to \((\X,\W_2,\S_1,\W_1)\). To see why this is true, first observe that by construction, it holds that for any \((x,w_2)\in\supp(\X,\W_2)\),
\[
\left((\X^\pr,\W_2^\pr,\S_1^\pr,\W_1^\pr)\mid\X^\pr=x,\W_2^\pr=w_2\right)\equiv\left((\X,\W_2,\S_1,\W_1\mid\X=x,\W_2=w_2)\right).
\]
Combining this with the fact that \((\X^\pr,\W_2^\pr)\) is \(2^g\eps_2\)-close to \((\X,\W_2)\) by construction, it is straightforward to verify that
\[
(\X^\pr,\W_2^\pr,\S_1^\pr,\W_1^\pr)\approx_{2^g\eps_2}(\X,\W_2,\S_1,\W_1).
\]
Thus, we have successfully constructed a sequence \((\X^\pr,\W_2^\pr,\S_1^\pr,\W_1^\pr)\) with all of the properties originally desired. Now, let's see how to use it.

Recall that we originally wanted to show that \(\S_1\oplus\S_2\) is \(\eps\)-close to an \((m,r)\)-source, and planned to do so by performing some fixings that force \(\S_1\) to be constant. The fixings that we will perform are exactly on the random variables \((\S_1,\W_1)\). To analyze the probability that \(\S_1\oplus\S_2\) is \(\eps\)-close to an \((m,r)\)-source under these fixings, we will use the above-constructed sequence for help. In more detail, let \(\xi\) and \(g^\pr\) be parameters that we will set later. Then, note that
\begin{align*}
&\Pr_{(s_1,w_1)\sim(\S_1,\W_1)}\left[\left(\S_1\oplus\S_2\mid\S_1=s_1,\W_1=w_1\right)\text{ is not \(\xi\)-close to an \((m,m-g^\pr)\)-source}\right]\\
=&\Pr_{(s_1,w_1)\sim(\S_1,\W_1)}\left[\left(s_1\oplus\S_2\mid\S_1=s_1,\W_1=w_1\right)\text{ is not \(\xi\)-close to an \((m,m-g^\pr)\)-source}\right]\\
=&\Pr_{(s_1,w_1)\sim(\S_1,\W_1)}\left[\left(\S_2\mid\S_1=s_1,\W_1=w_1\right)\text{ is not \(\xi\)-close to an \((m,m-g^\pr)\)-source}\right]\\
=&\Pr_{(s_1,w_1)\sim(\S_1,\W_1)}\left[\left(\sExt_2^\pr(\X,\W_2)\mid\S_1=s_1,\W_1=w_1\right)\text{ is not \(\xi\)-close to an \((m,m-g^\pr)\)-source}\right].
\end{align*}
Now, since we know that \((\S_1,\W_1,\X,\W_2)\) is \(2^{g}\eps_2\)-close to \((\S_1^\pr,\W_1^\pr,\X^\pr,\W_2^\pr)\), we can apply \cref{claim:hard-closeness-smoothness} to upper bound the above by
\begin{align}\label{eq:another-eq-here-six-am}
\leq \Pr_{(s_1,w_1)\sim(\S_1^\pr,\W_1^\pr)}\Big[(\sExt_2^\pr(\X^\pr,\W_2^\pr)\mid\S_1^\pr=s_1,&\W_1^\pr=w_1)\text{ is not }\\&\text{\(\xi/2\)-close to an \((m,m-g^\pr)\)-source}\Big]+4\cdot2^g\eps_2/\xi + 2^g\eps_2.\notag
\end{align}
In order to continue bounding this probability, we can now apply our fixing lemma (\cref{lem:fixing-randomness-against-block}) as follows. First, note that we are dealing with random variables \((\X^\pr,\W_2^\pr)\) and \((\S_1^\pr,\W_1^\pr)\), where \((\X^\pr,\W_2^\pr)\) is an \(((n,k),(d_2,d_2-g))\)-block source, and \(\S_1^\pr,\W_1^\pr\) are supported on sets of size \(2^m\) and \(2^{d_1}\), respectively. Furthermore, note that \((\W_2^\pr\mid\X^\pr=x,\W_1^\pr=w_1)\) and \((\S_1^\pr\mid\X^\pr=x,\W_1^\pr=w_1)\) are independent, for all fixed \(x,w_1\). Indeed, this is simply because we constructed \(\S_1^\pr\) to be constant upon any fixing of \(\X^\pr,\W_1^\pr\). Plugging these observations into \cref{lem:fixing-randomness-against-block}, we immediately get that
\[
\Pr_{(s_1,w_1)\sim(\S_1^\pr,\W_1^\pr)}\left[(\X^\pr,\W_2^\pr\mid\S_1^\pr=s_1,\W_1^\pr=w_1)\text{ is not }2\sqrt{\nu}\text{-close to an \(((n,k^\prpr),(d_2,\ell^\prpr))\)-block source}\right]\leq2\sqrt{\nu},
\]
where \(k^\prpr=k-(m+d_1+\log(1/\nu))\) and \(\ell^\prpr=d_2-(g+d_1+\log(1/\nu))\). Now, consider any \(((n,k^\prpr),(d_2,\ell^\prpr))\)-block source \((\A,\B)\), and think about what happens when you plug it into \(\sExt_2^\pr:\zo^n\times\zo^{d_2}\to\zo^m\), which also works as a seeded \((n,k_0)\to_{\eps_2^\pr}(m,m)\) condenser. Since every seeded condenser also works for block sources (\cref{lem:seeded-condensers-work-on-block-sources}), it follows that \(\sExt_2^\pr(\A,\B)\) is \((2^{d_2-\ell^\prpr}\cdot\eps_2^\pr)\)-close to a source with min-entropy at least \(m-(d_2-\ell^\prpr)\), provided that \(k^\prpr\geq k_0\). Moreover, as we saw before, if \((\A,\B)\) is \(\eta\)-close to an \(((n,k^\prpr),(d_2,\ell^\prpr))\)-block source, then \(\sExt_2^\pr(\A,\B)\) is still guaranteed to be \((2^{d_2-\ell^\prpr}\cdot\eps_2^\pr + \eta)\)-close to a source with min-entropy at least \(m-(d_2-\ell^\prpr)\). Put differently, if \(\sExt_2^\pr(\A,\B)\) were \emph{not} this close to such a high entropy source, then we know that \((\A,\B)\) is also not \(\eta\)-close to an \(((n,k^\prpr),(d_2,\ell^\prpr))\)-block source.

By the discussion above, we know that if we set \(\xi/2:=(2^{d_2-\ell^\prpr}\eps_2^\pr+\eta)\) and \(\eta=2\sqrt{\nu}\) and \(g^\pr=d_2-\ell^\prpr\), we can upper bound \cref{eq:another-eq-here-six-am} by
\[2\sqrt{\nu}+4\cdot2^g\eps_2/\xi+2^g\eps_2\leq2\sqrt{\nu}+5\cdot2^g\eps_2/\xi.
\]
In summary, we get that
\begin{align*}
\Pr_{(s_1,w_1)\sim(\S_1,\W_1)}\Big[\left(\S_1\oplus\S_2\mid\S_1=s_1,\W_1=w_1\right)\text{ is not \(\xi\)-close to an \((m,m-g^\pr)\)-source}\Big]\leq2\sqrt{\nu}+5\cdot2^g\eps_2/\xi.
\end{align*}
By a standard fact about convex combinations (\cref{fact:convex-combo-error-to-closeness}), it immediately follows that \(\S_1\oplus\S_2\) is \((2\sqrt{\nu}+5\cdot2^g\eps_2/\xi)\)-close to a convex combination of sources that are \(\xi\)-close to an \((m,m-g^\pr)\)-source. As such, it holds that \(\S_1\oplus\S_2\) is \((2\sqrt{\nu}+5\cdot2^g\eps_2/\xi+\xi)\)-close to a convex combination of \((m,m-g^\pr)\)-sources. And since a convex combination of \((m,m-g^\pr)\)-sources is, itself, an \((m,m-g^\pr)\)-source, we conclude that \(\S_1\oplus\S_2\) is \((2\sqrt{\nu}+5\cdot2^g\eps_2/\xi+\xi)\)-close to an \((m,m-g^\pr)\)-source. Finally, recall that
\begin{align*}
2\sqrt{\nu}+5\cdot2^g\eps_2/\xi+\xi&=6\sqrt{\nu}+\frac{5\cdot2^g\eps_2}{2^{d_2-\ell^\prpr}\eps_2^\pr+4\sqrt{\nu}}+2^{d_2-\ell^\prpr}\eps_2^\pr\\
&=6\sqrt{\nu}+\frac{5\cdot2^g\eps_2}{2^{g+d_1+\log(1/\nu)}\eps_2^\pr+4\sqrt{\nu}}+2^{g+d_1+\log(1/\nu)}\eps_2^\pr,
\end{align*}
where \(\nu>0\) can be taken as anything. Taking it to be \(\nu:=\sqrt{\eps_2^\pr}\) allows us to upper bound the above by
\[
\leq 6(\eps_2^\pr)^{1/4}+\frac{6\eps_2}{2^{d_1}\sqrt{\eps_2^\pr}}+2^{g+d_1}\sqrt{\eps_2^\pr}.
\]

Then, taking \(\eps_2^\pr=\eps_2\cdot2^{-2d_1}\) allows us to upper bound the above by \(\leq 2^{g+4}\cdot\eps_2^{1/4}\).

Furthermore, recall that
\[
g^\pr=d_2-\ell^\prpr=g+d_1+\log(1/\nu)=g+2d_1+\log(1/\eps_2)/2.
\]
Recall that to make everything work, we needed \(k^\prpr\geq k_0\), and plugging in our definition of \(k^\prpr\) from before, this requirement becomes
\[
k\geq k_0 + m + d_1 + \log(1/\nu)=k_0+m+2d_1+\log(1/\eps_2)/2.
\]
Thus, we get that the output of the non-malleable condenser is \(2^{g+4}\eps_2^{1/4}\)-close to an \((m,m-(g+2d_1+\log(1/\eps_2)/2))\)-source, as long as \(k\geq k_0+m+2d_1+\log(1/\eps_2)/2\) and \(\eps_2^\pr=\eps_2\cdot2^{-2d_1}\).
\end{proof}

\subsection{The main explicit condenser}

Using the tools developed above, we can now construct our main explicit condenser for CG sources.

\begin{theorem}[The main explicit condenser for CG sources - \cref{thm:main:intro:explicit-condensers-CG-sources}, restated]\label{thm:main:technical:explicit-condensers-CG-sources}
    For any \(\alpha>0\), there exists a constant \(C\geq1\) such that the following holds. For all \(t,n\in\N\) and \(\delta,\eps>0\), there exists an explicit condenser \(\cond:(\zo^n)^t\to\zo^{k^\pr+g^\pr}\) for \((t,n,k=\delta n = n-g)\)-CG sources which has output entropy \(k^\pr\geq(1-\alpha)kt\), output gap \(g^\pr\leq C\cdot(1/\delta)^C\cdot(g+\log(1/\eps))\), and error \(\eps\).
\end{theorem}

The proof proceeds via three steps. First, in \cref{subsubsec:building-non-malleable}, we explicitly construct a non-malleable condenser for CG sources (using our framework from \cref{subsec:non-malleable-from-seeded-extractors}). Then, in \cref{subsubsec:condensing-to-very-high-rate}, we present the main part of our condenser, which uses our new non-malleable condenser in the ``purification'' framework from \cref{subsec:purification} in order to condense CG sources to rate \(0.99\). Finally, in \cref{subsubsec:getting-the-rest-of-entropy-out}, we show how to get the remaining entropy out of the source, while maintaining a very small gap, by showing that the classical iterative condensing framework of Nisan and Zuckerman \cite{nisan1996randomness} can be extended to handle a correlated seed.

\subsubsection{Building a non-malleable condenser}\label{subsubsec:building-non-malleable}

We proceed to build our non-malleable condenser for CG sources. We prove the following.

\begin{theorem}[Explicit non-malleable condensers]\label{thm:main-non-malleable-condenser}
For every constant \(\alpha>0\), there exist constants \(C\geq1\) and \(\gamma>0\) such that the following holds. There exists an explicit non-malleable condenser (with advice) \(\nmCond:\zo^n\times\zo^n\times\zo^d\times[2]\to\zo^m\) for \(((n,k),(n,k),(d,(1-\gamma)d))\)-block sources with error \(\eps\), output length \(m=\lfloor(\frac{1}{2}-\alpha)k-C\log(n/\eps)- d\rfloor\), and output gap \(g^\pr \leq C\log(n/\eps)+ d\), provided that \(d\geq C\log(n/\eps)\).
\end{theorem}

Our construction will follow by simply plugging in known seeded extractors into our recipe from \cref{subsec:non-malleable-from-seeded-extractors}. We will use the following classical extractors of Guruswami, Umans, and Vadhan.

\begin{theorem}[Explicit seeded extractors \cite{guruswami2009unbalanced}]\label{theorem:guv-extractor}
For every constant \(\alpha>0\), there is a constant \(C>0\) such that the following holds. There exists an explicit \((k,\eps)\)-seeded extractor \(\sExt:\zo^n\times\zo^d\to\zo^m\) with output length \(m\geq(1-\alpha)k\), as long as \(d\geq C\log(n/\eps)\).
\end{theorem}

With this tool in hand, we are ready to construct our non-malleable condensers.

\begin{proof}[Proof of \cref{thm:main-non-malleable-condenser}]
We simply plug \cref{theorem:guv-extractor} into \cref{lem:non-malleable-from-seeded}, and pick parameters appropriately.

In more detail, we need to find extractors
\begin{itemize}
    \item \(\sExt_1:\zo^n\times\zo^{p_1}\to\zo^{d_1}\) a \((k_0,\eps_1)\)-seeded extractor,
    \item \(\sExt_1^\pr:\zo^n\times\zo^{d_1}\to\zo^m\) a \((k_0,\eps_1^\pr)\)-seeded extractor,
    \item \(\sExt_2:\zo^n\times\zo^{p_2}\to\zo^{d_2}\) a \((k_0,\eps_2)\)-seeded extractor,
    \item \(\sExt^\pr_2:\zo^n\times\zo^{d_2}\to\zo^m\) a \((k_0,\eps_2^\pr)\)-seeded extractor,    
\end{itemize}
with parameters \(p_1,d_1,p_2,d_2,d_1,m,d_2,k_0,\eps_1,\eps_1^\pr,\eps_2,\eps_2^\pr\) that result in the non-malleable condensers we claim (using \cref{lem:non-malleable-from-seeded}). To make this easy, we start by focusing on achieving error \(\eps\). In order to do so, we define \(g:=\gamma d\) (for some constant \(\gamma\) to be fixed later), and note that \cref{lem:non-malleable-from-seeded} says that we can just pick \(\eps_1\) such that \(2^{g+p_2+3}\eps_1^{1/4}\leq\eps/2\) and \(2^{g+4}\eps_2^{1/4}\leq\eps/2\). Moreover, it always requires that we have \(\eps_1=\eps_1^\pr\) and \(\eps_2^\pr=\eps_2\cdot2^{-2d_1}\). Thus we pick errors \(\eps_1=\eps^42^{-4(g+p_2+4)}\), \(\eps_1^\pr=\eps_1\), \(\eps_2=\eps^4\cdot2^{-(g+5)4}\), and \(\eps_2^\pr=\eps_2^{-2d_1}\). This satisfies the error requirement.

Now, in order to explicitly construct these extractors, we invoke \cref{theorem:guv-extractor} so that we can handle the smallest possible seed length. Thus, we pick

\begin{itemize}
    \item \(p_2=C\log(n/\eps_2)=4C\cdot(\log(n/\eps)+g+5)=O(\log(n/\eps)+g)\),
    \item \(p_1=C\log(n/\eps_1)=4C\cdot(\log(n/\eps)+g+p_2+4)=O(\log(n/\eps)+g)\),
    \item \(d_1=C\log(n/\eps_1^\pr)=4C\cdot(\log(n/\eps)+g+p_2+4)=O(\log(n/\eps)+g)\),
    \item \(d_2=C\log(n/\eps_2^\pr)=4C\cdot(\log(n/\eps)+g+5+d_1/2)=O(\log(n/\eps)+g)\).
\end{itemize}

To make this work, \cref{lem:non-malleable-from-seeded} also says that we need \(k_0 \leq k - (m + 2d_1+d_2+p_2+\log(1/\eps_1)+\log(1/\eps_2))\), and recall that the right hand side is at least \(k-m-O(\log(n/\eps)+g)\). Thus we pick \(k_0\) to be this value, and all that remains is to check that we didn't ask one of the extractors to output more bits than \((1-\alpha)k_0\). For this, we simply need that \(m\leq (1-\alpha)k_0 = (1-\alpha)(k-m-O(\log(n/\eps)+g))\), or rather that \(m\leq(1/2-\alpha)k - O(\log(n/\eps)+g)\). Furthermore, recall that \(p_1,p_2\) are prefixes of \(d\), so we need \(d\geq p_1,p_2 = O(\log(n/\eps)+g)\). Now that all the conditions are satisfied, we get from \cref{lem:non-malleable-from-seeded} that the entropy gap is \(O(g + \log(n/\eps))\). To conclude, recall that \(g=\gamma d\), and set \(\gamma\) to a sufficiently small constant.
\end{proof}

\subsubsection{Condensing to rate \(0.99\)}\label{subsubsec:condensing-to-very-high-rate}

Now that we have our non-malleable condensers, we are ready to construct the core component of our main explicit condenser for CG sources. In this section, we prove the following.

\begin{lemma}[Condensing to rate \(0.99\)]\label{lem:technical:main-condenser:condensing-to-high-rate}
For any constants \(\alpha,C_0>0\), there exist constants \(C_1,C_2,C_3\geq C_0\) such that the following holds. There exists an explicit condenser \(\cond:(\zo^n)^t\to\zo^m\) for \((t,n,\delta n)\)-CG sources with output length \(m\in[0.05\delta n \tau^\star,\delta n \tau^\star]\), output entropy \(k^\pr\geq(1-\alpha)m\), and error \(\eps\), provided
\[
t\geq \tau^\star:=C_1\cdot\left((1/\delta)^{C_2} + (1/\delta)^{C_3}\log(1/\eps)/n\right).
\]
\end{lemma}

As discussed, the key idea is to instantiate our purification framework from \cref{subsec:purification} with a baseline somewhere-condenser and a non-malleable condenser. For our non-malleable condenser, we'll use the new one constructed above. For the baseline somewhere-condenser, we'll use a classical construction due to Barak, Kindler, Shaltiel, Sudakov, and Wigderson \cite{barak2010simulating} and Raz \cite{raz2005extractors} (see also \cite[Theorem 3.2]{zuckerman2007linear}).

\begin{theorem}[\protect{Explicit somewhere-condensers \cite{barak2010simulating,raz2005extractors}}]\label{thm:baseline-somewhere-condenser-zuckerman}
For every constant \(\beta>0\), there exist constants \(C_1,C_2,C_3\geq1\) such that the following holds. For any \(\delta=\delta(n)>0\), there exists an explicit somewhere-\(k^\pr\)-condenser \(\sCond:\zo^n\to(\zo^m)^D\) for \((n,\delta n)\)-sources with output length \(m=\lfloor\delta^{C_1} n\rfloor\), output entropy \(k^\pr\geq(1-\beta)m\), error \(\eps=2^{-\delta^{C_2}n}\), and \(D=\lceil(1/\delta)^{C_3}\rceil\) rows.
\end{theorem}

We are now ready to condense CG sources to rate \(0.99\), and prove the core lemma of this paper.

\begin{proof}[Proof of \cref{lem:technical:main-condenser:condensing-to-high-rate}]

Let \(\X\sim(\zo^n)^t\) be a \((t,n,k:=\delta n)\)-CG source. The idea is to expand the last block of \(\X\) into a somewhere-random (SR) source (using \cref{thm:baseline-somewhere-condenser-zuckerman}), and then proceed in iterations. In each iteration, we will halve the number of rows in the SR source, using our non-malleable condenser (\cref{thm:main-non-malleable-condenser}) and our purification lemma (\cref{lem:purify}).

At a high level, in order for this to work, the \emph{row length} of the SR source must line up with the \emph{seed length} requirement of the non-malleable condenser, and the \emph{entropy rate} of the (good row of the) SR source must line up with the \emph{seed (entropy) rate} requirement of the non-malleable condenser (which is roughly \(0.99\)). Furthermore, after we have halved the number of rows in the SR source with one application of the purification lemma, we need to make sure that the new SR source has a row length and row entropy rate that is good enough for another application of the purification lemma. To make sure this happens, the output entropy rate of the first non-malleable condenser calls must be at least \(0.99\). But since the output gap of the non-malleable condenser is always a constant factor larger than the gap of its seed (see \cref{thm:main-non-malleable-condenser}), we must make sure that its output length is also a constant factor larger than the length of its seed. And to make this happen, we must make sure that each of the two input sources to the non-malleable condenser has enough min-entropy. This is possible by concatenating several blocks of the CG source into a single ``super-block.'' Finally, we will continue to iterate until there is just a single row left in the SR source.

Thus, the game plan is as follows. First, we fix an arbitrary constant \(\alpha>0\), which will represent the allowed missing entropy rate in the final output of the condenser. Then, we let \(\eps>0\) denote another parameter, which will represent the target final error of the condenser.\footnote{Note that \(\eps\) can depend on all other parameters arbitrarily, and thus need not be a constant.} We also set up intermediate error values \(\eps_0,\eps_1,\eps_2,\dots\), which will represent the allowed error in each iteration of the procedure outlined above. We make sure that these are in a decaying geometric series, so that they will sum up to our overall target error \(\eps\). Finally, we determine the number of blocks that must be concatenated at each iteration (before passing them into the non-malleable condenser to collapse the SR source) in order satisfy all the requirements mentioned above. At the end, we sum up the total number of blocks we needed to fully collapse the SR source, and define \(\tau\) to be exactly this value, in order to finish the proof.

More formally now, fix a parameter \(\beta>0\) to either \(\alpha\) (from the current lemma statement) or \(\gamma\) (from \cref{thm:main-non-malleable-condenser}, when its first parameter is fixed to \(0.01\)) - whichever is smaller.\footnote{Furthermore, if both \(\alpha,\gamma>1/2\), set \(\beta:=1/2\).} Then, let \(b_0\in\N\) be a ``block parameter'' that we will fix later, and let \(\sCond_0:\zo^{nb_0}\to(\zo^{m_0})^D\) be an explicit somewhere-\(k_0^\pr\)-condenser for \((nb_0,\delta nb_0)\)-sources with output length \(m_0=\lfloor \delta^{C_1}nb_0\rfloor\), output entropy \(k_0^\pr\geq(1-\beta)m_0\), error \(\eps_0=2^{-\delta^{C_2}nb_0}\), and \(D=\lceil(1/\delta)^{C_3}\rceil_2\) rows, where \(\lceil x\rceil_2\) denotes the rounding of \(x\) up to the closest power of \(2\). Such an explicit somewhere-condenser exists due to \cref{thm:baseline-somewhere-condenser-zuckerman}.\footnote{\cref{thm:baseline-somewhere-condenser-zuckerman} technically doesn't guarantee that the number of rows will be a power of \(2\), but we can easily make this happen by appending the appropriate number of dummy rows (each set to the all zeroes string) to the output of the somewhere-condenser.}

Next, suppose there exists a sequence of explicit functions \(\nmCond_1,\nmCond_2,\dots,\nmCond_d\), where each \(\nmCond_i\) is an explicit non-malleable condenser (with advice) for \(((nb_i,\kappa_i),(nb_i,\kappa_i),(m_{i-1},k_{i-1}^\pr))\)-block sources with error \(\eps_i\), output length \(m_i=\lfloor0.49\kappa_i - C_4\log(nb_i/\eps_i)-m_{i-1}\rfloor\), and output entropy \(k^\pr_i\geq m_i - C_4\log(nb_i/\eps_i)-m_{i-1}\), where:
\begin{itemize}
    \item \(\kappa_i:= kb_i - d - \log(1/\eps_i)\),
    \item \(C_4\) is the constant \(C\) from \cref{thm:main-non-malleable-condenser} (when the first constant in that theorem is set to \(0.01\)), and
    \item all other parameters (appearing above) not yet set will be set later.
\end{itemize}

If such a sequence of explicit functions \(\nmCond_1,\dots,\nmCond_d\) actually exists, our purification lemma (\cref{lem:purify}) immediately tells us that we can use them (along with \(\sCond_0\)) to iteratively create a sequence of explicit somewhere-condensers \(\sCond_1,\dots,\sCond_d\), where \(\sCond_d\) is in fact an explicit condenser for \((\tau,n,k)\)-CG sources with error
\[
\eps_0+\sum_{i=1}^d(4\sqrt{\eps_i}+\eps_i),
\]
output length \(m_d\), output entropy \(k_d^\pr\), and \(\tau=b_0+2\sum_{i=1}^db_i\).

Thus, all that remains is to set the error parameters \(\eps_0,\eps_1,\dots,\eps_d\) and block parameters \(b_0,b_1,\dots,b_d\) so that (1) the explicit non-malleable condensers (described above) actually exist, (2) the overall error is at most \(\eps\), and (3) the output length \(m:=m_d\) is in the range \(m_d\in[\delta n\tau/4,\delta n\tau]\), (4) the output entropy \(k^\pr:=k_d\) satisfies \(k_d^\pr\geq(1-\beta)m_d\), and (5) the threshold value \(\tau\) matches its value in the lemma statement.

Let's start by satisfying the error requirement, listed as item (2) above. Towards this end, recall that we actually already set \(\eps_0:=2^{-\delta^{C_2}nb_0}\) above, so we can only control the parameter \(\eps_0\) via the unset parameter \(b_0\). On the other hand, we have not yet set the other error parameters. We do so now, and set \(\eps_i:=(\frac{\eps}{10\cdot2^i})^2\) for every \(i\in[d]\), making the overall error of the condenser at most \(\eps_0+\eps/2\). Thus, in order to ensure that the overall error is at most \(\eps\), we just need that \(\eps_0=2^{-\delta^{C_2}nb_0}\leq\eps/2\), or rather that \(b_0\geq\frac{\log(2/\eps)}{\delta^{C_2}n}\). Thus, the only unset parameters remaining are the block parameters \(b_0,b_1,\dots,b_d\), and as long as we ultimately set \(b_0\) so that it satisfies the above inequality, then the error requirement will be satisfied.

We now turn towards satisfying requirement (1) from above, which states that the explicit non-malleable condensers \(\nmCond_1,\dots,\nmCond_d\) actually exist. In order for this to happen, we just need to make sure that each non-malleable condenser is given a long enough seed, and that this seed has a high enough entropy rate (as dictated by \cref{thm:main-non-malleable-condenser}). Towards this end, notice that for each \(i\in[d]\), the non-malleable condenser \(\nmCond_i\) defined above is given an \((m_{i-1},k_{i-1}^\pr)\)-source as a seed, where
\begin{align}\label{eq:seed-parameters:first-block}
m_0&=\lfloor\delta^{C_1}nb_0\rfloor,\\
k_0^\pr&=(1-\beta)m_0,
\end{align}
and for every \(i\in[2,d]\),
\begin{align}\label{eq:seed-parameters:later-blocks}
    m_{i-1}&=\left\lfloor0.49\kappa_{i-1}-C_4\log(nb_{i-1}/\eps_{i-1})-m_{i-2}\right\rfloor,\\
    k_{i-1}^\pr&\geq m_{i-1} - C_4\log(nb_{i-1}/\eps_{i-1})-m_{i-2},
\end{align}
where recall that we defined
\begin{align*}
\kappa_{i-1}&:=kb_{i-1}-d-\log(1/\eps_{i-1})\\
&=kb_{i-1}-\log\lceil(1/\delta)^{C_3}\rceil_2-\log(1/\eps_{i-1})\\
&\geq kb_{i-1}-C_3\log(1/\delta)-1-\log(1/\eps_{i-1}).
\end{align*}
Now, \cref{thm:main-non-malleable-condenser} tells us that in order for the non-malleable condensers to exist, we just need the following:
\begin{itemize}
    \item \textbf{Sufficient seed length:} \(m_{i-1}\geq C_4\log(nb_i/\eps_i)\) for all \(i\in[d]\).
    \item \textbf{Sufficient seed entropy:} \(k_{i-1}^\pr\geq(1-\beta)m_{i-1}\) for all \(i\in[d]\).
\end{itemize}

Notice that when \(i=1\), the sufficient seed entropy condition is already satisfied. And when \(i>1\), the sufficient seed entropy condition becomes \(m_{i-1}\geq \frac{C_4}{\beta}\log(nb_{i-1}/\eps_{i-1})+\frac{1}{\beta}m_{i-2}\), due to the known lower bound on \(k_{i-1}^\pr\) given earlier. Thus, we just need to set block parameters so that the following are satisfied:
\begin{align*}
    m_0&\geq C_4\log(nb_1/\eps_1),\\
    m_{i-1}&\geq C_4\log(nb_i/\eps_i)+C_4\log(nb_{i-1}/\eps_{i-1})/\beta+m_{i-2}/\beta, \text{for all \(i\in[2,d]\)}.
\end{align*}

In order to make sure the above inequalities are satisfied, let us make them easier to digest. To do so, recall that we previously set the intermediate error parameters so that \(1\geq\eps_1\geq\dots\geq \eps_d\), and we will later set block parameters so that \(2\leq b_1\leq\dots\leq b_d\).\footnote{This will allow us to use convenient estimates, such as \(\log(b_{i-1})\geq1\) and \(\log(\frac{b_ib_{i-1}}{\eps_i\eps_{i-1}})\leq2\log(b_i/\eps_i)\) for all \(i\geq2\).} Next, note that \(m_{i-2}\leq kb_{i-2}\) for all \(i\geq2\). Using these observations, it is straightforward to plug in the actual values for \(m_{i-1}\) (from \cref{eq:seed-parameters:first-block,eq:seed-parameters:later-blocks}) so that the conditions above (that we need to satisfy) are satisfied if both of the following hold:
\begin{align*}
    \delta^{C_1}nb_0&\geq 2C_4\log(nb_1/\eps_1),\\
    kb_{i-1}&\geq3C_3\log(1/\delta) + \frac{18C_4}{\beta}\log(\frac{nb_i}{\eps_i}) + \frac{6}{\beta}kb_{i-2},\text{ for all \(i\in[2,d]\)}.
\end{align*}

Now, recall that we previously defined the error parameters as \(\eps_i=(\frac{\eps}{10\cdot2^i})^2\), for all \(i\in[d]\). Thus we have \(\eps_1\geq(\eps/20)^2\), and since we previously defined \(d=\log\lceil(1/\delta)^{C_3}\rceil\), we get \(\eps_i\geq(\eps\delta^{C_3}/20)^2\) for all \(i\in[2,d]\). Thus the two conditions above are satisfied if both of the following hold:
\begin{align*}
    \delta^{C_1}nb_0&\geq24C_4\log(nb_1/\eps),\\
    kb_{i-1}&\geq\frac{39C_3C_4}{\beta}\log(1/\delta)+\frac{216C_4}{\beta}\log(nb_i/\eps)+\frac{6}{\beta}kb_{i-2},\text{ for all \(i\in[2,d]\)}.
\end{align*}
We can rewrite these in terms of block requirements as follows (recalling that \(k=\delta n\)):
\begin{align*}
    b_0&\geq\frac{24C_4}{\delta^{C_1}n}\log(nb_1/\eps),\\
    b_{i-1}&\geq \frac{39C_3C_4}{\beta \delta n}\log(1/\delta) + \frac{216C_4}{\beta\delta n}\log(nb_i/\eps)+\frac{6}{\beta}b_{i-2},\text{ for all \(i\in[2,d]\)}.
\end{align*}
Now, if we define the constant \(C_5:=256C_1C_2C_3C_4/\beta\), then the above conditions are satisfied if both
\begin{align}
    b_0&\geq C_5\log(nb_1/\eps)/(\delta^{C_5}n),\label{eq:block-parameter-condition-1}\\
    b_{i-1}&\geq C_5 \log(nb_i/\eps)/(\delta^{C_5}n) + C_5 b_{i-2},\label{eq:block-parameter-condition-2}\text{ for all \(i\in[2,d]\)}.
\end{align}
Furthermore, recall that in order for the overall condenser to have error \(\eps\), we needed \(b_0\geq\log(2/\eps)/(\delta^{C_2}n)\), and this is indeed implied by the first condition above. Thus, we have arrived at sufficient conditions on the block parameters \(b_0,\dots,b_d\) for the explicit non-malleable condensers \(\nmCond_1,\dots,\nmCond_d\) to actually exist, and for the overall error of the final condenser to be at most \(\eps\). In fact, using an almost identical argument to the one given above, it is also straightforward to show that the overall output length \(m_d\) is in the range \(m_d\in[0.4kb_d,0.49kb_d]\), and the overall output entropy is \(k^\pr_d\geq(1-\beta)m_d\), as long as
\begin{align}
    b_d\geq C_6\log(nb_d/\eps)/(\delta^{C_6}n)+C_6b_{d-1}\label{eq:block-parameter-condition-3}
\end{align}
for some constant \(C_6\geq1\).\footnote{Recall that we actually originally requested that \(m_d\geq k\tau/4\), instead of \(m\geq0.4kb_d\). However, we will soon show that this follows from our setting of \(\tau\).} Thus, we now wish to set block parameters so they satisfy \cref{eq:block-parameter-condition-1,eq:block-parameter-condition-2,eq:block-parameter-condition-3}.

Towards this end, we let \(A\in\N\) be a sufficiently large constant, and set block parameters as follows:
\begin{align*}
b_0&:=\left\lceil\frac{ \log(1/\eps)}{\delta^An}\right\rceil,\\
b_i&:=A\cdot b_{i-1},\text{ for all }i\in[d-1],\\
b_d&:=\left\lceil A^{C_3\log(1/\delta)+1}\right\rceil\cdot b_0
\end{align*}

It is straightforward to verify that for all sufficiently large \(A\) (as a function of the constants \(C_5,C_6\)), all of \cref{eq:block-parameter-condition-1,eq:block-parameter-condition-2,eq:block-parameter-condition-3} hold. Moreover, we make sure to pick \(A\geq C_0\).

All that remains is to check the total number of blocks used, and to ensure that the overall output length \(m_d\) is sufficiently large. Towards this end, the total number of blocks used is
\begin{align*}
\tau&:=b_0+2\sum_{i=1}^db_i\\
&\leq8A\cdot\left((1/\delta)^{C_3\log A} + (1/\delta)^{C_3\log A + A}\log(1/\eps)/n\right)\\
&=:\tau^\star,
\end{align*}
while the overall output length is in the range \(m_d\in[0.4kb_d,0.49kb_d]\), which means
\begin{align*}
    m_d&\geq 0.4kb_d\\
    &\geq0.2Ak\cdot\left((1/\delta)^{C_3\log A} + (1/\delta)^{C_3\log A + A}\log(1/\eps)/n\right)\\
    &\geq 0.025 k \tau^\star.
\end{align*}

Of course, the fact that \(m_d\leq0.49kb_d\) also implies that \(m_d\leq k\tau^\star\) (since we set \(\tau^\star\gg b_d\) above). Thus, to conclude, as long as our CG-source originally started off with
\[
t\geq\tau^\star:=8A\cdot\left((1/\delta)^{C_3\log A}+(1/\delta)^{C_3\log A + A}\log(1/\eps)/n\right)
\]
blocks, we can obtain \(m_d\in[0.01k\tau^\star,k\tau^\star]\) output bits that are \(\eps\)-close to min-entropy \(k_d^\pr\geq(1-\beta)m_d\).
\end{proof}

\subsubsection{Condensing the rest of the entropy out}\label{subsubsec:getting-the-rest-of-entropy-out}

In this final step, we show how to get the rest of the entropy out of the CG source, while maintaining the gap, via \emph{iterative condensing}. We prove the following, which will later be combined with our core lemma (\cref{lem:technical:main-condenser:condensing-to-high-rate}) in order to yield our main theorem (\cref{thm:main:technical:explicit-condensers-CG-sources}).

\begin{lemma}[Condensing the rest of the entropy out]\label{lem:technical:main-condenser:post-processing}
    For every constant \(\alpha>0\), there is a constant \(C>0\) such that there exists an explicit condenser \(\cond:\zo^{n_1}\times\dots\times\zo^{n_t}\to\zo^m\) for \(((n_1,k_1),\dots,(n_t,k_t))\)-block sources with output length \(m\geq(1-\alpha)k_1\), output gap \(g^\pr=g:=n_t-k_t\), and error \(\eps\), provided
    \[
    k_{i+1}\geq C(\log(n_i/\eps) + (t-i)+ g)
    \]
    for all \(i\in[t-1]\).
\end{lemma}

\begin{proof}
We simply plug the GUV extractor (\cref{theorem:guv-extractor}) into our iterative condensing framework (\cref{lem:iterated-condensing-framework}), recalling that an extractor is simply a condenser with output gap \(0\).

In more detail, if we define \(m_t:=n_t\), then by \cref{theorem:guv-extractor}, the following holds. There exists a sequence of explicit functions \(\sCond_1,\sCond_2,\dots,\sCond_{t-1}\), where each \(\sCond_i:\zo^{n_i}\times\zo^{m_{i+1}}\to\zo^{m_i}\) is a seeded \((n_i,k_i)\to_{\eps_i}(m_i,m_i)\) condenser with output length \(m_i\geq(1-\alpha)k_i\), as long as
\[
(1-\alpha)k_{i+1}\geq C_\guv \log(n_i/\eps_i)
\]
for every \(i\in[t-1]\) (where \(C_\guv\) is a constant depending only on \(\alpha\)).\footnote{Note that when \(i=t-1\), the requirement is actually \(m_{i+1}=n_{i+1}\geq C_\guv\log(n_i/\eps_i)\), which is weaker than what is written.} Since we may assume that \(\alpha<1/2\),\footnote{This is because the lemma statement only claims a lower bound on the output length \(m\).} this requirement is satisfied when \(k_{i+1}\geq C\log(n_i/\eps_i)\), where we have used \(C:=2C_\guv\). And if we set \(\eps_i:=\eps\cdot2^{-g-(t-i)}\) for all \(i\in[t-1]\), then the requirement is satisfied when
\[
k_{i+1}\geq C(\log(n_i/\eps)+(t-i)+g)
\]
for all \(i\in[t-1]\). Now, by our iterative condensing framework (\cref{lem:iterated-condensing-framework}), this sequence of explicit functions \(\sCond_1,\sCond_2,\dots,\sCond_{t-1}\) can be composed to create an explicit condenser for \(((n_1,k_1),\dots,(n_t,k_t))\)-block sources with output length \(m_1\geq (1-\alpha)k_1\), output gap \(g^\pr:=\sum_{i\in[t-1]}(m_i-m_i)+g=g\), and error
\[
\sum_{i\in[t-1]}\eps_i\cdot2^g=\eps\sum_{i\in[t-1]}2^{-(t-i)}\leq\eps,
\]
as desired.
\end{proof}

While the above lemma is quite general, the following corollary will be more useful for our purposes.

\begin{corollary}[Condensing a geometric block source with a high-rate final block]\label{cor:post-processing:simple}
    For any constants \(\alpha_0>0\) and \(C_0\geq1\), there exist constants \(\beta>0\) and \(C\geq1\) such that the following holds. There exists an explicit condenser for \(((n_1,k_1),\dots,(n_t,k_t))\)-block sources with output length \(m\geq(1-\alpha_0)k_1\), output gap \(g^\pr=g:=n_t-k_t\), and error \(\eps\), provided that all of the following hold:
    \begin{itemize}
        \item \(k_1\geq4k_2\geq4^2k_3\geq\dots\geq4^{t-1}k_t\).
        \item \(n_1\leq (C_0 n_2)^2\leq (C_0 n_3)^{2^2}\leq\dots\leq (C_0 n_t)^{2^{t-1}}\).
        \item \(k_t\geq(1-\beta)n_t\).
        \item \(n_t\geq C\log(1/\eps)+C\).
    \end{itemize}
\end{corollary}
\begin{proof}
Let \(C^\star\) be the second constant from \cref{lem:technical:main-condenser:post-processing}, when the first constant is set to \(\alpha\). It suffices to show
\[
k_{i+1}\geq C^\star(\log(n_i/\eps)+(t-i)+g)
\]
for all \(i\in[t-1]\). This is straightforward via a backward induction on \(i\) (using the bullet points).
\end{proof}

\subsubsection*{Putting everything together}

At last, with all of our ingredients in place, we are ready to prove our main theorem.

\begin{proof}[Proof of \cref{thm:main:technical:explicit-condensers-CG-sources}]

Recall that we wish to construct an explicit condenser \(\cond:(\zo^n)^t\to\zo^{k^\pr+g^\pr}\) for \((t,n,k=\delta n = n-g)\)-CG sources, which has output entropy \(k^\pr\geq(1-\alpha)kt\), output gap \(g^\pr\leq(1/\delta)^C\cdot(g+\log(1/\eps))\), and error \(\eps\). Towards this end, let \(\X=(\X_1,\dots,\X_t)\) be a \((t,n,k)\)-CG source. The main idea is to use \cref{lem:technical:main-condenser:condensing-to-high-rate} to condense the last few blocks in \(\X\) to a block with rate \(0.99\), and then to use this high-rate block to get the rest of the entropy out of the source, using \cref{cor:post-processing:simple}.

More formally, set the constants \(\alpha_0,C_0\) in \cref{cor:post-processing:simple} to \(\alpha/2\) and \(100/\alpha\) (respectively), and let \(\beta^\star,C^\star\) denote the constants \(\beta,C\) (in that theorem) that come out. Then, set the constants \(\alpha,C_0\) in \cref{lem:technical:main-condenser:condensing-to-high-rate} to \(\beta^\star\) and \(2C^\star\) (respectively), and let \(C_1,C_2,C_3\) be the constants that come out (corresponding to the same-named constants in that lemma statement). Finally, define a size parameter \(s\) as
\[
s:=\left\lceil C_1\cdot\left((1/\delta)^{C_2}+(1/\delta)^{C_3}\log(2/\eps)/n\right)\right\rceil,
\]
and let \(w\) be the largest integer such that \(s\cdot\sum_{i=1}^w\lceil4/\alpha\rceil^{w-i}\leq t\). Note that if \(w<2\), then the claimed gap in the theorem statement is trivial, in that it can be achieved simply by applying the identity function.

Now, henceforth assuming that \(w\geq2\), define (for every \(i\in[w]\))
\[
s_i:=\begin{cases}
    s\cdot\lceil4/\alpha\rceil^{w-i}&\text{ if }i>1,\\
    t - s\cdot\sum_{i=2}^w\lceil4/\alpha\rceil^{w-i}&\text{ if }i=1.
\end{cases}
\]
Note that \(s_1\in[s\cdot\lceil4/\alpha\rceil^{w-1},s\cdot\lceil4/\alpha\rceil^{w+1}]\), and define a new source \(\Y=(\Y_1,\dots,\Y_w)\), where \(\Y_1\) consists of the first \(s_1\) blocks of \(\X\), \(\Y_2\) consists of the next \(s_2\) blocks of \(\X\), and so on. Note that \(\Y\) is an \(((ns_1,ks_1),\dots,(ns_w,ks_w))\)-block source.

Now, by \cref{lem:technical:main-condenser:condensing-to-high-rate}, there exists an explicit function \(\cond_1\) such that \(\Z:=\cond_1(\Y_w)\) is \((\eps/2)\)-close to a source with length \(m_w\in[0.025\delta n s,\delta n s]\) and min-entropy \(k_w^\pr\geq(1-\beta^\star)m_w\), and moreover, this is true for every fixing of the random variables \(\Y_1,\dots,\Y_{w-1}\). Thus, \(\Y^\star:=(\Y_1,\dots,\Y_{w-1},\Z)\) is a \(((0,0),\dots,(0,0),(0,\eps/2))\)-almost \(((ns_1,ks_1),\dots,(ns_{w-1},ks_{w-1}),(m_w,(1-\beta^\star)m_w))\)-block source. Thus, by \cref{lem:almost-means-close}, \(\Y^\star\) is \((\eps/2)\)-close to an \(((ns_1,ks_1),\dots,(ns_{w-1},ks_{w-1}),(m_w,(1-\beta^\star)m_w))\)-block source, \(\Y^{\star\star}\). Now, it is straightforward to verify (given our setting of parameters) that \(\Y^{\star\star}\) satisfies the requirements of \cref{cor:post-processing:simple}, and thus there is an explicit function \(\cond_2\) such that \(\cond_2(\Y^{\star\star})\) is \(\eps/2\)-close to a source of length \(m\geq(1-\alpha/2)ks_1\) and gap \(g^\pr\leq\beta^\star m_w\), and thus the data-processing inequality tells us that \(\cond_2(\Y^{\star})\) is \(\eps\)-close to a source of length \(m\geq(1-\alpha/2)ks_1\) and gap \(g^\pr\leq\beta^\star m_w\). Furthermore, note that by our setting of \(s_i\), we have \(m\geq(1-\alpha)kt\), and gap
\[
g^\pr\leq \beta^\star m_w\leq m_w\leq\delta ns\leq \delta n\cdot2 C_1\left((1/\delta)^{C_2}+(1/\delta)^{C_3}\log(2/\eps)/n\right),
\]
which is at most
\begin{align}\label{eq:final-gap-equation}
C\cdot(1/\delta)^C\cdot(n+\log(1/\eps))
\end{align}
for some constant \(C\geq1\). Finally, we  may assume that the original gap was \(g>\beta^\star n\), since otherwise we could easily obtain an output gap of the form \(g^\pr\leq C\cdot(1/\delta)^C\cdot(g+\log(1/\eps))\), simply by replacing \(\cond_1\) with the identity function. Thus, we can upper bound \cref{eq:final-gap-equation} by
\[
\frac{C}{\beta^\star}\cdot(1/\delta)^C\cdot(g+\log(1/\eps)),
\]
which is again at most \(C^\pr\cdot(1/\delta)^{C^\pr}\cdot(g+\log(1/\eps))\) for a slightly larger constant \(C^\pr\), as desired.
\end{proof}

\dobib

\dobib

\section{Existential results}\label{sec:existential}

In this section, we present and prove all our existential results. We start by showing that a random function is a good seedless condenser for any small family (\cref{thm:main:intro:existential-results:seedless-condenser:small-family}). Then, we instantiate this result to get improved parameters for non-explicit seeded condensers (\cref{thm:main:intro:existential-results:seeded-condenser}). Finally, we plug the latter existential result into the iterative condensing framework in to get our existential results for CG and block sources (\cref{thm:main:intro:existential-results:CG-sources}).

\subsection{A random function is a seedless condenser (for any small family)}\label{subsec:seedless-condenser-any-small-family}

In order to show that a random function is a good seedless condenser for any small family, we show that a random function is (with high probability) a good condenser for a single source. We prove the following, which can be viewed as the condenser version of the classic observation that a random function is a good extractor \cite[Proposition 6.12]{vadhan2012pseudorandomness}. (In fact, we will see that it generalizes it.)

\begin{theorem}[A random function is a condenser for a single source]\label{thm:main-existential-condenser:technical}
There exist universal constants \(C,c>0\) such that the following holds. Let \(\X\) be an arbitrary \((n,k)\)-source. For any \(\ell\in[0,k]\) and \(g>0\) such that \(m:=k-\ell+g\) is an integer, and any \(\eps\in(0,1]\), the following holds. If \(f:\zo^n\to\zo^m\) is a uniformly random function, then
\[
\Pr_f\left[H_\infty^\eps(f(\X))<k-\ell\right]<C\cdot2^{-c\eps K\psi},
\]
where
\begin{align*}
\psi:=\max\left\{g-\frac{1}{\lfloor L\rfloor}\log(1/\eps)-C,\quad g-\frac{1}{\lfloor L\rfloor}\log(C2^gg/\eps)\cdot\frac{C2^g}{g}\right\}.
\end{align*}
\end{theorem}

Note that \(\psi\) evaluates to the first argument when the gap exceeds a sufficiently large constant, and the second argument for all other \(g>0\) (where \(2^g\) becomes a constant). In all applications, one should set the gap \(g\) so that \(\psi=\Omega(g)\) or \(\psi=1\).

Before we continue, we take a moment to make some remarks about the above theorem. First, we emphasize that it works for \emph{any} \((n,k)\)-source, not just flat ones. This is crucial to showing the existence of good seedless condensers for small families, since (unlike in the seeded setting) you cannot assume such families only contain flat sources.\footnote{This is because such existential results proceed by counting the number of sources in the family \(\mathcal{X}\), and arguing that there are not too many. And while it is true that every \((n,k)\)-source is a convex combination of flat sources, it is not true that it is a convex combination of flat sources \emph{in that family}, which is the collection whose size was actually estimated. The family \(\mathcal{X}^\pr\) of flat sources that arises by decomposing each \(\X\in\mathcal{X}\) into a convex combination of flat sources may have size much larger \(|\mathcal{X}|\).} We also note that the above  strictly generalizes the classic result that a random function is a good extractor (i.e., condenser with \(g=0\)) with probability \(1-2^{-\Omega(\eps^2 K)}\). This is because we can instantiate our theorem with gap \(g=\eps/2\) and error \(\eps/2\), since a source with gap \(g\) is \(g\)-close to a source with gap \(0\). Moreover, our generalization reveals that the well-known required loss of \(\ell=2\log(1/\eps)\) for extractors generalizes to roughly \(\ell=2\log(1/g)\), meaning that the loss is primarily due to the gap, not the error. Furthermore, the success probability generalizes to \(1-2^{-\Omega(g\eps K)}\). Overall, this means that even if you are in the regime \(g<1\) (which is close to the extractor regime of \(g=\eps/2\)), you can benefit by applying the condenser result instead of the extractor result.

Next, we record the following corollary, which is immediate via the probabilistic method.\footnote{In particular, apply \cref{thm:main-existential-condenser:technical} to each \(\X\in\mathcal{X}\) and use a union bound.}

\begin{corollary}[A random function is a condenser for any small family]\label{cor:technical:existential-condenser-small-family}
There exist universal constants \(C,c>0\) such that the following holds. Let \(\mathcal{X}\) be a family of \((n,k)\)-sources. For any \(\ell\in[0,k]\) and \(g>0\) such that \(m:=k-\ell+g\) is an integer, and any \(\eps\in(0,1]\), the following holds. If
\[
|\mathcal{X}|\leq c\cdot2^{c\eps K\psi},
\]
where \(\psi\) is as defined in \cref{thm:main-existential-condenser:technical}, then there exists a condenser \(\cond:\zo^n\to\zo^m\) for \(\mathcal{X}\) with loss \(\ell\), gap \(g\), and error \(\eps\).
\end{corollary}

We now proceed to prove \cref{thm:main-existential-condenser:technical}. First, we prove it in the extractor (small gap) regime, via \cref{thm:main-existential-condenser:technical:extractor-regime}. Then, we prove it in the much more challenging condenser (large gap) regime, via \cref{thm:main-existential-condenser:technical:condenser-regime}. Combined, these two theorems immediately yield \cref{thm:main-existential-condenser:technical}. We briefly note that from here onwards, we often simplify notation and use \([N]\) to represent \(\zo^n\), and \([M]\) to represent \(\zo^m\). Furthermore, we always use \(\mu\) to represent the density of a set \(S\), \emph{not} the mean of a random variable (though they will often coincide). The set to which \(\mu\) corresponds will always be clear from context.

\subsubsection{The extractor regime: small gap, large loss}\label{subsubsec:extractor-regime}

We start by proving our existential result for the extractor (small gap) regime.

\begin{theorem}[\cref{thm:main-existential-condenser:technical}, Part I]\label{thm:main-existential-condenser:technical:extractor-regime}
    Let \(\X\) be an arbitrary \((n,k)\)-source. For any \(\ell\in[0,k]\) and \(g>0\) such that \(m:=k-\ell+g\) is an integer, and any \(\eps\in(0,1]\), the following holds. If \(f:\zo^n\to\zo^m\) is a uniformly random function, then
    \[
\Pr_f\left[H_\infty^\eps\left(f(\X)\right)<k-\ell\right]<2^{-\frac{\eps K}{2}(g-\frac{1}{L}\frac{3G}{g}\log(\frac{2Gg}{\eps}))}.
    \]
\end{theorem}

Note that this result is most useful in the extractor regime, i.e., when the gap is a constant or even in the range \(g\in(0,1]\). (Recall that the exact extractor regime is when \(g=\eps/2\).) In this regime, the above bound is of the form \(2^{-\frac{\eps K}{2}(g-O(\frac{1}{L}\frac{1}{g}\log(g/\eps)))}\). Now, in order to prove \cref{thm:main-existential-condenser:technical:extractor-regime}, we use the following proposition, which is just a restatement of (one direction of) \cref{cor:characterization-corollary}.

\begin{proposition}[Necessary condition for condensing failure]\label{prop:necessary-conditions-failure-to-condense-extractor-regime}
For any fixed function \(f:\zo^n\to\zo^m\), any \((n,k)\)-source \(\X\), and any \(k^\pr\in[0,m],\eps>0\), and \(g:=m-k^\pr\),
\[
H_\infty^\eps(f(\X))<k^\pr\implies\exists S\subseteq[M] \text{ of density \(\mu:=|S|/M\) such that } \Pr[f(\X)\in S]>\mu G+\eps.
\]
\end{proposition}

In particular, we'll use the following corollary.

\begin{corollary}\label{cor:necessary-conditions-failure-to-condense-extractor-regime}
If \(H_\infty^\eps(f(\X))<k^\pr\), then for any threshold value \(\tau\in[M]\), one of the following must hold:
\begin{itemize}
    \item \(\exists S\subseteq[M]\) of size \(|S|<\tau\) and density \(\mu:=|S|/M\) such that \(\Pr[f(\X)\in S]>\mu G+\eps\).
    \item \(\exists S\subseteq[M]\) of size \(|S|=\tau\) and density \(\mu:=|S|/M\) such that \(\Pr[f(\X)\in S]>\mu G\).
\end{itemize}
\end{corollary}
\begin{proof}
    By \cref{prop:necessary-conditions-failure-to-condense-extractor-regime}, there exists some set \(S\subseteq[M]\) of density \(\mu\) such that \(\Pr[f(\X)\in S]>\mu G+\eps\). If \(|S|<\tau\), then the first bullet holds. If \(|S|\geq\tau\), then let \(S^\star\) denote the \(\tau\) elements in \(S\) hit by \(f(\X)\) with the highest probability (breaking ties arbitrarily), and let \(\mu^\star\) denote the density of \(S^\star\). Then
    \[
    \Pr[f(\X)\in S^\star]\geq\frac{|S^\star|}{|S|}\cdot\Pr[f(\X)\in S]=\frac{\mu^\star}{\mu}\Pr[f(\X)\in S]>\mu^\star(G+\eps/\mu)\geq\mu^\star G,
    \]
    and the second bullet holds, as desired.
\end{proof}

Now, the idea is to eventually pick some threshold \(\tau\) so that for a random function, both bullets happen with low (and close to the same) probability. We start with the first bullet.

\begin{claim}\label{cl:failure-via-meets-and-exceeds}
Let \(f:\zo^n\to\zo^m\) be a uniformly random function, let \(\X\) be an \((n,k)\)-source, and let \(S\subseteq[M]\) be a set of density \(\mu:=|S|/M\). Then for any \(\eps>0\) and \(G\geq0\),
\[
\Pr_f\left[\Pr_\X\left[f(\X)\in S\right]\geq\mu G+\eps\right]\leq G^{-\eps K}.
\]
\end{claim}
\begin{proof}
We may assume that \(\mu>0\), since the claim trivially holds if \(\mu=0\) (as this implies \(S\) is empty).

Now, for each \(x\in\zo^n\), define the random variable
\[
\Z_x:=1[f(x)\in S]\cdot\Pr[\X=x]\cdot K.
\]
Note that its randomness comes from \(f\), and it is supported on the interval \([0,1]\), since \(H_\infty(\X)\geq k\). Furthermore, if we define \(\Z:=\sum_x\Z_x\), it is easy to verify that \(\Z=\Pr_\X[f(\X)\in S]\cdot K\), and we also have \(\E[\Z]=K|S|/M=\mu K\). Combining these observations with the Chernoff bound (\cref{thm:chernoff-bound}), we have
\begin{align*}
\Pr_f\left[\Pr_\X[f(\X)\in S]\geq\mu G+\eps\right]&=\Pr_f\left[\Pr_\X[f(\X)\in S]\cdot K\geq\mu GK + \eps K\right]\\
&=\Pr\left[\Z\geq\E[\Z]\cdot G + \eps K\right]\\
&=\Pr\left[\Z\geq(G+\eps/\mu)\E[\Z]\right]\\
&\leq\left(\frac{e^{G-1+\eps/\mu}}{(G+\eps/\mu)^{G+\eps/\mu}}\right)^{\mu K}\stepcounter{equation}\tag{\theequation}\label{eq:intermediate-chernoff-in-proof}\\
&=\exp\left(-\eps K\left((1+\alpha)(\ln G+\ln(1+1/\alpha)-1)+\alpha/G\right)\right)
\end{align*}
for \(\alpha:=\mu G/\eps\).\footnote{Here and henceforth, we may assume that \(\eps>0\), since the claim trivially holds if \(\eps=0\).} Finally, using routine calculus, it is straightforward to verify that the function
\[
\phi(\alpha,g):=(1+\alpha)(\ln G + \ln(1+1/\alpha)-1)+\alpha/G
\]
is \(\geq g\ln 2\) for all \(g\geq0,\alpha>0\). The result follows.
\end{proof}

Next, we bound the probability that bullet two in \cref{cor:necessary-conditions-failure-to-condense-extractor-regime} occurs for a uniformly random function. Using the same parameters and objects as defined in \cref{cl:failure-via-meets-and-exceeds}, we have the following.

\begin{claim}\label{cl:failure-via-meets}
    \[
    \Pr_f\left[\Pr_\X\left[f(\X)\in S\right]>\mu G\right]\leq\exp\big(-\mu GK(\ln G-1+1/G)\big).
    \]
\end{claim}
\begin{proof}
The claim is immediate via the proof of \cref{cl:failure-via-meets-and-exceeds} up to \cref{eq:intermediate-chernoff-in-proof}, setting \(\eps=0\).
\end{proof}

Using the above claims, we can show that the necessary conditions for condensing failure (\cref{cor:necessary-conditions-failure-to-condense-extractor-regime}) happen with low probability, allowing us to prove that a random function is a good condenser (\cref{thm:main-existential-condenser:technical:extractor-regime}).

\begin{proof}[Proof of \cref{thm:main-existential-condenser:technical:extractor-regime}]

Let \(k^\pr:=k-\ell\), \(g:=m-k^\pr\), and suppose that \(H_\infty^\eps(f(\X))<k^\pr\). By \cref{cor:necessary-conditions-failure-to-condense-extractor-regime}, we know that for any threshold value \(\tau\in[M]\) (to be set momentarily), one of the following must hold:
\begin{itemize}
    \item \(\exists S\subseteq[M]\) of size \(|S|<\tau\) and density \(\mu:=|S|/M\) such that \(\Pr[f(\X)\in S]>\mu G+\eps\).
    \item \(\exists S\subseteq[M]\) of size \(|S|=\tau\) and density \(\mu:=|S|/M\) such that \(\Pr[f(\X)\in S]>\mu G\).
\end{itemize}
By combining this with \cref{cl:failure-via-meets-and-exceeds} and \cref{cl:failure-via-meets}, we get the following.
\begin{align*}
    \Pr_f\left[H_\infty^\eps(f(\X))<k^\pr\right]&\leq \Pr_f\left[\exists S\subseteq[M], |S|<\tau : \Pr[f(\X)\in S]>\mu G+\eps\right]\\
    &+\Pr_f\left[\exists S\subseteq[M],|S|=\tau : \Pr[f(\X)\in S]>\mu G\right]\\
    &\leq\binom{M}{<\tau}2^{-g\eps K}+\binom{M}{\tau}\exp\left(-\tau L(\ln G-1+1/G)\right).
\end{align*}
Now, consider the quantity \(\phi:=(\ln G - 1 + 1/G)\log e\). If \(g\eps K^\pr/\phi<1\), then we set \(\tau:=\lceil g\eps K^\pr/\phi\rceil\). Otherwise, we set \(\tau:=\lfloor g\eps K^\pr/\phi\rfloor\). Notice that in the first case, the probability that produced the first term in the above sum would have actually realized to \(0\). And in the second case, observe that \(g\eps K\geq\tau L \phi\). Thus the above expression can be bounded by
\begin{align}\label{eq:should-be-easiest-inequality}
\leq\binom{M}{\leq\tau}2^{-\tau L \phi}\leq2^{-\tau L(\phi - \frac{1}{L}\log(eM/\tau))}\leq2^{-\frac{g\eps K}{2\phi}(\phi-\frac{1}{L}\log(\frac{2eG\phi}{g\eps}))}.
\end{align}
Finally, it is straightforward to verify that
\[
\frac{2}{g\ln 2}\leq \frac{g}{\phi}\leq\frac{2G}{g\ln 2},
\]
and using this observation, we can upper bound \cref{eq:should-be-easiest-inequality} by
\[
2^{-\frac{\eps K}{2}(g-\frac{1}{L}\frac{1}{g}\frac{2G}{\ln 2})\log(\frac{g\cdot eG\ln 2}{\eps})}
\]
as desired.
\end{proof}

\subsubsection{The condenser regime: large gap, small loss}\label{subsubsec:condenser-regime}

Next, we turn to prove our existential result for the much more challenging condenser regime.

\begin{theorem}[\cref{thm:main-existential-condenser:technical}, Part II]\label{thm:main-existential-condenser:technical:condenser-regime}

Let \(\X\) be an arbitrary \((n,k)\)-source. For any \(\ell\in[0,k]\) and \(g>0\) such that \(m:=k-\ell+g\) is an integer, and any \(\eps\in(0,1]\), the following holds. If \(f:\zo^n\to\zo^m\) is a uniformly random function, then
\[
\Pr_f\left[H_\infty^\eps(f(\X))<k-\ell\right]\leq4\cdot2^{-\frac{\eps K}{6}(g-\frac{1}{\lfloor L\rfloor}\log(1/\eps) - 16)}.
\]
\end{theorem}

As in the previous section, we consider several conditions which are necessary for condenser failure, and show that each happens with small probability. Similar to before, these conditions have to do with whether certain sets \(S\subseteq[M]\) are assigned too much probability. This time, however, we're in a regime where we want to be able to handle very small loss, by paying some price in the gap, and thus the output length. As a result, it will be too expensive to check tests \(S\subseteq[M]\) by choosing them from the set \([M]\), which may now be very large. Instead, we'll have to implicitly specify them using their preimages.

The above plan would work well if \(\X\) had a small support size, which would be the case if \(\X\) were flat. However, we don't want to make any such assumption, and therefore need a new idea. Our idea is to split the support of \(\X\) into two sets: one which is small (and therefore easy to choose sets from), and one which is big (but is guaranteed to have better ``local'' entropy). Then, we ultimately check whether \(f(\X)\) fails the appropriate tests \(S\subseteq[M]\) by specifying them through their preimages in these sets.

We now proceed to present the formal conditions we're looking for that indicate condenser failure.

\begin{proposition}[Necessary conditions for condensing failure]\label{prop:necessary-conditions-failure-to-condense-condenser-regime}
Let \(f:\zo^n\to\zo^m\) be a fixed function, and let \(\X\) be an \((n,k)\)-source whose support is partitioned into sets \(X_1,X_2\). Fix any \(\ell\in[0,k]\) and \(\eps>0\), and define \(k^\pr:=k-\ell\) and \(g:=m-k^\pr\). If \(H_\infty^\eps(f(\X))<k^\pr\), there must exist some set \(S\subseteq[M]\) of density \(\mu\) such that at least one of the following holds:
    \begin{enumerate}
    \item \textbf{\emph{\(X_1\) has bad smooth min-entropy:}} \(\Pr[f(\X)\in S\land\X\in X_1]>\mu G + \eps/3\).
    \item \textbf{\emph{\(X_2\) has bad smooth min-entropy:}} \(\Pr[f(\X)\in S\land\X\in X_2]>\mu G/L + \eps/3\).
    \item \textbf{\emph{\(X_1,X_2\) have bad ``joint'' smooth min-entropy:}} Both of the following hold:
    \begin{itemize}
    
    \item \(\Pr[f(\X)=v\land \X\in X_1]>\frac{L-1}{L}\cdot\frac{1}{K^\pr}\) for all \(v\in S\).        
        \item \(\Pr[f(\X)\in S\land\X\in X_2]>\eps/3\).
    \end{itemize}
\end{enumerate}
\end{proposition}
\begin{proof}
By definition of smooth min-entropy, we know that if \(H_\infty^\eps(f(\X))<k^\pr\), then there is some set \(S\subseteq\zo^m\) of density \(\mu\) such that \(\Pr[f(\X)\in S]>\mu G+\eps\), by \cref{prop:necessary-conditions-failure-to-condense-extractor-regime}. Partition \(S\) into sets \(S_1,S_2\) such that \(S_1\) contains all elements \(v\in\zo^m\) satisfying
\[
\Pr[f(\X)=v\land\X\in X_1]>\frac{L-1}{L}\cdot\frac{1}{K^\pr}.
\]
Suppose that neither the first nor third case in the proposition hold. Then
\begin{align*}
\Pr[f(\X)\in S_1]&=\Pr[f(\X)\in S_1\land \X\in X_1]+\Pr[f(\X)\in S_1\land\X\in X_2]\\
&\leq \frac{|S_1|}{M}G+\eps/3 + \eps/3\\
&=\frac{|S_1|}{M}G+2\eps/3.
\end{align*}
Furthermore, if the second case also does not hold, then
\begin{align*}
\Pr[f(\X)\in S_2]&=\Pr[f(\X)\in S_2\land\X\in X_1]+\Pr[f(\X)\in S_2\land\X\in X_2]\\
&\leq\frac{L-1}{L}\cdot\frac{|S_2|}{K^\pr}+\frac{|S_2|}{M}G/L+\eps/3\\
&=\frac{|S_2|}{M}G+\eps/3.
\end{align*}
But this implies that \(\Pr[f(\X)\in S]\leq\mu G + \eps\), contradicting our original assumption.
\end{proof}

We now show that each of these three events happens with low probability, starting with the second one.

\subsubsection*{Case 2: The subdistribution on \(X_2\) has bad smooth min-entropy.}

We prove the following, which bounds the probability that the second bullet in \cref{prop:necessary-conditions-failure-to-condense-condenser-regime} can occur.

\begin{lemma}[A random function condenses the subdistribution on \(X_2\)]\label{lem:condenses-subdistribution-on-X2}
Let \(\X\sim\zo^n\) be a source, and let \(X\subseteq\supp(\X)\) be a set with \(\max_{x\in X}\Pr[\X=x]\leq1/K\). For any \(\ell\in[0,k]\) and \(g\geq0\) such that \(m:=k-\ell+g\) is an integer, and any \(\eps\in(0,1]\), the following holds. If \(f:\zo^n\to\zo^m\) is a uniformly random function, then
    \[
    \Pr_f\left[\exists S\subseteq[M] : \Pr_\X[f(\X)\in S\text{ and }\X\in X]>\mu G + \eps\right]\leq2^{-\frac{\eps K}{2}(g-\frac{1}{L}\log(2eG/\eps)-\log e)}.
    \]
\end{lemma}

Just as in the proof of \cref{thm:main-existential-condenser:technical:extractor-regime}, we will upper bound the above probability by splitting the event in two, as prescribed by \cref{cor:necessary-conditions-failure-to-condense-extractor-regime}. To help us with this, we need subdistribution versions of \cref{cl:failure-via-meets-and-exceeds} and \cref{cl:failure-via-meets}, which we prove next.

\begin{claim}[\cref{cl:failure-via-meets-and-exceeds}, subdistribution version]\label{cl:failure-via-m-and-e-subdistribution-version}

Let \(f:\zo^n\to\zo^m\) be a uniformly random function, let \(\X\sim\zo^n\) be a source, and let \(X\subseteq\supp(\X)\) be a set with \(\max_{x\in X}\Pr[\X=x]\leq1/K\). Then for any set \(S\subseteq[M]\) of density \(\mu:=|S|/M\), and any \(\eps>0\) and \(G\geq0\),

\[
\Pr_f\left[\Pr_\X\left[f(\X)\in S\text{ and }\X\in X\right]>\mu G+\eps\right]\leq G^{-\eps K}.
\]

\end{claim}
\begin{proof}
We may assume that \(\mu>0\), since the claim trivially holds if \(\mu=0\) (as this implies \(S\) is empty).

Now, for each \(x\in X\), define the random variable
\[
\Z_x:=1[f(x)\in S]\cdot\Pr[\X=x]\cdot K.
\]
Note that its randomness comes from \(f\), and it is supported on the interval \([0,1]\), since \(\max_{x\in X}\leq1/K\). Furthermore, if we define \(\Z:=\sum_{x\in X}\Z_x\), it is easy to verify that \(\Z=\Pr_\X[f(\X)\in S\text{ and }\X\in X]\cdot K\) and \(\E[\Z]=(K|S|/M)\Pr[\X\in X]=\mu K\Pr[\X\in X]\leq\mu K\). Using these observations, we have
\begin{align*}
\Pr_f\left[\Pr_\X[f(\X)\in S\text{ and }\X\in X]\geq\mu G+\eps\right]&=\Pr_f\left[\Pr_\X[f(\X)\in S\text{ and }\X\in X]\cdot K\geq\mu GK+\eps K\right]\\
&=\Pr_f[\Z\geq(G+\eps/\mu)\mu K]\\
&\leq\left(\frac{e^{G-1+\eps/\mu}}{(G+\eps/\mu)^{G+\eps/\mu}}\right)^{\mu K}\stepcounter{equation}\tag{\theequation}\label{eq:intermediate-chernoff-in-proof-second-time},
\end{align*}
where the last inequality follows from the fact that the Chernoff bound (\cref{thm:chernoff-bound}) can be used with just an upper bound \(\mu K\) on the expectation \(\E[\Z]\). The remainder of the proof is now identical to the proof of \cref{cl:failure-via-meets-and-exceeds}, following \cref{eq:intermediate-chernoff-in-proof}.
\end{proof}

Next, using the same parameters and objects as described in the claim above, we prove the following.

\begin{claim}[\cref{cl:failure-via-meets}, subdistribution version]\label{cl:failure-via-just-meets-subdistribution-version}
    \[
    \Pr_f\left[\Pr_\X[f(\X)\in S\text{ and }\X\in X]>\mu G\right]\leq\exp\left(-\mu GK(\ln G-1+1/G)\right).
    \]
\end{claim}
\begin{proof}
    The claim is immediate via the proof of \cref{cl:failure-via-m-and-e-subdistribution-version} up to \cref{eq:intermediate-chernoff-in-proof-second-time}, setting \(\eps=0\).
\end{proof}

With these claims in hand, it is now easy to prove \cref{lem:condenses-subdistribution-on-X2}.

\begin{proof}[Proof of \cref{lem:condenses-subdistribution-on-X2}]
Just as in the proof to \cref{thm:main-existential-condenser:technical:extractor-regime} (substituting in \cref{cl:failure-via-m-and-e-subdistribution-version} for \cref{cl:failure-via-meets-and-exceeds} and \cref{cl:failure-via-just-meets-subdistribution-version} for \cref{cl:failure-via-meets}), we have
\begin{align}\label{eq:penultimate-almost-done-ineq}
    \Pr_f\left[\exists S\subseteq[M] : \Pr_\X[f(\X)\in S\text{ and }\X\in X]>\mu G+\eps\right]\leq2^{-\frac{g\eps K}{2\phi}(\phi-\frac{1}{L}\log(\frac{2eG\phi}{g\eps}))},
\end{align}
where \(\phi:=(\ln G - 1 + 1/G)\log e\). It is straightforward to verify that for all \(g\geq0\), we have
\[
g-\log e\leq\phi\leq g.
\]
Using this observation, we can upper bound \cref{eq:penultimate-almost-done-ineq} by
\[
2^{-\frac{\eps K}{2}(g-\frac{1}{L}\log(\frac{2eG}{\eps})-\log e)}
\]
as desired.
\end{proof}

\subsubsection*{Case 1: The subdistribution on \(X_1\) has bad smooth min-entropy.}

Next, we upper bound the probability that the first bullet in \cref{prop:necessary-conditions-failure-to-condense-condenser-regime} can occur.

\begin{lemma}[A random function condenses the subdistribution on \(X_1\)]\label{lem:condenses-subdistribution-on-X1}
Let \(\X\) be an \((n,k)\)-source, and let \(X\subseteq\supp(\X)\) be an arbitrary set. For any \(\ell\in[0,k]\) and \(g\geq0\) such that \(m:=k-\ell+g\) is an integer, and any \(\eps\in(0,1]\), the following holds. If \(f:\zo^n\to\zo^m\) is a uniformly random function, then
    \[
    \Pr_f\left[\exists S\subseteq[M] : \Pr[f(\X)\in S\text{ and }\X\in X]>\mu G + \eps\right]\leq2^{-\frac{\eps K}{2}(g-\frac{1}{L}\log(\frac{|X|}{\eps K})-5.886)}.
    \]
\end{lemma}

As before, we will upper bound this event by splitting it in two. This time, however, we will \emph{not} ultimately specify the sets \(S\) by picking them from \([M]\). Instead, we will specify them implicitly, via their preimages. To do this, it will be useful to define a notion of ``superlevel sets.'' Given an \((n,k)\)-source \(\X\) and element \(v\in\zo^n\), we let \(\SL_v\) denote its superlevel set, defined as follows:
\[
\SL_v:=\{x\in\zo^n : \Pr[\X=x]\geq\Pr[\X=v]\}.
\]
Given this definition, we are ready to prove the preimage versions of the key claims we have been using.

\begin{claim}[\cref{cl:failure-via-meets-and-exceeds}, preimage version]\label{cl:failure-via-meets-and-exceeds:preimage-version}

Let \(\X\) be an \((n,k)\)-source. For any \(\ell\in[0,k]\) and \(g\geq0\) such that \(m:=k-\ell+g\) is an integer, any \(\eps\in(0,1]\), and any \(S\subseteq\zo^n\) with \(\mu:=|S|/M\), the following holds. If \(f:\zo^n\to\zo^m\) is a uniformly random function, then
    \[
    \Pr_f\left[|f(S)|=|S|\text{ and }\Pr_\X\left[\exists v\in S : f(\X)=f(v)\text{ and }\X\in\mathsf{SL}_v\right]>\mu G + \eps\right]\leq\left(\frac{e\mu}{\eps}\right)^{\eps K}
    \]
\end{claim}
\begin{proof}

Let \(f:\zo^n\to\zo^m\) be a uniformly random function. For a fixed function \(h:S\to\zo^m\), let \(f_h:\zo^n\to\zo^m\) be a function such that \(f_h(x)=h(x)\) for all \(x\in S\), and \(f_h(x)\) is an independent, uniformly random value from \(\zo^m\) for all other \(x\). By the law of total probability, there exists a worst-case fixing \(h^\star\) that is injective on \(S\) such that
    \begin{align*}
    &\Pr_f\left[|f(S)|=|S|\text{ and }\Pr_\X\left[\exists v\in S : f(\X)=f(v)\text{ and }\X\in\mathsf{SL}_v\right]>\mu G + \eps\right]\\
    \leq\ &\Pr_{f_{h^\star}}\left[\Pr_\X[\exists v\in S:f_{h^\star}(\X)=f_{h^\star}(v)\text{ and }\X\in\SL_v]>\mu G+\eps\right].
    \end{align*}
For ease of notation, we will henceforth use \(f^\pr\) to denote \(f_{h^\star}\).

    Now, for all \(v\in S\) and \(x\in\SL_v\setminus S\), define the random variable
    \[
    \Z_{x,v}:=1[f^\pr(x)=f^\pr(v)]\cdot\Pr[\X=x]\cdot K.
    \]
    Then, for all \(x\in(\cup_{v\in S}\SL_v)\setminus S\), define
    \[
    \Z_x:=\sum_{v\in S : x\in\SL_v}\Z_{x,v},
    \]
    and finally let \(\Z:=\sum_{x\in(\cup_{v\in S}\SL_v)\setminus S}\Z_x\). Let us now make some observations about these random variables.

    First, note that the randomness in these random variables comes exclusively from \(f^\pr\), and each random variable \(\Z_x\) is supported on \([0,1]\), since \(H_\infty(\X)\geq k\) and since \(1[f^\pr(x)=f^\pr(v)]\) can only equal \(1\) for at most one value \(v\in S\) (since \(f^\pr\) is injective on \(S\)). Furthermore, observe that
    \[
    \Z=\sum_{v\in S,x\in\SL_v\setminus S}\Z_{x,v}=K\cdot\Pr_\X[\exists v\in S:f^\pr(\X)=f^\pr(v)\text{ and }\X\in\SL_v\setminus S].
    \]
    Looking back at the probability we must analyze, it would be much more convenient if the expression above had the condition \(\X\in\SL_v\) instead of \(\X\in\SL_v\setminus S\). Luckily, it is easy to verify that
    \begin{align*}
    &\Pr_\X[\exists v\in S:f^\pr(\X)
    =f^\pr(v)\text{ and }\X\in\SL_v]\\
    =&\Pr_\X[\exists v\in S:f^\pr(\X)=f^\pr(v)\text{ and }\X\in\SL_v\setminus S]+\Pr_\X[\X\in S]\\
    =&\frac{1}{K}\cdot\Z+\Pr_\X[\X\in S]\\
    \leq&\frac{1}{K}(\Z+|S|)\\
    \leq&\frac{1}{K}(\Z+\mu GK),
    \end{align*}
    where the penultimate inequality is because \(\X\) has min-entropy at least \(k\), and the final inequality is because \(G\geq M/K\). Next, we can upper bound the expected value of \(\Z\) as follows:
    \[
    \E[\Z]=\sum_{v\in S,x\in\SL_v\setminus S}\E[\Z_{x,v}]=\frac{K}{M}\sum_{v\in S,x\in\SL_v\setminus S}\Pr[\X=x]=\frac{K}{M}\sum_{v\in S}\Pr[\X\in\SL_v\setminus S]\leq\frac{K}{M}\cdot|S|=\mu K.
    \]
    Finally, since each \(\Z_x\) is independent, and \(\Z=\sum_x\Z_x\), we are ready to apply a Chernoff bound to upper bound our desired probability. In particular, we have
    \begin{align*}
        &\Pr_{f^\pr}\left[\Pr_\X[\exists v\in S:f^\pr(\X)=f^\pr(v)\text{ and }\X\in\SL_v]>\mu G+\eps\right]\\
        &=\Pr_{f^\pr}\left[\Pr_\X[\exists v\in S:f^\pr(\X)=f^\pr(v)\text{ and }\X\in\SL_v]\cdot K>\mu GK+\eps K\right]\\
        &\leq\Pr_{f^\pr}\left[\Z+\mu GK>\mu GK+\eps K\right]\\
        &=\Pr_{f^\pr}\left[\Z>(\eps/\mu)\mu K\right].
    \end{align*}
    Since we showed above that \(\mu K \geq \E[\Z]\), the Chernoff bound (\cref{thm:chernoff-bound}) tells us that the above is
    \[
    \leq\left(\frac{e^\delta}{(1+\delta)^{1+\delta}}\right)^{\mu K}\leq\left(\frac{e}{1+\delta}\right)^{(1+\delta)\mu K}
    \]
    for \(\delta:=\eps/\mu-1\).\footnote{Note that we may assume \(\delta>0\), since otherwise \(\mu/\eps\geq1\) and the bound in the claim trivially holds.} Plugging this value of \(\delta\) into the expression above yields
    \[
    \leq\left(e\mu/\eps\right)^{\eps K},
    \]
    as desired.
\end{proof}

Next, using the same parameters as in the claim above (with \(k^\pr:=k-\ell\)), we prove the following.

\begin{claim}[\cref{cl:failure-via-meets}, preimage version]\label{cl:failure-via-meets:preimage-version}
        \[
    \Pr_f\left[|f(S)|=|S|,\text{ and for all \(v\in S\), }\Pr_\X[f(\X)=f(v)\text{ and }\X\in\SL_v]>1/K^\pr\right]\leq\left(\frac{4e}{G}\right)^{\mu GK}.
    \]
\end{claim}
\begin{proof}

As in the proof of \cref{cl:failure-via-meets-and-exceeds:preimage-version}, it suffices to show the claimed upper bound on the quantity
\begin{align}\label{eq:hardest-probability-to-bound}
\Pr_{f^\pr}\left[\forall v\in S:\Pr_\X[f^\pr(\X)=f^\pr(v)\text{ and }\X\in\SL_v]>1/K^\pr\right],
\end{align}
where \(f^\pr\) is some function that is injective (and fixed to constants) on \(S\), and uniformly random on all other inputs. Now, let us once again proceed with defining random variables so that we can upper bound this quantity via a Chernoff bound. We must be a little more careful this time.

Towards this end, for all \(v\in S\) and \(x\in\SL_v\setminus S\), we once again want to define a random variable \(\Z_{x,v}\). But this time, we base the definition on just how likely \(x\) is to be hit. In particular, let \(X^\star\) denote the \(2K\) most probable elements in \(\supp(\X)\), breaking ties arbitrarily. Then, define
\[
\Z_{x,v}:=\begin{cases}
    1[f^\pr(x)=f^\pr(v)]&\text{ if }x\in X^\star,\\
    1[f^\pr(x)=f^\pr(v)]\cdot\Pr[\X=x]\cdot2K&\text{ otherwise.}
\end{cases}
\]
Now, as before, for all \(x\in(\cup_{v\in S}\SL_v)\setminus S\), define
\[
\Z_x:=\sum_{v\in S:x\in\SL_v}\Z_{x,v},
\]
and let \(\Z:=\sum_{x\in(\cup_{v\in S}\SL_v)\setminus S}\Z_x\). Let us now make some observations about these random variables.

First, the randomness in these random variables comes exclusively from \(f^\pr\). Next, we claim that each random variable \(\Z_x\) is supported on \([0,1]\). To see why, observe that only one term \(\Z_{x,v}\) in the sum that defines \(\Z_x\) can be nonzero, since \(f^\pr\) is injective on \(S\). Then, note that such a nonzero term \(\Z_{x,v}\) is always in the range \([0,1]\): for the first definition of \(\Z_{x,v}\), this is clear. For the second definition, simply note that all elements \(x\in\supp(\X)\) that are not among the \(2K\) most probable must be hit with probability \(<1/(2K)\), because otherwise the sum of \(\Pr[\X=x]\) over all elements \(x\) will exceed \(1\) - a contradiction.

Next, observe that
\begin{align*}
\E[\Z]&=\sum_{v\in S,x\in(\SL_v\setminus S)\cap X^\star}\E[\Z_{x,v}]+\sum_{v\in S,x\in(\SL_v\setminus S)\setminus X^\star}\E[\Z_{x,v}]\\
&\leq|S||X^\star|/M+2K|S|/M\\
&=4\mu K.
\end{align*}

Finally, for our last step before applying the Chernoff bound, we must relate \(\Z\) to the event in \cref{eq:hardest-probability-to-bound}. Towards this end, fix some \(v\in S\) and suppose the following \emph{inequality} holds (note that the \emph{equality} always holds, by the injectivity of \(f^\pr\) on \(S\)):
\[
\Pr_\X[f^\pr(\X)=f^\pr(v)\text{ and }\X\in\SL_v]=\Pr_\X[f^\pr(\X)=f^\pr(v)\text{ and }\X\in\SL_v\setminus S]+\Pr_\X[\X=v]>1/K^\pr.
\]
Then, since \(\X\) has min-entropy \(\geq k\), and \(1/K^\pr=L/K\), \(f^\pr\) must send at least \(L+1\) elements from \(SL_v\) to \(f^\pr(v)\). We consider two cases. First, if at least \(L\) of these elements occur in \(X^\star\), then there must be at least \(L\) elements that \(f^\pr\) maps from \((\SL_v\setminus S)\cap X^\star\) to \(f^\pr(v)\). As such, we have
\[
\sum_{x\in\SL_v\setminus S}\Z_{x,v}\geq\sum_{x\in(\SL_v\setminus S)\cap X^\star}\Z_{x,v}\geq L=K\cdot 1/K^\pr.
\]
On the other hand, if less than \(L\) of these elements occur in \(X^\star\), then there must be at least \(2\) elements that \(f^\pr\) maps from \((\SL_v\setminus S)\setminus X^\star\) to \(f^\pr(v)\). In this case, by definition of \(\SL_v\), we have that
\begin{align*}
\sum_{x\in\SL_v\setminus S}\Z_{x,v}&=\sum_{x\in(\SL_v\setminus S)\cap X^\star}\Z_{x,v}+\sum_{x\in(\SL_v\setminus S)\setminus X^\star}\Z_{x,v}\\
&\geq K\cdot\Pr[f^\pr(\X)=f^\pr(v)\text{ and }\X\in(\SL_v\setminus S)\cap X^\star]\\
&+2K\cdot\Pr[f^\pr(\X)=f^\pr(v)\text{ and }\X\in(\SL_v\setminus S)\setminus X^\star]\\
&= K\cdot\Pr[f^\pr(\X)=f^\pr(v)\text{ and }\X\in(\SL_v\setminus\{v\})\cap X^\star]\\
&+2K\cdot\Pr[f^\pr(\X)=f^\pr(v)\text{ and }\X\in(\SL_v\setminus\{v\})\setminus X^\star]\tag{by injectivity of \(f^\pr\) on \(S\)}\\
&\geq K\cdot\Pr[f^\pr(\X)=f^\pr(v)\text{ and }\X\in(\SL_v\setminus\{v\})\cap X^\star]\\
&+K\cdot\Pr[f^\pr(\X)=f^\pr(v)\text{ and }\X\in(\SL_v\setminus\{v\})\setminus X^\star]\\
&+K\cdot\Pr[f^\pr(\X)=f^\pr(v)\text{ and }\X=v]\\
&=K\cdot\Pr[f^\pr(\X)=f^\pr(v)\text{ and }\X\in\SL_v]\\
&>K\cdot\frac{1}{K^\pr}.
\end{align*}
Combining these two cases, we get that
\[
\Pr_\X[f^\pr(\X)=f^\pr(v)\text{ and }\X\in\SL_v]>1/K^\pr\implies\sum_{x\in\SL_v\setminus S}\Z_{x,v}\geq K\cdot 1/K^\pr,
\]
and moreover,
\[
\Pr_\X[f^\pr(\X)=f^\pr(v)\text{ and }\X\in\SL_v]>1/K^\pr\emph{ for all \(v\in S\) }\implies\Z=\sum_{v\in S}\sum_{x\in\SL_v\setminus S}\Z_{x,v}\geq K|S|/K^\pr.
\]

With all of these observations in hand, we are finally ready to apply a Chernoff bound to upper bound our desired probability. Towards this end, we have
\begin{align*}
    &\Pr_{f^\pr}\left[\forall v\in S : \Pr_\X[f^\pr(\X)=f^\pr(v)\text{ and }\X\in\SL_v]>1/K^\pr\right]\\
    &\leq\Pr_{f^\pr}\left[\Z\geq K|S|/K^\pr\right]\\
    &=\Pr_{f^\pr}[\Z\geq (4\mu K)(G/4)].
\end{align*}
Since we showed that \(4\mu K\geq\E[\Z]\), the Chernoff bound (\cref{thm:chernoff-bound}) tells us that the above is
\[
\leq\left(\frac{e^\delta}{(1+\delta)^{1+\delta}}\right)^{4\mu K}\leq\left(\frac{e}{1+\delta}\right)^{(1+\delta)4\mu K}
\]
for \(\delta:=G/4-1\).\footnote{Note that we may assume \(\delta>0\), since otherwise \(4/G\geq1\) and the bound in the claim trivially holds.} Plugging this value of \(\delta\) into the expression above yields
\[
(4e/G)^{\mu GK},
\]
as desired.
\end{proof}

Using these claims, it is easy to prove \cref{lem:condenses-subdistribution-on-X1}.

\begin{proof}[Proof of \cref{lem:condenses-subdistribution-on-X1}]

Fix a function \(f:\zo^n\to\zo^m\), and suppose there is a set \(S\subseteq[M]\) with density \(\mu:=|S|/M\) such that \(\Pr[f(\X)\in S\text{ and }\X\in X]>\mu G + \eps\). We may assume each \(v\in S\) has a preimage in \(X\), since otherwise we could remove \(v\) from \(S\), while keeping the probability guarantee. For the same reason, we may assume that \(\Pr[f(\X)=v\text{ and }\X\in X]>1/K^\pr\) for each \(v\in S\).

Now, let \(\tau\in[M]\) be a threshold we will set later. Observe that one of the following must hold:
\begin{itemize}
    \item \(\exists S\subseteq[M]\) with size \(<\tau\) and density \(\mu:=|S|/M\) such that \(\Pr[f(\X)\in S\text{ and }\X\in X]>\mu G+\eps\).
    \item \(\exists S\subseteq[M]\) with size \(\tau\) and density \(\mu:=\tau/M\) such that \(\Pr[f(\X)=v\text{ and }\X\in X]>1/K^\pr\), \(\forall v\in S\).
\end{itemize}
Indeed, this follows immediately from the discussion above, since if the original set \(S\) had size \(<\tau\), then the first bullet holds, and if it had size \(\geq\tau\), then any subset of \(S\) of size \(\tau\) satisfies the second bullet.

Next, regardless of which bullet holds, we let \(S\subseteq[M]\) denote the set referred to in that bullet, and define a new set \(S^\star\subseteq X\) as follows. First, for each \(v\in S\), let \(v^\star\) denote the element in \(f^{-1}(v)\cap X\) that receives the least probability under \(\X\). Then, define the set \(S^\star:=\{v^\star : v\in S\}\), and observe the following:
\begin{itemize}
\item If \(S\) originally referred to the first bullet above, then all of the following hold:
\begin{itemize}
    \item \(S^\star\subseteq X\) and \(|S^\star|<\tau\).
    \item \(|f(S^\star)|=|S^\star|\).
    \item \(\Pr_\X[\exists v^\star\in S^\star : f(\X)=f(v^\star)\text{ and }\X\in\SL_{v^\star}]=\Pr_\X[f(\X)\in S\text{ and }\X\in X]>\mu G+\eps\).
\end{itemize}
\item If \(S\) originally referred to the second bullet above, then all of the following must hold:
\begin{itemize}
    \item \(S^\star\subseteq X\) and \(|S^\star|=\tau\).
    \item \(|f(S^\star)|=|S^\star|\).
    \item \(\Pr_\X[f(\X)=f(v^\star)\text{ and }\X\in\SL_{v^\star}]=\Pr_\X[f(\X)=f(v^\star)\text{ and }\X\in X]>1/K^\pr, \forall v^\star\in S^\star\).
\end{itemize}
\end{itemize}
By combining these observations with \cref{cl:failure-via-meets-and-exceeds:preimage-version} and \cref{cl:failure-via-meets:preimage-version}, we get the following.
\begin{align*}
    &\Pr_f\left[\exists S\subseteq[M] : \Pr[f(\X)\in S\text{ and }\X\in X]>\mu G + \eps\right]\\
    &\leq\Pr_f\left[\exists S\subseteq X,|S|<\tau : |f(S)|=|S|\text{ and }\Pr_\X[\exists v\in S : f(\X)=f(v)\text{ and }\X\in\SL_v]>\mu G+\eps\right]\\
    &+\Pr_f\left[\exists S\subseteq X, |S|=\tau:|f(S)|=|S|\text{ and }\Pr_\X[f(\X)=f(v)\text{ and }\X\in\SL_v]>1/K^\pr, \forall v\in S\right]\\
    &\leq\binom{|X|}{<\tau}\left(\frac{e\tau}{\eps M}\right)^{\eps K} + \binom{|X|}{\tau}\left(\frac{4e}{G}\right)^{\tau L}.
\end{align*}

Finally, we check if \(\eps K/L<1\). If this holds, we set \(\tau=\lceil\eps K/L\rceil\leq2\eps K/L\), and observe that the probability that produced the term \(\binom{|X|}{<\tau}\left(\frac{e\tau}{\eps M}\right)^{\eps K}\) would have actually been \(0\). If \(\eps K/L\geq1\), we set \(\tau=\lfloor\eps K/L\rfloor\geq(\eps K/L)/2\), and observe that \(\left(\frac{e\tau}{\eps M}\right)^{\eps K}\leq\left(\frac{e}{G}\right)^{\tau L}\). In either case, we can upper bound the above sum by
\[
\leq\binom{|X|}{\leq\tau}\left(\frac{4e}{G}\right)^{\tau L}\leq2^{-\tau L(g-\frac{1}{L}\log(\frac{2eL}{\eps}\cdot\frac{|X|}{K})-\log(4e))}\leq2^{-\frac{\eps K}{2}(g-\frac{1}{L}\log(\frac{|X|}{\eps K})-5.886)},
\]
as desired.
\end{proof}

\subsubsection*{Case 3: The subdistributions on \(\X_1,\X_2\) have bad joint smooth min-entropy.}

Finally, we upper bound the probability that the third bullet in \cref{prop:necessary-conditions-failure-to-condense-condenser-regime} can occur.

\begin{lemma}[A random function jointly condenses the subdistributions on \(X_1,X_2\)]\label{lem:condenses-subdistribution-on-X1-and-X2}
Let \(\X\) be an \((n,k)\)-source. For any \(\ell\in[0,k]\) and \(g\geq0\) such that \(m:=k-\ell+g=k^\pr+g\) is an integer, and any \(\eps\in(0,1]\) the following holds. Suppose the support of \(\X\) is partitioned into sets \(X_1,X_2\), where \(X_1\) contains the \(\min\{\lceil4KL\rceil,N\}\) highest probability elements, and \(X_2\) the rest. If \(f:\zo^n\to\zo^m\) is a uniformly random function, then
    \begin{align*}
    \Pr_f\left[\exists S\subseteq[M]:\Pr_\X[f(\X)=v\land\X\in X_1]>\frac{L-1}{L}\cdot\frac{1}{K^\pr} \forall v\in S,\text{ and }\Pr_\X[f(\X)\in S\land\X\in X_2]>\eps\right]\\
    \leq2\cdot2^{-\frac{\eps K}{2}(g-\frac{1}{\lfloor L\rfloor}\log(1/\eps)-11)}.
    \end{align*}
\end{lemma}
\begin{proof}

We start by claiming that we can assume \(L\geq2\), since otherwise the result is easy to prove. To see why, suppose that \(L<2\) (and thus \(\lfloor L\rfloor=1\)). In order for the bad event (in the probability expression above) to hold, the random function \(f\) must map \(>\eps\) weight from \(X_2\) into the set \(f(X_1)\). But here, the size of \(X_1\) is at most \(\lceil4KL\rceil<\lceil8K\rceil\), and thus the size of \(f(X_1)\) is also \(<\lceil8K\rceil\). Since \(f\) acts independently on \(X_1,X_2\) (as they are disjoint), we get that the bad event above holds with probability at most
\[
\Pr_f\left[\Pr_\X[f(\X)\in S^\star\land\X\in X_2]>\eps\right],
\]
where \(f\) is a uniformly random function, and \(S^\star\) is an (adversarially) fixed set of size \(<\lceil8K\rceil\). Now, by definition of \(X_2\), each \(x\in X_2\) is hit by \(\X\) with probability at most \(1/(4K)\). Thus, applying \cref{cl:failure-via-m-and-e-subdistribution-version} (setting parameters appropriately), we get
\[
\Pr_f\left[\Pr_\X[f(\X)\in S^\star\land\X\in X_2]>\eps\right]\leq2^{-2\eps K(g-\log(36/\eps))}=2^{-2\eps K(g-\frac{1}{\lfloor L\rfloor}\log(36/\eps))},
\]
as desired. Thus, we can henceforth assume that \(L\geq2\).

Now, let \(\mathcal{E}\) denote the (bad) event in the lemma statement, and let \(\tau\in[M]\) be a threshold value that we will set later. Since \(f\) acts independently on \(X_1,X_2\) (as they are disjoint), observe that
\begin{align*}
\Pr_f[\mathcal{E}]&\leq\Pr_f\left[\exists S\subseteq[M],|S|=\tau : \Pr_\X[f(\X)=v\land\X\in X_1]>\frac{L-1}{L}\cdot\frac{1}{K^\pr},\forall v\in S\right]\\
&+\Pr_f\left[\Pr_\X[f(\X)\in S^\star\land \X\in X_2]>\eps\right],
\end{align*}
where \(S^\star\subseteq[M]\) is an arbitrary fixed set of size \(\tau-1\).\footnote{Notice that the second term realizes to \(0\) if \(\tau=1\).} By \cref{cl:failure-via-m-and-e-subdistribution-version} (setting parameters appropriately), and the fact that each \(x\in X_2\) is hit by \(\X\) with probability at most \(1/(4KL)\), we have
\[
\Pr_f\left[\Pr_\X[f(\X)\in S^\star\land \X\in X_2]>\eps\right]\leq2^{-2\eps KL\log(\frac{\eps M}{2\tau})}.
\]
Finally, consider any fixed set \(S\subseteq[M]\) of size \(\tau\). Then, by \cref{cl:failure-via-meets:preimage-version} (used in a similar manner as in the proof to \cref{lem:condenses-subdistribution-on-X1}), we have
\begin{align*}
&\Pr_f\left[\exists S\subseteq[M],|S|=\tau:\Pr_\X[f(\X)=v\land\X\in X_1]>\frac{L-1}{L}\cdot\frac{1}{K^\pr},\forall v\in S\right]\\
&\leq \binom{|X_1|}{\tau}\left(\frac{4eL}{G(L-1)}\right)^{\tau (L-1)}\\
&\leq\left(\frac{e\lceil 4KL\rceil}{\tau}\right)^{\tau}\cdot\left(\frac{8e}{G}\right)^{\tau (L-1)}\\
&\leq2^{-\tau L(\frac{L-1}{L}(g-\log(8e))-\frac{1}{L}\log(8eKL/\tau))}.
\end{align*}
And thus, we have
\[
\Pr_f[\mathcal{E}]\leq2^{-2\eps KL\log(\frac{\eps M}{2\tau})}+2^{-\tau L( \frac{L-1}{L}(g-\log(8e))-\frac{1}{L}\log(8eKL/\tau))}.
\]
Finally, setting \(\tau:=\lceil\eps^{(L-1)/L}K^\pr\rceil\) yields
\[
\Pr_f\left[\mathcal{E}\right]\leq2\cdot2^{-\frac{\eps K}{2}(g-\frac{1}{L}\log(1/\eps)-11)},
\]
as desired.
\end{proof}

\subsubsection*{Putting everything together}

By combining the necessary conditions for condensing failure (\cref{prop:necessary-conditions-failure-to-condense-condenser-regime}) with the fact that each such condition happens with low probability (\cref{lem:condenses-subdistribution-on-X2,lem:condenses-subdistribution-on-X1,lem:condenses-subdistribution-on-X1-and-X2}), we are finally able to prove that a random function is a good condenser (\cref{thm:main-existential-condenser:technical:condenser-regime}).

\begin{proof}[Proof of \cref{thm:main-existential-condenser:technical:condenser-regime}]

Before we start, we note that we may assume \(\ell\leq g/4\). This is because if \(\ell>g/4\), then combining \cref{prop:necessary-conditions-failure-to-condense-extractor-regime} and \cref{lem:condenses-subdistribution-on-X2} (observing that \(\log(G)/L\leq\log(4\ell)/L\leq 2\)) yields the result.

Now that we may assume \(\ell\leq g/4\), the result is almost immediate, via the sketch above. In particular, we first set \(k^\pr:=k-\ell\), and let \(X_1\) denote the heaviest \(\min\{\lceil4KL\rceil,N\}\) elements in \(\supp(X)\), and \(X_2\) the rest. Then, by \cref{lem:condenses-subdistribution-on-X1}, we know that the first condition in \cref{prop:necessary-conditions-failure-to-condense-condenser-regime} holds with probability at most
\begin{align*}
&2^{-\frac{\eps K}{6}(g-\frac{1}{L}\log(\frac{3\lceil4KL\rceil}{\eps K})-5.886)}\\
\leq &\ 2^{-\frac{\eps K}{6}(g-\frac{1}{L}\log(\frac{1}{\eps})-16)}.
\end{align*}
Next, define \(\tilde{G}=G/L\), \(\tilde{\eps}=\eps/3\), \(\tilde{K}=4KL\), and \(\tilde{L}=\tilde{K}\tilde{G}/M=4L\). Since each \(x\in X_2\) is hit with probability at most \(1/(4KL)=1/\tilde{K}\), \cref{lem:condenses-subdistribution-on-X2} tells us that the second condition in \cref{prop:necessary-conditions-failure-to-condense-condenser-regime} holds with probability at most
\begin{align*}
&2^{-\frac{\tilde{\eps}\tilde{K}}{2}(\tilde{g}-\frac{1}{\tilde{L}}\log(2e\tilde{G}/\tilde{\eps})-\log e)}\\
\leq\ &2^{-\frac{4\eps KL}{6}(g-\ell-\frac{1}{4L}\log(6eG/\eps)-\log e)}\\
\leq\ &2^{-\frac{4\eps KL}{6}(3g/4-\frac{1}{4L}\log(6eG/\eps)-\log e)}\tag{since we assumed \(\ell\leq g/4\)}\\
\leq\ &2^{-\frac{\eps KL}{6}(2g-\frac{1}{L}\log(6e/\eps)-4\log e)}\\
\leq\ &2^{-\frac{\eps KL}{6}(2g-\frac{1}{L}\log(1/\eps)-10)}.
\end{align*}
Finally, by \cref{lem:condenses-subdistribution-on-X1-and-X2}, the third condition in \cref{prop:necessary-conditions-failure-to-condense-condenser-regime} holds with probability at most
\begin{align*}
&2\cdot2^{-\frac{\eps K}{6}(g-\frac{1}{\lfloor L\rfloor}\log(3/\eps)-11)}\\
\leq\ &2\cdot2^{-\frac{\eps K}{6}(g-\frac{1}{\lfloor L\rfloor}\log(1/\eps)-13)}.
\end{align*}
Thus, by a simple union bound, one of the conditions in \cref{prop:necessary-conditions-failure-to-condense-condenser-regime} holds with probability at most
\[
4\cdot2^{-\frac{\eps K}{6}(g-\frac{1}{\lfloor L\rfloor}\log(1/\eps)-16)}.
\]
By the statement of \cref{prop:necessary-conditions-failure-to-condense-condenser-regime}, the result follows.
\end{proof}

\dobib

\subsection{A random function is a seeded condenser}

Using our main existential result from the previous section, it is now straightforward to obtain our existential results for seeded condensers.

\begin{theorem}[A random function is a seeded condenser]\label{thm:existential-result:seeded-condenser}
    There exists a universal constant \(C\geq1\) such that for any \(\ell\in[0,k+d]\) and \(g\geq0\) such that \(m:=k+d-\ell+g\) is an integer, and any \(\eps\in(0,1]\), the following holds. If \(d\geq\log\left(\frac{n-k}{\eps}\right)+C\) and \(g\geq\frac{1}{\lfloor L\rfloor}\log\left(\frac{1}{\eps}\right)+C\), then there exists a seeded condenser \(\sCond:\zo^n\times\zo^d\to\zo^m\) for \((n,k)\)-sources with loss \(\ell\), gap \(g\), error \(\eps\), and seed length \(d\).
\end{theorem}
\begin{proof}
This is an immediate corollary of our main existential result (\cref{thm:main-existential-condenser:technical}), by considering the family \(\mathcal{X}\) of sources of the form \((\X,\Y)\), where \(\X\) is an \((n,k)\)-source and \(\Y\sim\zo^d\) is a uniform independent seed.
\end{proof}

We remark that a more general theorem can be established (that allows for gap \(g\in[0,C]\) and recovers known existential results for seeded extractors), but we only record the one above for simplicity.

\dobib

\subsection{Existential condensers for block sources}\label{subsec:existential-block-source-condensers}

In this section, we show our existential results for Chor-Goldreich sources, and ultimately prove \cref{thm:main:intro:existential-results:CG-sources} from the introduction. As a reminder, we cannot simply invoke our black box result on the existence of seedless condensers for any small family (\cref{cor:technical:existential-condenser-small-family}), because the family of CG sources is not small. Indeed, a rough estimate would indicate that the number of \((t,n,k)\)-CG sources is roughly \(\binom{N}{K}^{K^0+K^1+\dots+K^{t-1}}\approx2^{gK^t}\). However, since each such source contain \(kt\) bits of min-entropy, applying \cref{cor:technical:existential-condenser-small-family} would only work if we allowed the gap blow-up by a factor of at least \(1/\eps\). Here, we aim to do much better, and in fact prove such results for the more general setting of block sources.

\subsubsection{Two blocks (via seeded condensers)}

We start by showing existential results for condensing block sources that contain only two blocks. As a reminder, we let \(g_i:=n_i-k_i\) denote the entropy gap in the \(i^\text{th}\) block of the input block source.

\begin{theorem}[Existential results for block sources with two blocks]\label{thm:main:technical:existential-results:CG-sources:two-blocks}

There is a universal constant \(C\geq1\) such that the following holds. There exists a (non-explicit) condenser \(\cond:\zo^{n_1}\times\zo^{n_2}\to\zo^m\) for \(((n_1,k_1),(n_2,k_2))\)-block sources with output length \(m=k_1+k_2-\ell+g\), error \(\eps\), loss \(\ell\), and gap
\[
g\leq g_2 + \frac{1}{\lfloor L\rfloor}(g_2+\log(1/\eps))+C,
\]
provided that \(k_2\geq\log(g_1/\eps)+C\).
\end{theorem}
\begin{proof}
This is an immediate corollary of our existential result for seeded condensers (\cref{thm:existential-result:seeded-condenser}), combined with fact that seeded condensers work for CG-correlated seeds (\cref{lem:seeded-condensers-work-on-block-sources}).
\end{proof}

Before we move on to the multi-block setting, a few remarks are in order. First, note that the first bullet in \cref{thm:main:intro:existential-results:CG-sources} is an immediate corollary of the above, since CG sources are less general than block sources. Next, we note that when there are not too many blocks (say, \(t=O(1)\), and they all have similar lengths), the above result will give the best parameters. This is because one may simply group together the first \(t-1\) blocks into a single block, and this will only add about \(\log(t)\) onto the min-entropy requirement, which is not bad when \(t\) is small. Finally, we mention that using this idea and the above result, one may recover the parameters of the explicit condensers in \cite{doron2023almost} (for constant-sized blocks), by brute-force searching for an excellent block-source condenser (using the above existential result), which condenses to rate \(0.99\). Then, one can apply the explicit instantiation of the iterative condensing framework, instantiated with the GUV extractor (as in \cref{subsubsec:getting-the-rest-of-entropy-out}).

\subsubsection{More than two blocks (via iterative condensing)}\label{subsec:existential:multi-block:iterative-condensing}

We now turn to prove our existential result for the multi-block setting. As above, we do so by combining our existential seeded condensers with the fact that such condensers can handle correlated seeds. This time, however, we'll need to iterate, and apply a sequence of several condensers. We present our main existential result for the multi-block setting below, and remind the reader that we always use \(g_i:=n_i-k_i\) to denote the entropy gap in the \(i^\text{th}\) block.

\begin{theorem}[Existential results for block sources with many blocks]\label{thm:main:technical:existential-results:CG-sources:more-than-two-blocks}
There is a universal constant \(C\geq1\) such that the following holds. There exists a (non-explicit) condenser \(\cond:\zo^{n_1}\times\dots\times\zo^{n_t}\to\zo^m\) for \(((n_1,k_1),\dots,(n_t,k_t=:n_t-g))\)-block sources with output length \(m=(\sum_{i\in[t]}k_i)-\ell+g^\pr\), error \(\eps\), loss \(\ell\), and gap
\[
g^\pr\leq g + \exp\left(\frac{6\lceil\frac{4t^2}{\ell+1}\rceil}{\lfloor L^{\frac{1}{2t}}\rfloor}\right)\cdot\left(\frac{6\lceil\frac{4t^2}{\ell+1}\rceil}{\lfloor L^{\frac{1}{2t}}\rfloor}\right)\cdot\left(g_t+\log(1/\eps)+Ct\right)+Ct
\]
provided that \(k_{i+1}\geq \log(g_i/\eps)+\ell/t+C\) for all \(i\in[t-1]\).
\end{theorem}

While the above theorem is quite general and can work for nearly any block source, the parameters may be a bit difficult to digest. Soon, we will show exactly what this theorem can yield for the less general (and more standard) setting of CG sources (in \cref{cor:main-existential-many-block-CG-result}, \cref{cor:main-existential-many-block-CG-result:lossless-regime}, and \cref{cor:main-existential-many-block-CG-result:small-gap-regime}). But first, we present its proof, which relies on the following lemma (allowing for a more careful fine-tuning of parameters).

\begin{lemma}[Existential results for block sources with many blocks]\label{lem:main:technical:existential-results:CG-sources:more-than-two-blocks}

There is a universal constant \(C\geq1\) such that for any (not necessarily constant) parameters \(\ell\geq0\) and \(\tau\geq1\), the following holds. There exists a (non-explicit) condenser \(\cond:\zo^{n_1}\times\dots\times\zo^{n_t}\to\zo^m\) for \(((n_1,k_1),\dots,(n_t,k_t))\)-block sources with output length \(m=(\sum_{i\in[t]}k_i)-\ell^\star+g^\star\), error \(\eps\), loss \(\ell^\star\leq\ell t + \lfloor (t-2)/\tau\rfloor t\), and gap
\[
g^\star\leq g_t + e^{\frac{6\tau}{\lfloor L\rfloor}}\cdot\frac{6\tau}{\lfloor L\rfloor}\cdot(g_t+\log(1/\eps)+Ct)+Ct
\]
provided that \(k_{i+1}\geq\log(g_i/\eps)+\ell+\lfloor\frac{t-(i+1)}{\tau}\rfloor+C\) for all \(i\in[t-1]\).
\end{lemma}

Given this lemma, it is straightforward to prove our main existential result for block sources with many blocks (\cref{thm:main:technical:existential-results:CG-sources:more-than-two-blocks}). Indeed, it just involves picking the best settings of the parameters \(\ell,\tau\).

\begin{proof}[Proof of \cref{thm:main:technical:existential-results:CG-sources:more-than-two-blocks}]
Let \(\ell_0 := \ell/(2t),\tau_0:=\lceil\frac{4t^2}{\ell+1}\rceil\), and set these as the first two parameters in \cref{lem:main:technical:existential-results:CG-sources:more-than-two-blocks}.
\end{proof}

At last, we are ready to prove our core lemma. We do so, below.

\begin{proof}[Proof of \cref{lem:main:technical:existential-results:CG-sources:more-than-two-blocks}]
Let \(\sCond_1,\sCond_2,\dots,\sCond_{t-1}\) be a sequence of functions, where each \(\sCond_i:\zo^{n_i}\times\zo^{m_{i+1}}\to\zo^{m_i}\) is a seeded \((n_i,k_i)\to_{\eps_i}(m_i,m_i-g_i^\pr)\) condenser. Then, define \(m_t:=n_t\) and
\[
m_i:=k_i+m_{i+1}-\ell_i+g_i^\pr
\]
for every \(i\in[t-1]\), where \(\ell_i\) is some parameter to be set later. Our existential result for seeded condensers (\cref{thm:existential-result:seeded-condenser}) says that such condensers must exist, provided that each \(m_i\) is a positive integer and both of the following hold, for every \(i\in[t-1]\) (where \(C>0\) is a universal constant):
\begin{itemize}
    \item \textbf{Seed length requirement:} \(m_{i+1}\geq\log\left(g_i/\eps_i\right)+C\).
    \item \textbf{Output gap requirement:} \(g_i^\pr\geq\frac{1}{\lfloor L_i\rfloor}\log(1/\eps_i)+C\).
\end{itemize}
Moreover, our iterative condensing framework (\cref{lem:iterated-condensing-framework}) says that given such seeded condensers, there exists a condenser \(\cond:\zo^{n_1}\times\dots\times\zo^{n_t}\to\zo^{m_1}\) for \(((n_1,k_1),\dots,(n_t,k_t))\)-block sources with output length \(m_1\), output gap \(g^\pr=g_t + \sum_{i\in[t-1]}g_i^\pr\), and error \(\eps^\pr=\sum_{i\in[t-1]}\eps_i\cdot2^{g_t+\sum_{j\in(i,t-1]}g_j^\pr}\). Thus, our goal is to set parameters \(\eps_i,\ell_i,g_i^\pr\) for every \(i\in[t-1]\) such that each seeded condenser \(\sCond_i\) exists, and so that the final condenser \(\cond\) achieves the parameters claimed in the theorem statement.

We start by introducing some intermediate parameters, which will help keep our calculations tidy. In particular, we define \(\ell_t:=0\), and for every \(i\in[t-1]\), we define
\begin{align*}
k_{\geq i} &:= \sum_{j\in[i,t]}k_j,\\
\ell_{\geq i} &:= \sum_{j\in[i,t]}\ell_j,\\
g_{\geq i}^\pr&:=g_t+\sum_{j\in[i,t-1]}g_j^\pr.
\end{align*}
Using these definitions, it is easy to verify that each output length parameter \(m_i,i\in[t]\) takes the form
\[
m_i = k_{\geq i} -\ell_{\geq i} + g^\pr_{\geq i}.
\]
Thus, the final condenser \(\cond\) will have output length \(m_1 = k_{\geq 1} - \ell_{\geq 1} + g^\pr_{\geq 1}\), output gap \(g^\pr = g^\pr_{\geq 1}\), and error \(\eps^\pr = \sum_{i\in[t-1]}\eps_i\cdot2^{g^\pr_{\geq i+1}}\). With these observations in hand, we are ready to start setting parameters.

To start, we focus on setting the error parameters \(\eps_i\). We would like to set them so that the overall error \(\eps^\pr\) is at most some target error \(\eps\). Looking at the expression for \(\eps^\pr\) above, this can be done by setting \(\eps_i\) to a geometric series. In particular, for every \(i\in[t-1]\), we define
\[
\eps_i:=\eps\cdot2^{-(t-i)}\cdot2^{-g_{\geq i+1}^\pr}.
\]
In doing so, it is straightforward to verify that the overall error \(\eps^\pr\) is at most \(\eps\), as desired.

Next, before we set each \(\ell_i,g_i^\pr\), let's see how the setting of \(\eps_i\) affected the seed length and output length requirements of the seeded condensers. First, plugging in our value of \(\eps_i\) (and using our observation about the form of each \(m_i\)), our seed length requirement becomes the following, for every \(i\in[t-1]\):
\[
k_{\geq i+1}-\ell_{\geq i+1}\geq\log(g_i/\eps)+(t-i)+C.
\]
In fact, by incrementing the universal constant \(C\) by \(1\), it suffices to satisfy the following, for every \(i\in[t-1]\):
\begin{align}\label{eq:seed-length-requirement-multi-block-almost-done}
k_{i+1}-\ell_{i+1}\geq\log(g_i/\eps)+C.
\end{align}
Let's see how our output gap requirement changed. Plugging in our \(\eps_i\), it becomes, for every \(i\in[t-1]\):
\[
g_i^\pr\geq\frac{1}{\lfloor L_i\rfloor}(\log(1/\eps)+t-i+g^\pr_{\geq i+1})+C.
\]
Moreover, if we add \(g^\pr_{\geq i+1}\) to both sides, the output gap requirement becomes:
\begin{align}\label{eq:output-gap-requirement-multi-block-almost-done}
g^\pr_{\geq i}\geq\frac{1}{\lfloor L_i\rfloor}(\log(1/\eps)+t-i)+(1+\frac{1}{\lfloor L_i\rfloor})g^\pr_{\geq i+1}+C.
\end{align}
Finally, recall that each \(m_i=k_{\geq i}-\ell_{\geq i}+g_{\geq i}^\pr\) must be a positive integer.

Now, let's turn to setting the loss parameters \(\ell_i\). We would like to set them so that the overall loss is not too high, but also so that the output gap requirement (which depends on \(1/L_i\)) stays low. Looking ahead, the final gap \(g^\pr_{\geq 1}\) will depend roughly on the sum of the terms \(1/L_i\), and thus we set the loss parameters so \(\{1/L_i\}\) forms a geometric series. We give ourselves some freedom over the shape of this geometric series, using the parameters \(\ell\geq0\) and \(\tau\geq1\) from the theorem statement. Then, for every \(i\in[t-1]\) we define
\[
\ell_i:=\ell + \left\lfloor\frac{t-(i+1)}{\tau}\right\rfloor.
\]
\(\tau\) should be thought of as a controller for how much additional loss (between \([0,1]\)) should be experienced by each successive seeded condenser. Notice that all \(\tau>t-2\) yield an additional loss of zero.

Given this setting of loss parameters, observe that the total loss of the final condenser will be
\[
\ell^\star=\ell_{\geq 1}=\sum_{i\in[t-1]}\left(\ell+\left\lfloor\frac{t-(i+1)}{\tau}\right\rfloor\right)\leq\ell t+\left\lfloor\frac{t-2}{\tau}\right\rfloor t,
\]
as desired. Furthermore, observe that our seed length requirement (\cref{eq:seed-length-requirement-multi-block-almost-done}) is satisfied if
\[
k_{i+1}\geq\log(g_i/\eps) + \ell + \left\lfloor\frac{t-(i+1)}{\tau}\right\rfloor+C
\]
for every \(i\in[t-1]\), as provided in the theorem statement.

Thus, all that remains is to set the gap parameters \(g_i^\pr\) for all \(i\in[t-1]\). Towards this end, we pick the smallest values satisfying \cref{eq:output-gap-requirement-multi-block-almost-done}, and so that each \(m_i=k_{\geq i}-\ell_{\geq i}+g_{\geq i}^\pr\) is a positive integer. By rounding up, notice that the latter requirement can always be satisfied as long as the former requirement is satisfied with the universal constant \(C\) incremented by \(1\), and so we can safely ignore it. Thus, we henceforth focus on picking the smallest values \(g_i^\pr\) satisfying \cref{eq:output-gap-requirement-multi-block-almost-done}. That is, we define each \(g_i^\pr,i\in[t-1]\) so that
\begin{align*}
g_{\geq i}^\pr=\frac{1}{\lfloor L_i\rfloor}(\log(1/\eps)+t-i) + (1+\frac{1}{\lfloor L_i\rfloor})g_{\geq i+1}^\pr+C
\end{align*}
Then, we observe the following inequality.
\[
g_{\geq i}^\pr\leq\frac{1}{\lfloor L_i\rfloor}(\log(1/\eps)+t-1)+(1+\frac{1}{\lfloor L_i\rfloor})(g_{\geq i+1}^\pr+C)
\]

Finally, we just need to upper bound \(g^\star\leq g^\pr_{\geq 1}\). Recalling that \(g_{\geq t}^\pr=g_t\), we solve the recurrence above.
\begin{align*}
    g_{\geq1}^\pr&\leq\left(-1+\prod_{i\in[t-1]}(1+\frac{1}{\lfloor L_i\rfloor})\right)\left(\log(1/\eps)+t-1\right)+\left(\prod_{i\in[t-1]}(1+\frac{1}{\lfloor L_i\rfloor})\right)(g_t + C(t-1))\\
    &\leq\left(e^{\sum_{i\in[t-1]}\frac{1}{\lfloor L_i\rfloor}}-1\right)(\log(1/\eps)+t)+\left(e^{\sum_{i\in[t-1]}\frac{1}{\lfloor L_i\rfloor}}\right)(g_t + Ct)\\
    &\leq\left(e^{\sum_{i\in[t-1]}\frac{1}{\lfloor L_i\rfloor}}-1\right)\log(1/\eps)+\left(e^{\sum_{i\in[t-1]}\frac{1}{\lfloor L_i\rfloor}}\right)(g_t + C^\pr t),
\end{align*}
where the last step set \(C^\pr:=C+1\). Now, plugging in our parameter setting \(\ell_i:=\ell+\lfloor\frac{t-(i+1)}{\tau}\rfloor\) (and recalling the convention \(L_i=2^{\ell_i}\)), we can bound the term in the exponent as follows.
\begin{align*}
    \sum_{i\in[t-1]}\frac{1}{\lfloor L_i\rfloor}&\leq\frac{1}{\lfloor L\rfloor}\sum_{i\in[t-1]}\frac{1}{2^{\lfloor\frac{t-(i+1)}{\tau}\rfloor}}\\
    &\leq\frac{2}{\lfloor L\rfloor}\sum_{i\in[t-1]}2^{\frac{i+1-t}{\tau}}\\
    &\leq\frac{4}{\lfloor L\rfloor}\sum_{i\in[t-1]}2^{-\frac{i}{\tau}}\\
    &=\frac{4}{\lfloor L\rfloor}\cdot\frac{1-2^{-(t-1)/\tau}}{2^{1/\tau}-1}\\
    &\leq \frac{4}{\lfloor L\rfloor}\cdot\frac{\tau}{\ln 2}\\
    &\leq\frac{6\tau}{\lfloor L\rfloor}.
\end{align*}
Plugging this expression back into our bound for \(g_{\geq 1}^\pr\), we get
\[
g_{\geq1}^\pr\leq(e^{6\tau/\lfloor L\rfloor}-1)\log(1/\eps)+e^{6\tau/\lfloor L\rfloor}(g_t + C^\pr t).
\]
Now, since \(e^x-1\leq e^xx\) for all \(x\geq0\), we get
\[
g^\star\leq g_{\geq 1}^\pr \leq g_t+e^{6\tau/\lfloor L\rfloor}\cdot\frac{6\tau}{\lfloor L\rfloor}(\log(1/\eps)+g_t+C^\pr t) + C^\pr t,
\]
as desired. This completes the proof.
\end{proof}

\subsubsection*{Corollaries for Chor-Goldreich sources}

Now that we have proven our existential result for multi-block sources, we are ready to see what parameters it yields in the more well-behaved CG-source setting. We present our main existential result for multi-block CG sources, and note that \(\log^\ast()\) denotes the extremely slow-growing iterated logarithm function.

\begin{corollary}[Existential results for CG sources with many blocks]\label{cor:main-existential-many-block-CG-result}
There is a universal constant \(C\geq1\) such that the following holds. There exists a (non-explicit) condenser \(\cond:(\zo^n)^t\to\zo^m\) for \((t,n,k=:n-g)\)-CG sources with output length \(m=kt - \ell + g^\pr\), error \(\eps\), loss \(\ell\), and gap
\[
g^\pr\leq g + \exp\left(\frac{6\lceil\frac{4(\log^*t)^2}{\ell+1}\rceil}{\lfloor L^{\frac{1}{2\log^*t}}\rfloor}\right)\cdot\left(\frac{6\lceil\frac{4(\log^*t)^2}{\ell+1}\rceil}{\lfloor L^{\frac{1}{2\log^*t}}\rfloor}\right)\cdot\left(g+\log(1/\eps)+C\log^*t\right)+C\log^*t
\]
provided that \(k\geq\log(g/\eps)+\ell/\log^*t + C\).
\end{corollary}

Before we present its proof, we take some time to digest its parameters. In particular, we list two immediate corollaries, which are presented as bullet two in \cref{thm:main:intro:existential-results:CG-sources}. In the first corollary, we show what happens to the gap if one asks for a \emph{lossless} condenser for CG sources. In the second, we show that if one is willing to lose a very small amount of min-entropy, the gap can be very well maintained.

\begin{corollary}[Existential results for CG sources with many blocks - lossless regime]\label{cor:main-existential-many-block-CG-result:lossless-regime}
    There is a universal constant \(C\geq1\) such that the following holds. There exists a (non-explicit) condenser \(\cond:(\zo^n)^t\to\zo^m\) for \((t,n,k=:n-g)\)-CG sources with output length \(m=kt+g^\pr\), error \(\eps\), loss \(\ell=0\), and gap
    \[
    g^\pr\leq g+\exp(C(\log^\ast t)^2)\cdot(g+\log(1/\eps)+C\log^\ast t),
    \]
    provided that \(k\geq\log(g/\eps)+C\).
\end{corollary}

\begin{corollary}[Existential results for CG sources with many blocks - small gap regime]\label{cor:main-existential-many-block-CG-result:small-gap-regime}
    There is a universal constant \(C\geq1\) such that the following holds. There exists a (non-explicit) condenser \(\cond:(\zo^n)^t\to\zo^m\) for \((t,n,k=:n-g)\)-CG sources with output length \(m=kt-\ell+g^\pr\), error \(\eps\), loss \(\ell\leq 2(\log^*t)^2\), and gap
    \[
    g^\pr\leq g + C\cdot2^{-\log^\ast t}\cdot(g+\log(1/\eps))+C\log^\ast t,
    \]
    provided that \(k\geq\log(g/\eps)+2\log^\ast t + C\).
\end{corollary}

With these results in hand, we turn to prove \cref{cor:main-existential-many-block-CG-result}.

\begin{proof}[Proof of \cref{cor:main-existential-many-block-CG-result}]
Let \(t^\pr\in\N\) and \(b_1,\dots,b_{t^\pr}\in\N\) be parameters that we will set later, so that \(\sum_ib_i=t\). Then, define \(n_1,\dots,n_{t^\pr}\) and \(k_1,\dots,k_{t^\pr}\) such that \(n_i:=nb_i\) and \(k_i:= kb_i\). Notice that any \((t,n,k)\)-CG source is automatically an \(((n_1,k_1),\dots,(n_t,k_t))\)-block source, simply by grouping the blocks into buckets.

The goal is to find the smallest number of buckets \(t^\pr\) that we can divide the CG source into, while maintaining a relatively modest entropy requirement. In particular, recall that in order to get the strong upper bound on the final gap \(g^\pr\) provided in \cref{thm:main:technical:existential-results:CG-sources:more-than-two-blocks}, the min-entropy of the block source must satisfy
\[
k_{i+1}\geq\log(g_i/\eps)+\ell/t^\pr+C
\]
for all \(i\in[t^\pr-1]\), where \(g_i:=n_i-k_i\). Using our block parameters \(b_1,\dots,b_{t^\pr}\) and the relations described above, this min-entropy requirement becomes
\begin{align}\label{eq:min-entropy-iterated-log}
kb_{i+1}\geq\log(gb_i/\eps) + \ell/t^\pr+C,
\end{align}
for all \(i\in[t^\pr-1]\).

Now, define the parameter \(t^\pr\) and block parameters \(b_1,\dots,b_{t^\pr}\) such that the following hold:\footnote{Note that we may assume that we started off with \(t>2\) blocks, for otherwise this result holds via \cref{thm:main:technical:existential-results:CG-sources:two-blocks}.}
\begin{itemize}
    \item \(b_{t^\pr}:=2\),
    \item \(b_i\leq2^{b_{i+1}}\) for every \(i\in[t^\pr-1]\),
    \item \(b_{t^\pr}\leq b_{t^\pr-1}\leq\dots\leq b_1\),
    \item \(b_1+\dots+b_{t^\pr}=t\),
    \item \(t^\pr\in\N\) is the smallest integer for which there exist \(b_1,\dots,b_{t^\pr}\) satisfying the above constraints.
\end{itemize}
Notice that for such parameters, the min-entropy requirement (given in \cref{eq:min-entropy-iterated-log}) is satisfied if
\[
kb_{i+1}\geq\log(g/\eps)+b_{i+1}+\ell/t^\pr+C,
\]
or rather
\[
b_{i+1}(k-1)\geq\log(g/\eps)+\ell/t^\pr+C
\]
for every \(i\in[t^\pr-1]\). But observe that if we simply require
\[
k\geq\log(g/\eps)+\ell/t^\pr+C,
\]
then all of these conditions must hold, as the above implies that \((k-1)\geq k/2\) (when \(k\geq2\)), and we know from our constraints that \(b_{i+1}\geq2\).

Thus for any \((t,n,k)\)-CG source and parameters \(b_1,\dots,b_{t^\pr}\) satisfying the above constraints, we know that we can condense (with an output gap as promised in \cref{thm:main:technical:existential-results:CG-sources:more-than-two-blocks}) as long as \(k\geq\log(g/\eps)+\ell/t^\pr+C\). All that remains is to check how big \(t^\pr\) can be, and in particular provide an upper bound on it. Towards this end, looking at the constraints on our parameters \(b_i\) and the minimality of \(t^\pr\), it is straightforward to verify that \(t^\pr\) cannot exceed the iterated logarithm of \(t\). In other words, \(t^\pr\leq\log^*t\), as desired.
\end{proof}

To conclude this section, we note that one may wish for an existential result for CG sources with many blocks, where the output gap has \emph{no dependence} on the number of blocks \(t\). It is straightforward to combine the above ideas to obtain such a result, albeit with significantly more loss. In particular, one can instantiate the iterative condensing framework with optimal seeded \emph{extractors}, instead of seeded condensers, so that the output gap is \emph{exactly equal to} the input gap \(g\), but the loss becomes roughly \(O((\log^*t)(\log^*t + g + \log(1/\eps)))\), and more importantly the required starting min-entropy (per block) becomes roughly \(k\geq\log(n/\eps) + 0.99n\). This required starting min-entropy can then be reduced to \(k\geq C\log(n/\eps)\) (for some constant \(C\)) by adding in (at the beginning) a \emph{single} call to an optimal seeded condenser with seed length that has dependence \(1\log(1/\eps)\) on the error. This will not significantly affect the overall loss, and the final gap will be of the form \(g+O(1)\).

\dobib

\dobib

\section{Impossibility results}\label{sec:impossibility}

We conclude the technical portion of the paper with simple, but useful, impossibility results.

\subsection{An impossibility result for condensing general sources}

First, we show a condenser version of the classic extractor impossibility result.

\begin{theorem}[There do not exist condensers for general sources]
    Fix any function \(f:\zo^n\to\zo^m\) and gap \(g\) such that \(0\leq g\leq n\). Then for any \(0\leq\eps<1\) there exists a source \(\X\sim\zo^n\) with min-entropy gap \(g\) such that
    \[
    H_\infty^\eps(f(\X))\leq \min\{n,m\}-\min\{m,g\}+\log\left(\frac{1}{1-\eps}\right).
    \]
\end{theorem}

The term \(c_\eps:=\log(\frac{1}{1-\eps})\) is merely an artifact of the definition of smooth min-entropy (see \cref{subsec:entropy}).

\begin{proof}
Let \(g^\pr:=\min\{m,g\}\). By definition of probability, there must be a prefix \(\sigma\in\zo^{g^\pr}\) such that \(\Pr[f(\U_n)_{[g^\pr]}=\sigma]\geq2^{-g^\pr}\). Thus there is a set \(X\subseteq\zo^n\) of density exactly \(2^{-g^\pr}\) such that \(f(X)_{[g^\pr]}=\{\sigma\}\). Let \(S=f(X)\) be the image of this set, and note it has size \(|S|\leq2^{\min\{n,m\}-g^\pr}\), since \(S\) is the image of a set of size \(2^{n-g^\pr}\), and since \(S\) is a subset of \(\zo^m\) where all prefixes of length \(g^\pr\) are the same (leaving at most \(m-g^\pr\) coordinates unfixed). Now, by the characterization of smooth min-entropy (\cref{cor:characterization-corollary}),
\begin{align*}
1&=\Pr[f(\X)\in S]\\
&\leq|S|\cdot2^{-H_\infty^\eps(f(\X))}+\eps\\
&=2^{\min\{n,m\}-g^\pr-H_\infty^\eps(f(\X))}+\eps.
\end{align*}
Solving for \(H_\infty^\eps(f(\X))\) completes the proof.
\end{proof}

\subsection{An impossibility result for condensing block sources}

Finally, we extend the above argument to show that it is impossible to condense a CG source without the gap of one of the input blocks showing up in the output.

\begin{theorem}[Condensers for CG sources must maintain the gap]
Fix any function \(f:(\zo^n)^t\to\zo^m\) and gap \(g\) such that \(0\leq g\leq n\). Then for any \(0\leq\eps<1\) there exists a \((t,n,n-g)\)-CG source \(\X\sim(\zo^n)^t\) such that
\[
H_\infty^\eps(f(\X))\leq m-g+\log\left(\frac{1}{1-\eps}\right).
\]
\end{theorem}
\begin{proof}
        By induction. By the proof above, we know that for any function \(f:\zo^n\to\zo^m\) there is a set \(X\subseteq\zo^n\) of size \(2^{n-g^\pr}\) such that the \(g^\pr\)-prefix of the set \(f(X)\) is a constant \(\sigma\). Consider now a function \(f:(\zo^n)^t\to\zo^m\) and all of its restrictions \(f_\alpha:=f(\alpha,\cdot)\). By induction, for each \(\alpha\) there is a \((t-1,n,n-g^\pr)\)-CG source \(\X_\alpha\) such that the \(g^\pr\)-prefix of \(f(\alpha,\X_\alpha)\) is a constant \(\sigma\). By averaging, this constant \(\sigma\) must be the same for some \(2^{-g^\pr}\) fraction of \(\alpha\)'s. Let \(\A\) be uniform over these, and consider the \((t,n,n-g^\pr)\) source \((\A,\X_\A)\). By construction, the prefix of \(f\) is constantly \(\sigma\) on \((\A,\X_\A)\). Moreover, if we define \(S\) as the image of this source, we know it has size at most \(2^{m-g^\pr}\), since its \(g^\pr\)-prefix is fixed. We also know that it has size at most \(2^{t(n-g^\pr)}\), given the entropy of \((\A,\X_\A)\). Thus
    \begin{align*}
    1&=\Pr[f(\A,\X_\A)\in S]\\
    &\leq|S|\cdot2^{-H_\infty^\eps(f(\A,\X_\A))}+\eps\\
    &\leq2^{\min\{m-g^\pr,t(n-g^\pr)\}-H_\infty^\eps(f(\A,\X_\A))}+\eps.
    \end{align*}
    Solving for \(H_\infty^\eps(f(\A,\X_\A))\) completes the proof.
\end{proof}

\dobib

\section{Open problems}\label{sec:conclusions}

The most attractive open problem is to get better explicit seeded condensers. If one could explicitly construct such condensers with seed length that has dependence \(1\log(1/\eps)\) on the error (and a reasonably small output gap), then it would become trivial to condense CG sources with even better parameters than in this paper. Indeed, all of the work behind our CG source condensers goes into creating a single block of entropy rate \(0.99\), and any good enough seeded condenser (i.e., with the above parameters) can do this in a single step.\footnote{As a reminder, see \cref{lem:seeded-condensers-work-on-block-sources} for how the parameters of a seeded condenser translate to its performance on CG sources. It is worth noting that for this application, we would also be more than happy with a seeded condenser that is quite lossy.}

Even if such seeded condensers remain out of reach, other natural questions remain about condensing CG sources. For example, while we were able to construct explicit condensers for CG sources with very low entropy, we could only do so while blowing up the gap by a polynomial factor.\footnote{This blow-up is due to the number of rows produced by the somewhere-condensers used in our constructions.} It would be great to see if one could explicitly condense CG sources whose blocks have min-entropy (say) \(n^{0.99}\), while keeping the gap blow-up to just a constant factor. This would seem to require completely new techniques.

Finally, it would be interesting to study other natural classes of sources for which we cannot deterministically extract, but \emph{can} deterministically condense, and try to construct the corresponding explicit condensers. Chor-Goldreich sources are just one family in this new category of sources, and we hope that the study of other such families will lead to a long line of fruitful research.

\dobib

\bibliographystyle{alpha}
\bibliography{references}

\end{document}